\pdfoutput=1

\newcommand*{\ATLASLATEXPATH}{atlasPackage/latex/} 

\PassOptionsToPackage{usenames,dvipsnames}{xcolor}

\documentclass[cernpreprint,UKenglish,texlive=2011,txfonts=true,PAPER]{\ATLASLATEXPATH atlasdoc}

\usepackage[biblatex=false]{\ATLASLATEXPATH atlaspackage}

\usepackage{\ATLASLATEXPATH atlascontribute}

\usepackage{\ATLASLATEXPATH atlasphysics}

\graphicspath{{logos/}{figures/}}

\usepackage{dpsPaper-defs}

\AtlasTitle{Study of hard double-parton scattering in four-jet events in $pp$ collisions at $\sqrt{s} = 7\TeV$ with the ATLAS experiment}

\renewcommand\Authands{, } 
\renewcommand\Affilfont{\itshape\small} 

\AtlasRefCode{STDM-2015-08}

\PreprintIdNumber{CERN-EP-2016-183}

\AtlasJournal{JHEP}
\AtlasJournalRef{JHEP 11 (2016) 110}
\AtlasDOI{10.1007/JHEP11(2016)110}

\AtlasAbstract{%
    Inclusive four-jet events produced in proton--proton collisions at a centre-of-mass energy of $\sqrt{s} = 7 \tev$ are analysed for the presence of hard double-parton scattering 
    using data corresponding to an integrated luminosity of $37.3~\mathrm{pb}^{-1}$, collected with the ATLAS detector at the LHC.
    The contribution of hard double-parton scattering to the production of four-jet events is extracted using an artificial neural network,    
    assuming that hard double-parton scattering can be approximated by an uncorrelated overlaying of dijet events.
    For events containing at least four jets with transverse momentum \ptGe{20} and pseudorapidity \etaLe{4.4}, and at least one having \ptGe{42.5}, 
    the contribution of hard double-parton scattering is estimated to be $\fdps = \measFdps$.
    After combining this measurement with those of the inclusive dijet and four-jet cross-sections in the appropriate phase space regions, 
    the effective cross-section, \sigeff, was determined to be $\sigeff = \measSigeff$.
    This result is consistent within the quoted uncertainties with previous measurements of \sigeff, performed at centre-of-mass energies between 63~\gev\xspace and 8~\tev\xspace 
    using various final states, and it corresponds to \fracInelXsecPer of the total inelastic cross-section measured at $\sqrt{s} = 7 \tev$.
    The distributions of the observables sensitive to the contribution of hard double-parton scattering, corrected for detector effects, are also provided.
}
\hypersetup{breaklinks={true}, colorlinks={true}, pdfpagemode={UseNone}, pdffitwindow=true, pdfstartview={FitB}, linkcolor={BlueViolet}, menucolor={BlueViolet}, citecolor={Blue}, filecolor={OliveGreen}, urlcolor={MidnightBlue}, pdftitle={Hard double-parton scattering in four-jet events with the ATLAS experiment}, pdfauthor={The ATLAS Collaboration}, pdfsubject={Double-Parton Scattering}, pdfkeywords={ATLAS,Jets,DPS,DPI,MPI,LHC}}

\begin{document}

\maketitle

\clearpage


\section{Introduction\label{sec:intro}}

Interactions involving more than one pair of incident partons in the same collision have been discussed on theoretical grounds since the introduction of the parton model 
to the description of particle production in hadron--hadron collisions~\cite{PhysRevD.18.3344,1979PhRvL..43.1296T,PhysRevD.22.2789}.  
These first studies were followed by the generalization of the Altarelli--\-Parisi evolution equations to the case of multi-parton states in Refs.~\cite{Kirschner1979266,Shelest:1982dg} 
and a discussion of possible correlations in the colour and spin degrees of freedom of the incident partons~\cite{Mekhfi:1985dv}.
In the first phenomenological studies of such effects, the most prominent role was played by processes known as \DPS (\dps), 
which is the simplest case of multi-parton interactions (MPI),
leading to final states such as four leptons, four jets, three jets plus a photon, or a leptonically decaying gauge boson accompanied by two jets~\cite{
  Paver:1982yp,Paver:1983hi,Mekhfi:1983az,Humpert:1983pw,Humpert:1983fy,
  Humpert:1984ay,Ametller:1985tp,Halzen:1986ue,Godbole:1989ti}.
These studies have been supplemented by experimental measurements of \dps effects in hadron collisions at different centre-of-mass energies, 
which now range over two orders of magnitude, from 63~\gev\xspace to 8~\tev~\cite{
  Akesson:173908,Alitti1991145,PhysRevD.47.4857,Abe:1997bp,PhysRevD.56.3811,
  Abazov:2009gc,Aaij:2012dz,Aad:2013bjm,Chatrchyan:2013xxa,Abazov:2014fha,Abazov:2014qba,Aad:2014kba,Aaij:2015wpa,Abazov:2015fbl,Abazov:2015nnn},
and which have firmly established the existence of this mechanism.
The abundance of MPI phenomena at the LHC and their importance for the full picture of hadronic collisions have reignited the phenomenological interest in \dps 
and have led to a deepening of its theoretical understanding~\cite{
  Blok:2010ge,Gaunt:2011xd,Blok:2011bu,Diehl:2011yj,Kasemets:2012pr,
  Blok:2013bpa,Diehl:2014vaa,Gaunt:2014ska,Gaunt:2014rua}.
Despite this progress, 
quantitative measurements of the effect of \dps on distributions of observables sensitive to it are affected by large systematic uncertainties. 
This is a clear indication of the experimental challenges and of the complexity of the analysis related to such measurements.
Therefore, the cross-section of \dps continues to be estimated by ignoring the likely existence of complicated correlation effects.  
For a process in which a final state $\text{A}+\text{B}$ is produced at a hadronic centre-of-mass energy $\sqrt{s}$, 
the simplified formalism of Refs.~\cite{Humpert:1984ay,Ametller:1985tp} yields
\begin{equation}
  \done\hat{\sigma}^{(\dps)}_{\!\text{A}+\text{B}}(s)\,=\,
  \frac{1}{1 + \delta_{\!\text{AB}}}\,
  \frac{\done\hat{\sigma}_{\!\text{A}}(s)\done\hat{\sigma}_{\text{B}}(s)}{\sigeff(s)}\,.
\label{eq:dpsXsecDiff}
\end{equation}
The quantity $\delta_{\!\text{AB}}$ is the Kronecker delta used to construct a symmetry factor such that for identical final states with identical phase space, 
the \dps cross-section is divided by two.
The \sigeff, usually referred to as the effective cross-section, 
is a purely phenomenological parameter describing the effective overlap of the spatial distribution of partons in the plane perpendicular to the direction of motion.
In hadronic collisions it was typically found to range between 10 and 25 mb~\cite{
  Akesson:173908,Alitti1991145,PhysRevD.47.4857,Abe:1997bp,PhysRevD.56.3811,
  Abazov:2009gc,Aaij:2012dz,Aad:2013bjm,Chatrchyan:2013xxa,Abazov:2014fha,Abazov:2014qba,Aad:2014kba,Aaij:2015wpa,Abazov:2015fbl,Abazov:2015nnn}.
In \autoref{eq:dpsXsecDiff}, the various $\hat{\sigma}$ are the parton-level cross-sections, either for the \dps events, indicated by the subscript $\text{A}+\text{B}$, 
or for the production of a final state $\text{A}$ or $\text{B}$ in a single parton scatter (\sps), given by
\begin{equation}
  \done\hat{\sigma}_{\!\text{A}}(s)\,=\,
  \frac{1}{2s}\sum\limits_{ij}
  \int\done x_1 \done x_2 \, f_i(x_1, \mu_{\mathrm{F}}) \, f_j(x_2, \mu_{\mathrm{F}}) \,
  \done\Phi_{\!\text{A}} \, |\mathcal{M}_{ij \to \text{A}}(x_1x_2s, \mu_{\mathrm{F}}, \mu_{\mathrm{R}})|^2\,.
\label{eq:spsXsec}
\end{equation}
Here the functions $f_i(x,\,\mu_{\mathrm{F}})$ are the single parton distribution functions (PDFs) which at leading order parameterize the probability 
of finding a parton $i$ at a momentum fraction $x$ at a given factorization scale $\mu_{\mathrm{F}}$ in the incident hadron; 
$\done\Phi_\text{A}$ is the invariant differential phase-space element for the final state $\text{A}$;
$\mathcal{M}$ is the perturbative matrix element for the process $ij\to \text{A}$;
and $\mu_{\mathrm{R}}$ is the renormalization scale at which the couplings are evaluated.
To constrain the phase space to that allowed by the energy of each incoming proton, a simple two-parton PDF is defined as
\begin{equation}
    f_{ij}(\vect{b},x_i,\,x_j,\,\mu_{\mathrm{F}})\,=\,
    \Gamma(\vect{b}) \, f_{i}(x_i,\mu_{\mathrm{F}}) \, f_{j}(x_j,\mu_{\mathrm{F}}) \, \Theta(1-x_i-x_j)\,,
\label{eq:pdfAnsatz}
\end{equation}
where $\Theta(x)$ is the Heaviside step function, 
$\Gamma(\vect{b})$ the area overlap function, and the $x$ and scale dependence of the PDF are assumed to be independent of the impact parameter $\vect{b}$.
\Autoref{eq:pdfAnsatz} reflects the omission of correlations between the partons in the proton.
At high energy, \autoref{eq:dpsXsecDiff} can be derived using \autoref{eq:pdfAnsatz} by neglecting the contribution of the step function.

Typically,
the main challenge in measurements of \dps is to determine if the $\text{A}+\text{B}$ final state was produced in an \sps via the $2 \to 4$ process 
or in \dps through two independent $2 \to 2$ interactions.
In one of the first studies of \dps in four-jet production at hadron colliders~\cite{Humpert:1983pw} 
the kinematic configuration in which there is a pairwise balance of the transverse momenta (\pt) of the jets was identified as increasing the contribution of the \dps mechanism 
relative to the perturbative QCD production of four jets in \sps.
The idea is that in typical $2 \to 2$ scattering processes the two outgoing particles -- here the partons identified as jets -- 
are oriented back-to-back in transverse plane such that their net transverse momentum is zero.
Corrections to this simple picture include initial- and final-state radiation as well as fragmentation and hadronization.
In addition, recoil against the underlying event can modify the four-momentum of the overall final-state particle configuration. 
In attempting to describe all of these features, Monte Carlo (MC) event generators form an integral part, 
providing a link between the experimentally observed jets and the simple partonic picture of \dps as two almost independent $2 \to 2$ scatters.

An analysis of inclusive four-jet events produced in proton--proton collisions at a centre-of-mass energy of $\sqrt{s} = 7 \tev$ at the LHC 
and collected during 2010 with the ATLAS detector is presented here.
The topology of the four jets is exploited to construct observables sensitive to the \dps contribution.
The \dps contribution to the four-jet final state is estimated 
and combined with the measured inclusive dijet and four-jet cross-sections in the appropriate phase space regions to determine \sigeff.
The normalized differential four-jet cross-sections as a function of \dps-sensitive observables are measured and presented here as well.

\section{Analysis strategy\label{sec:strategy}}

To extract \sigeff in the four-jet final state, \autoref{eq:dpsXsecDiff} is rearranged as follows. 
The differential cross-sections in \autoref{eq:dpsXsecDiff} are rewritten for the four-jet and dijet final states 
and integrated over the phase space defined by the selection requirements of the dijet phase space regions $\text{A}$ and $\text{B}$.
This yields the following expression for the \dps cross-section in the four-jet final state:
\begin{equation}
    \sigma^{\mathrm{\dps}}_{\text{4j}} = 
    \frac{1}{1 + \delta_{\!\text{AB}}}\frac{ \sigma_{\text{2j}}^{\text{A}} \sigma_{\text{2j}}^{\text{B}} }{ \sigeff } \, ,
\label{eq:dpsXsecFourJets}
\end{equation}
where $\sigma_{\text{2j}}^{\text{A}}$ and $\sigma_{\text{2j}}^{\text{B}}$ are the cross-sections for dijet events in the phase space regions labelled $\text{A}$ and $\text{B}$ respectively.
The assumed dependence of the cross-sections and \sigeff on $s$ is omitted for simplicity.
The \dps cross-section may be expressed as
\begin{equation}
    \sigma^{\mathrm{\dps}}_{\text{4j}} = \fdps \cdot \sigma_{\text{4j}} \, ,
    \label{eq:fdpsDefinition}
\end{equation}
where $\sigma_{\text{4j}}$ is the inclusive cross-section for four-jet events in the phase-space region $\text{A} \oplus \text{B}$, 
including all four-jet final states, namely both the \sps and \dps topologies, and where \fdps represents the fraction of \dps events in these four-jet final states.
The expression for \sigeff then becomes,
\begin{equation}
    \sigeff = 
    \frac{1}{1 + \delta_{\!\text{AB}}} \frac{1}{\fdps} \frac{ \sigma_{\text{2j}}^{\text{A}} \sigma_{\text{2j}}^{\text{B}} }{ \sigma_{\text{4j}} } \, .
\label{eq:sigEffFourJet}
\end{equation}
To extract \sigeff, it is therefore necessary to measure three cross-sections, $\sigma_{\text{2j}}^{\text{A}}$, $\sigma_{\text{2j}}^{\text{B}}$ and $\sigma_{\text{4j}}$, 
and estimate \fdps.

The four-jet and dijet final states are defined inclusively~\cite{Treleani:2007gi,Bahr:2013gkj} such that at least four jets or two jets respectively are required in the event, 
while no restrictions are applied to additional jets.
When measuring the cross-section of $n$-jet events, the leading (highest-\pt) $n$ jets in the event are considered.
The general expression for the measured four-jet and dijet cross-sections may be written as
\begin{equation}
    \sigma_{n\text{j}} = \frac{N_{n\text{j}}}{ \mathcal{C}_{n\text{j}} \mathcal{L}_{n\text{j}} } \;,
\label{eq:measuredXsec}
\end{equation}
where the subscript $n\text{j}$ denotes either dijet ($\text{2j}$) or four-jet ($\text{4j}$) topologies. 
For each $n\text{j}$ channel, $N_{n\text{j}}$ is the number of observed events, 
$\mathcal{C}_{n\text{j}}$ is the correction for detector effects, particularly due to the jet energy scale and resolution, 
and $\mathcal{L}_{n\text{j}}$ is the corresponding proton--proton integrated luminosity.

The \dps model contributes in two ways to the production of events with at least four jets, leading to two separate event classifications. 
In one contribution, the secondary scatter produces two of the four leading jets in the event; such events are classified as \cDPS (\cDps).  
In the second contribution of \dps to four-jet production, three of the four leading jets are produced in the hardest scatter, and the fourth jet is produced in the secondary scatter; 
such events are classified as \sDPS (\sDps).
The \dps fraction is therefore rewritten as $\fdps = \fcDps + \fsDps$, and \fcDps and \fsDps\hspace{-0.15em} are both determined from data.
The dijet cross-sections in \autoref{eq:sigEffFourJet} do not require any modification since they are all inclusive cross-sections, 
\ie, the three-jet cross-section accounting for the production of an \sDps event is already included in the dijet cross-sections.

Denoting the observed cross-section at the detector level by
\begin{equation}
    \mathcal{S}_{n\text{j}} = \frac{ N_{n\text{j}} }{ \mathcal{L}_{n\text{j}} } \, ,
\label{eq:observedXsec}
\end{equation}
and the ratio of the corrections for detector effects by 
\begin{equation}
    \alpha_{\text{2j}}^{\text{4j}} = \frac{ \mathcal{C}_{\text{4j}} }{ \mathcal{C}_{\text{2j}}^{\text{A}} \mathcal{C}_{\text{2j}}^{\text{B}} } \, ,
\label{eq:alphaDef}
\end{equation}
yields the expression from which \sigeff is determined,
\begin{equation}
    \sigeff = \frac{1}{1 + \delta_{\!\text{AB}}} \frac{\alpha_{\text{2j}}^{\text{4j}}}{\fcDps + \fsDps}
    \frac{ \mathcal{S}_{\text{2j}}^{\text{A}} \mathcal{S}_{\text{2j}}^{\text{B}} }{ \mathcal{S}_{\text{4j}} } \, .
\label{eq:reducedSigEff}
\end{equation}
The main challenge of the measurement is the extraction of $\fdps = \fcDps + \fsDps$ from optimally selected measured observables.
An artificial neural network (NN) is used for the classification of events~\cite{Michie94}, using as input various observables sensitive to the contribution of \dps.
The differential distributions of these observables are also presented here.


\section{The ATLAS detector\label{sec:ATLASdetector}}

The ATLAS detector is described in detail in Ref.~\cite{Aad:2008zzm}. 
In this analysis, the tracking detectors are used to define candidate collision events by constructing vertices from tracks, and the calorimeters are used to reconstruct jets.

The inner detector used for tracking and particle identification has complete azimuthal coverage and spans the pseudorapidity region $|\eta|<2.5$.%
\footnote{ATLAS uses a right-handed coordinate system with its origin at the nominal interaction point (IP) in the centre of the detector and the $z$-axis along the beam pipe. 
    The $x$-axis points from the IP to the centre of the LHC ring, and the $y$-axis points upward. 
    Cylindrical coordinates $(r,\phi)$ are used in the transverse plane, $\phi$ being the azimuthal angle around the beam pipe, referred to the $x$-axis. 
    The pseudorapidity is defined in terms of the polar angle $\theta$ with respect to the beamline as $\eta=-\ln\tan(\theta/2)$.  
    When dealing with massive jets and particles, the rapidity $\y = \frac{1}{2} \ln \left( \frac{E + p_z}{E - p_z} \right)$ is used, 
    where $E$ is the jet energy and $p_{z}$ is the $z$-component of the jet momentum.}
It consists of layers of silicon pixel detectors, silicon microstrip detectors, and transition-radiation tracking detectors, 
surrounded by a solenoid magnet that provides a uniform axial field of 2~T.

The electromagnetic calorimetry is provided by the liquid argon (LAr) calorimeters that are split into three regions: 
the barrel ($|\eta|<1.475$) and the endcap ($1.375<|\eta|<3.2$) regions which consist LAr/Pb calorimeter modules, and the forward (FCal: $3.1<|\eta|<4.9$) region which utilizes LAr/Cu modules. 
The hadronic calorimeter is divided into four distinct regions: the barrel ($|\eta|<0.8$), the extended barrel ($0.8<|\eta|<1.7$), both of which are scintillator/steel sampling calorimeters, 
the hadronic endcap ($1.5<|\eta|<3.2$), which has LAr/Cu calorimeter modules, and the hadronic FCal (same $\eta$-range as for the EM-FCal) which uses LAr/W modules.  
The calorimeter covers the range $|\eta|<4.9$.

The trigger system for the ATLAS detector consists of a hardware-based level-1 trigger (L1) and the software-based high-level trigger (HLT)~\cite{Aad:2012xs}. 
Jets are first identified at L1 using a sliding-window algorithm from coarse granularity calorimeter towers.
This is refined using jets reconstructed from calorimeter cells in the HLT.
Three different triggers are used to select events for this measurement:
the minimum-bias trigger scintillators, the central jet trigger ($\left|\eta\right| < 3.2$) and the forward jet trigger ($3.1 < \left|\eta\right| < 4.9$).
The jet triggers require at least one jet in the event.


\section{Monte Carlo simulation\label{sec:MC}}

Multi-jet events were generated using fixed-order QCD matrix elements ($2 \to n$, with $n = 2,3,4,5,6$) 
with \alpgen~2.14~\cite{Mangano:2002ea} utilizing the CTEQ6L1 PDF set~\cite{Pumplin:2002vw}, 
interfaced to \jimmy~\cite{Butterworth:1996zw} and \herwig~6.520~\cite{Corcella:2000bw}. 
The events were generated using the AUET2~\cite{ATL-PHYS-PUB-2011-008} set of parameters (tune), 
optimized to describe underlying-event distributions obtained from a subsample of the 2010, 7~\tev\xspace ATLAS data as well as from the Tevatron and LEP experiments.
The MLM~\cite{Mangano:2001xp} matching scale, 
which divides the parton emission phase space into regions modelled either by the perturbative matrix-element calculation or by the shower resummation, was set to~15 \gev\xspace.
The implication of this choice is that partons with \ptG{15} in the final state originate from matrix elements, and not from the parton shower.
Event-record information was used to extract a sample of \sps candidate events from the sample generated with the \alpgen + \herwig + \jimmy MC combination (\ahj).
A sample of candidate \dps events was also extracted from \ahj in order to study the topology of such events and validate the measurement methodology.

An additional \ahj sample was available that differed only in its use of the earlier AUET1~\cite{ATLAS:2010osr} tune.
Because this sample contained three times as many events, it was used to derive the corrections for detector effects in all differential distributions in the data.

Tree-level matrix elements with up to five outgoing partons were used to generate a sample of multi-jet events without multi-parton interactions using 
\sherpa~1.4.2~\cite{Krauss:2001iv,Gleisberg:2008ta} 
with the CT10 PDF set~\cite{Lai:2010vv} and the default \sherpa tune.
The CKKW~\cite{Catani:2001cc,Krauss:2002up} matching scale, similarly to the MLM one, was set to 15 \gev\xspace.
This \sps sample was compared to the \sps sample extracted from the \ahj sample for validation purposes.

In addition, a sample of multi-jet events was generated with \pythia~6.425~\cite{Sjostrand:2006za} 
using a $2 \to 2$ matrix element at leading order with additional radiation modelled in the leading-logarithmic approximation by \pt-ordered parton showers. 
The sample was generated utilizing the modified leading-order PDF set MRST~LO*~\cite{Sherstnev:2007nd} with the AMBT1~\cite{Aad:2010ac} tune.

To account for the effects of multiple proton--proton interactions in the LHC (pile-up), 
the multi-jet events were overlaid with inelastic soft QCD events generated with \pythia~6.423 using the MRST~LO* PDF set with the AMBT1 tune.
All the events were processed through the ATLAS detector simulation framework~\cite{Aad:2010ah}, which is based on \geant{4}~\cite{Agostinelli:2002hh}. 
They were then reconstructed and analysed by the same program chain used for the data.


\section{Cross-section measurements\label{sec:xsec}}

\subsection{Data set and event selection}

The measurement presented here is based on the full ATLAS 2010 data sample from proton--proton collisions at \sqs.
The trigger conditions evolved during the year with changing thresholds and prescales.
A full description of the trigger strategy, developed and used for the measurement of the dijet cross-section using 2010 data, is given in Ref.~\cite{Aad:2011fc}.
For the events in the samples used in this study, the trigger was fully efficient.
In total, the data used correspond to a luminosity of $37.3~\mathrm{pb}^{-1}$, with a systematic uncertainty of 3.5\%~\cite{Aad:2013ucp}. 
This data set was chosen because it has a low number of proton--proton interactions per bunch crossing, averaging to approximately~0.4.
It was therefore possible to collect multi-jet events with low \pt thresholds 
and to efficiently select events with exactly one reconstructed vertex (single-vertex events),
thereby removing any contribution from pile-up collisions to the four-jet final-state topologies.

To reject events initiated by cosmic-ray muons and other non-collision backgrounds, events were required to have at least one reconstructed primary vertex, 
defined as a vertex that is consistent with the beam spot and is associated with at least five tracks with transverse momentum $\pt^{\mathrm{track}} > 150 \mev$.
The efficiency for collision events to pass these requirements was over 99\%, while the contribution from fake vertices was negligible~\cite{Aad:2011fc,Aad:2011he}.

Jets were identified using the \AKT jet algorithm~\cite{Cacciari:2008gp}, implemented in the \fastjet~\cite{Cacciari:2005hq} package, with radius parameter $R = 0.6$.
The inputs to jet reconstruction are the energies in three-dimensional topological clusters~\cite{EndcapTBelectronPion2002,TopoClusters} built from calorimeter cells, 
calibrated at the electromagnetic (EM) scale.%
\footnote{The electromagnetic scale is the basic calorimeter signal scale to which the ATLAS calorimeters are calibrated. 
It was established using test-beam measurements for electrons and muons to give the correct response for the energy deposited by electromagnetic showers, 
while it does not correct for the lower response to hadrons.}
A jet energy calibration was subsequently applied at the jet level, relating the jet energy measured with the ATLAS calorimeter to the true energy of the stable particles entering the detector.
A full description of the jet energy calibration is given in Ref.~\cite{Aad:2011he}.
For the MC samples, particle jets were built from particles with a lifetime longer than 30~ps in the Monte Carlo event record, excluding muons and neutrinos.

For the purpose of measuring \sigeff in the four-jet final state, three samples of events were selected, two dijet samples and one four-jet sample. 
The former two samples have at least two, and the latter at least four, jets in the final state, where each jet was required to have \ptGe{20} and \etaLe{4.4}.
In each event, jets were sorted in decreasing order of their transverse momenta.
The transverse momentum of the $i^{\, \mathrm{th}}$ jet is denoted by $\pt^{i}$ and the jet with the highest \pt ($\pt^{1}$) is referred to as the leading jet.
To ensure 100\% trigger efficiency, the leading jet in four-jet events was required to have $\pt^{1} \ge 42.5$ \gev\xspace.

The selection requirements for the dijet samples were dictated by those used to select four-jet events.
In one class of dijet events, 
the requirement on the transverse momentum of the leading jet must be equivalent to the requirement on the leading jet in four-jet events, $\pt^{1} \ge 42.5$ \gev\xspace.
The other type of dijet event corresponds to the sub-leading pair of jets in the four-jet event, with a requirement of \ptGe{20}.
In the following, the cross-section for dijets selected with $\pt^{1} \ge 20$ \gev\xspace is denoted by $\sigma_{\text{2j}}^{\text{A}}$ 
and the cross-section for dijets with $\pt^{1} \ge 42.5$ \gev\xspace is denoted by $\sigma_{\text{2j}}^{\text{B}}$.

\begin{figure}[htbp]
\begin{center}
\subfloat[]{\label{fig:kinematics:pt}\includegraphics[trim=0mm 0mm 0mm 10mm,clip,width=0.5\textwidth]{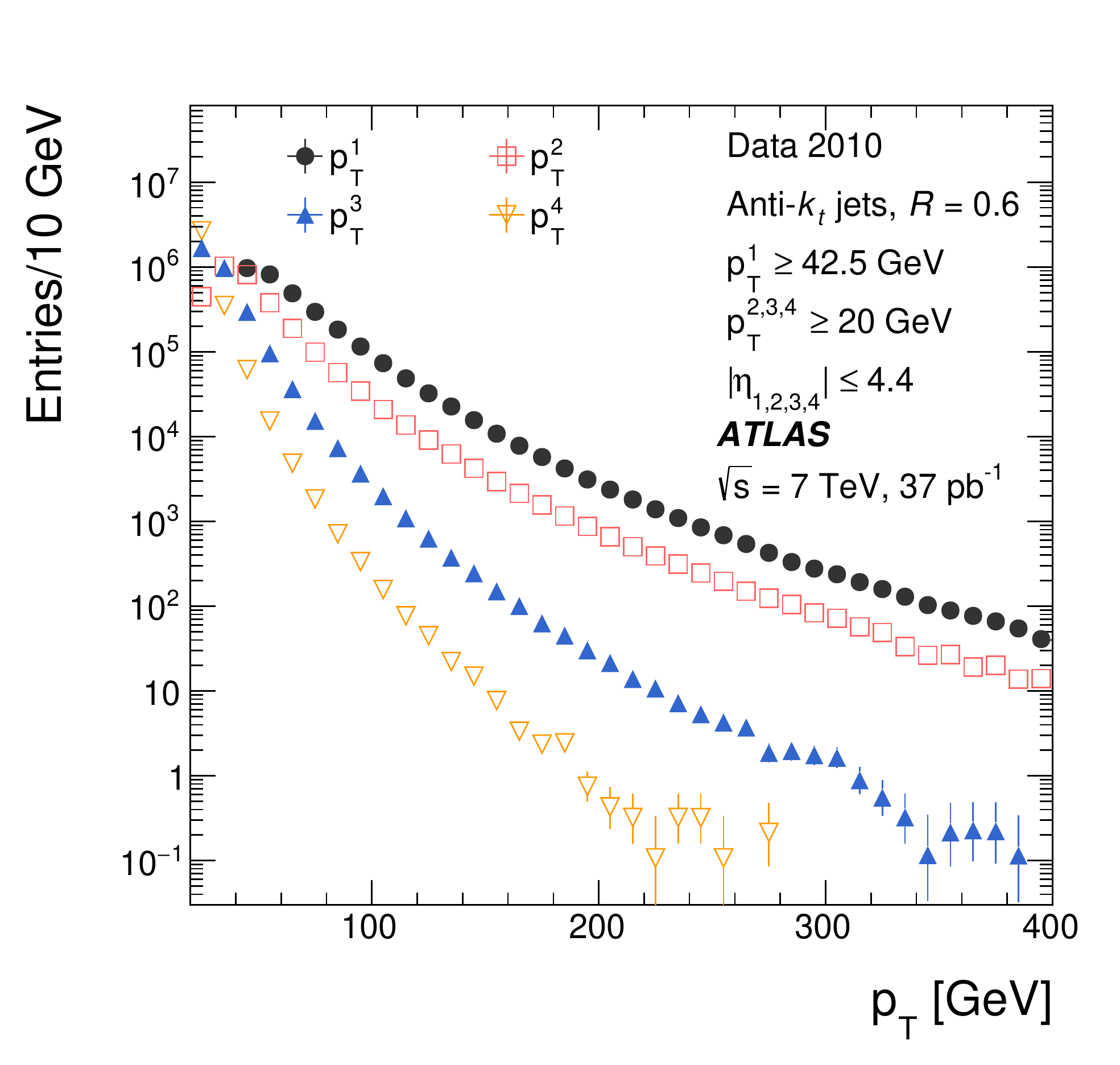}}
\subfloat[]{\label{fig:kinematics:y}\includegraphics[trim=0mm 0mm 0mm 10mm,clip,width=0.5\textwidth]{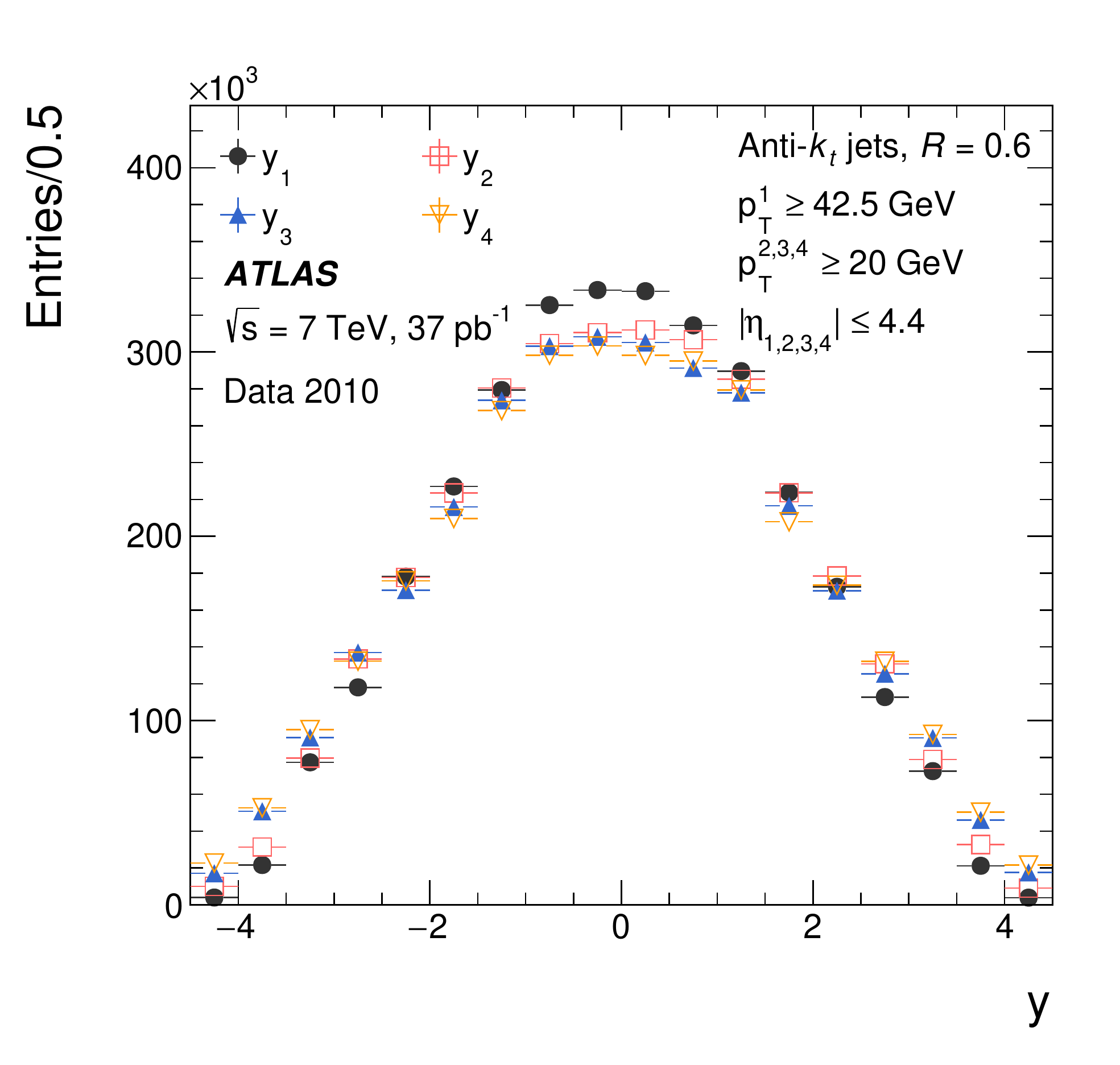}}
\caption{\label{fig:kinematics}
    Distributions of the \Subref{fig:kinematics:pt} transverse momentum, \pt, and \Subref{fig:kinematics:y} rapidity, \y, 
    of the four highest-\pt jets, denoted by $\pt^{1,2,3,4}$ and $\y_{1,2,3,4}$, in four-jet events in data selected in the phase space as defined in the legend.
}
\end{center}
\end{figure}

To summarize, the measurement was performed using the dijet $\text{A}$ sample and its two subsamples (dijet $\text{B}$ and four-jet), selected using the following requirements:
\begin{equation}
  \renewcommand{\arraystretch}{1.75}
  \begin{array}{cccccc}
      \text{Dijet} \, \text{A:}  & N_{\mathrm{jet}} \ge 2 \;,& \pt^{1} \ge 20 \gev   \;, & \pt^{2} \ge 20 \gev     \;, & |\eta_{1,2}| \le 4.4 \;, \\
      \text{Dijet} \, \text{B:}  & N_{\mathrm{jet}} \ge 2 \;,& \pt^{1} \ge 42.5 \gev \;, & \pt^{2} \ge 20 \gev     \;, & |\eta_{1,2}| \le 4.4 \;, \\
      \text{Four-jet:}           & N_{\mathrm{jet}} \ge 4 \;,& \pt^{1} \ge 42.5 \gev \;, & \pt^{2,3,4} \ge 20\GeV  \;, & |\eta_{1,2,3,4}| \le 4.4 \;,
  \end{array}
\label{eq:eventSelection}
\end{equation}
where $N_{\mathrm{jet}}$ denotes the number of reconstructed jets.
All of the selected events were corrected for jet reconstruction and trigger inefficiencies, the corrections ranging from 2\%--4\% for low-\pt jets to less than~1\% for jets with \ptGe{60}.
The observed distributions of the \pt and \y of the four leading jets in the events are shown in \autorefs{fig:kinematics:pt}~and~\ref{fig:kinematics:y} respectively.

\subsection{Correction for detector effects\label{sec:Acceptance}}

\enlargethispage{1\baselineskip}

The correction for detector effects was estimated separately for each class of events using the \pythia{6} MC sample.
The same restrictions on the phase space of reconstructed jets, defined in \autoref{eq:eventSelection}, were applied to particle jets.
The correction is given by
\begin{equation}
    \mathcal{C}_{n\text{j}}^{\text{A,B}} = \frac{ N_{n\text{j}}^{\text{A,B}\;\mathrm{reco}} }{ N_{n\text{j}}^{\text{A,B}\;\mathrm{particle}} } \;,
\label{eq:acceptanceDefinition}
\end{equation}
where $N_{n\text{j}}^{\text{A,B}\;\mathrm{reco}}$ ($N_{n\text{j}}^{\text{A,B}\;\mathrm{particle}}$) is the number of $n$-jet events passing the 
$\text{A}$-or-$\text{B}$ selection requirements using reconstructed (particle) jets.

This correction is sensitive to the migration of events into and out of the phase space of the measurement.
Due to the very steep jet-\pt spectrum in dijet and four-jet events, 
it is crucial to have good agreement between the jet \pt spectra in data and in MC simulation close to the selection threshold before calculating the correction.
Therefore, the jet \pt threshold was lowered to 10~\gev\xspace and the fiducial $|\eta|$ range was increased to 4.5 for both the reconstructed and particle jets, 
and the MC events were reweighted such that the jet \pt--\y distributions reproduced those measured in data.
The value of $\alpha_{\text{2j}}^{\text{4j}}$ (see \autoref{eq:alphaDef}), as determined from the reweighted MC events, is
\begin{equation}
    \alpha_{\text{2j}}^{\text{4j}} = 0.93 \pm 0.01 \stat \;,
\label{eq:alpha}
\end{equation}
where the uncertainty is statistical. The systematic uncertainties are discussed in \autoref{sec:uncertainties}.


\section{Determination of the fraction of \dps events\label{sec:methodology}}

The main challenge in the measurement of \sigeff is to estimate the \dps contribution to the four-jet data sample.
It is impossible to extract \cDps and \sDps candidate events on an event-by-event basis. 
Therefore, the usual approach adopted is to fit the distributions of variables sensitive to \cDps and \sDps in the data 
to a combination of templates for the expected \sps, \cDps and \sDps contributions.
The template for the \sps contribution is extracted from the \ahj MC sample, while the \cDps and \sDps templates are obtained by overlaying two events from the data.
In addition to assuming that the two interactions producing the four-jet final state in a \dps event are kinematically decoupled, 
the analysis relies on the assumption that the \sps template from \ahj properly describes the expected topology of four-jet production in a single interaction.
The latter assumption is supported by the observation of good agreement between various distributions in the \sps samples in \ahj and in \sherpa.
To exploit the full spectrum of variables sensitive to the various contributions and their correlations, the classification was performed with an artificial neural network.

\subsection{Template samples\label{sec:templateSamples}}

Differences were observed when comparing the \pt and \y distributions in data with those in \ahj.
Therefore, before extracting template samples, 
the events in the four-jet \ahj sample selected with the requirements detailed in \autoref{eq:eventSelection} are reweighted such that they reproduce the distributions in data.

In events generated in \ahj, the outgoing partons can be assigned to the primary interaction from the \alpgen generator or to a secondary interaction, 
generated by \jimmy, based on the MC generator's event record.
The former are referred to as primary-scatter partons and the latter as secondary-scatter partons.
The \pt of secondary-scatter partons was required to be \ptGe{15} in order to match the minimum \pt of primary-scatter partons set by the MLM matching scale in \ahj.
Once the outgoing partons were classified, the jets in the event were matched to outgoing partons and the event was classified as an \sps, \cDps or \sDps event.

The matching of jets to partons is done in the $\phi$--$\y$ plane by calculating the angular distance, $\Delta R_{\mathrm{parton-jet}}$, between the jet and the outgoing parton as
\begin{equation}
    \Delta R_{\mathrm{parton-jet}} = \sqrt{ (\y_{\mathrm{parton}} - \y_{\mathrm{jet}})^2 + (\phi_{\mathrm{parton}} - \phi_{\mathrm{jet}})^2 } \, .
\label{eq:deltaRpartonJet}
\end{equation}
For~99\% of the primary-scatter partons, the parton can be matched to a jet within $\Delta R_{\mathrm{parton-jet}} \leq 1.0$,
which was therefore used as a requirement for the matching of jets and partons.
Jets were first matched to primary-scatter partons and the remaining jets were matched to secondary-scatter partons.

Events in which none of the leading four jets match a secondary-scatter parton were assigned to the \sps sample.
This method of obtaining an \sps sample is preferred over turning off the MPI module in the generator since it retains all of the soft MPI and underlying activity in the selected \sps events.
Events were classified as \cDps events if two of the four leading jets match primary-scatter partons and the other two match secondary-scatter partons. 
Events in which three of the leading jets match primary-scatter partons and the fourth jet matches a secondary-scatter parton were classified as \sDps events.

Four-jet \dps events were modelled by overlaying two different events.
To reduce any dependence of the measurement on the modelling of jet production, this construction used events from data rather than MC simulation.
Complete-\dps events were built using dijet events from the $\text{A}$~and~$\text{B}$ samples selected from data (see \autoref{eq:eventSelection}).
To build \sDps events, two other samples were selected with the following requirements:
\begin{equation}
  \renewcommand{\arraystretch}{1.75}
  \begin{array}{cccccc}
      \text{One-jet:}   & N_{\mathrm{jet}} \ge 1 \;,& \multicolumn{2}{c}{ \pt^{1} \ge 20 \gev \;, }       & |\eta_{1}| \le 4.4 \;, \\
      \text{Three-jet:} & N_{\mathrm{jet}} \ge 3 \;,& \pt^{1} \ge 42.5 \gev \;, & \pt^{2,3} \ge 20 \gev   \;, & |\eta_{1,2,3}| \le 4.4 \;.
  \end{array}
\label{eq:sDpsSamples}
\end{equation}
The overlay was performed at the reconstructed jet level. 
When constructing \cDps and \sDps events the following conditions were imposed for a given pair of events to be overlaid:
\begin{itemize}
    \item none of the four jets contains the axis of one of the other jets, \ie, $\Delta R_{\,\mathrm{jet-jet}} > 0.6$;
    \item the vertices of the two overlaid events are no more than 10~mm apart in the $z$ direction;
    \item when building \cDps events, each of the overlaid events contributes two jets to the four leading jets in the constructed event;
    \item when building \sDps events, one of the overlaid events contributes three jets to the four leading jets in the constructed event and the other contributes one jet.
\end{itemize}
The first condition ensures that none of the jets would be merged if the four-jet event had been reconstructed as a real event;
the second condition avoids possible kinematic bias due to events where two jet pairs originate from far-away vertices;
the last two conditions enforce the appropriate composition of the four leading jets in the constructed event.

As is discussed in \autoref{sec:validation}, 
the topology of \cDps and \sDps events constructed by overlaying two events is compared to the topology of \cDps and \sDps events extracted from the \ahj sample respectively.

\subsection{Kinematic characteristics of event classes}

In \cDps, double dijet production should result in pairwise \pt-balanced jets with a distance $\left|\phi_{1}-\phi_{2}\right| \approx \pi$ between the jets in each pair.
In addition, the azimuthal angle between the two planes of interactions is expected to have a uniform random distribution.
In \sps, the pairwise \pt balancing of jets is not as likely; therefore the topology of the four jets is expected to be different for \cDps and \sps.

The topology of three of the jets in \sDps events would resemble the topology of the jets in \sps interactions.
The fourth jet initiated by the primary interaction in an \sps is expected to be closer, in the $\phi$--$\y$ plane, to the other three jets originating from that interaction.
In an \sDps event, the jet produced in the secondary interaction would be emitted in a random direction relative to the other three jets.

In constructing possible differentiating variables, three guiding principles were followed:
\begin{enumerate}
    \item use pairwise relations that have the potential to differentiate between \sps and \cDps topologies;
    \item include angular relations between all jets in light of the expected topology of \sDps events;
    \item attempt to construct variables least sensitive to systematic uncertainties.
\end{enumerate}
The first two guidelines encapsulate the different characteristics of \sps and \dps events.
The third guideline led to the usage of ratios of \pt in order to avoid large dependencies on the jet energy scale (JES) uncertainty.
Various studies, including the use of a principal component analysis~\cite{PCA}, led to the following list of candidate variables for distinguishing event topologies:
\begin{equation}
\renewcommand{\arraystretch}{1.5} 
\begin{array}{ccccccc}
    \dpt{ij} = \dfrac{\left|\vec{p}_{\text{T}}^{\,i}+\vec{p}_{\text{T}}^{\,j}\right|}{\pt^{i}+\pt^{j}} ; &
    \dphi{ij} = \left|\phi_{i}-\phi_{j}\right|                                                             ; & 
    \dy{ij} = \left|\y_{i}-\y_{j}\right|                                                                   ; & \\[8mm]
    \left|\phi_{1 + 2} - \phi_{3 + 4}\right| ; & 
    \left|\phi_{1 + 3} - \phi_{2 + 4}\right| ; & 
    \left|\phi_{1 + 4} - \phi_{2 + 3}\right| ; & \\
\end{array}
\label{eq:inputVariablesPCA} 
\end{equation}
where $\pt^{i}$, $\,\vec{p}_{\text{T}}^{\,i}$, $\,\y_{i}$ and $\,\phi_{i}$ stand for the scalar and vectorial transverse momentum, the rapidity and the azimuthal angle of jet $i$ respectively, 
with $i=1,2,3,4$.
The variables with the subscript $ij$ are calculated for all possible jet combinations.
The term $\phi_{i + j}$ denotes the azimuthal angle of the four-vector obtained by the sum of jets $i$~and~$j$.

In the following, the pairing notation $\{ \langle i,j \rangle \langle k,l \rangle \}$ is used to describe a \cDps event 
in which jets~$i$~and~$j$ originate from one interaction and jets~$k$~and~$l$ originate from the other.
In around~85\% of \cDps events, the two leading jets originate from one interaction and jets~3~and~4 originate from the other.

\begin{figure}[tbp]
\begin{center}
\subfloat[]{\label{fig:inputVariables:dpt12}\includegraphics[trim=0mm 0mm 0mm 10mm,clip,width=0.5\textwidth]{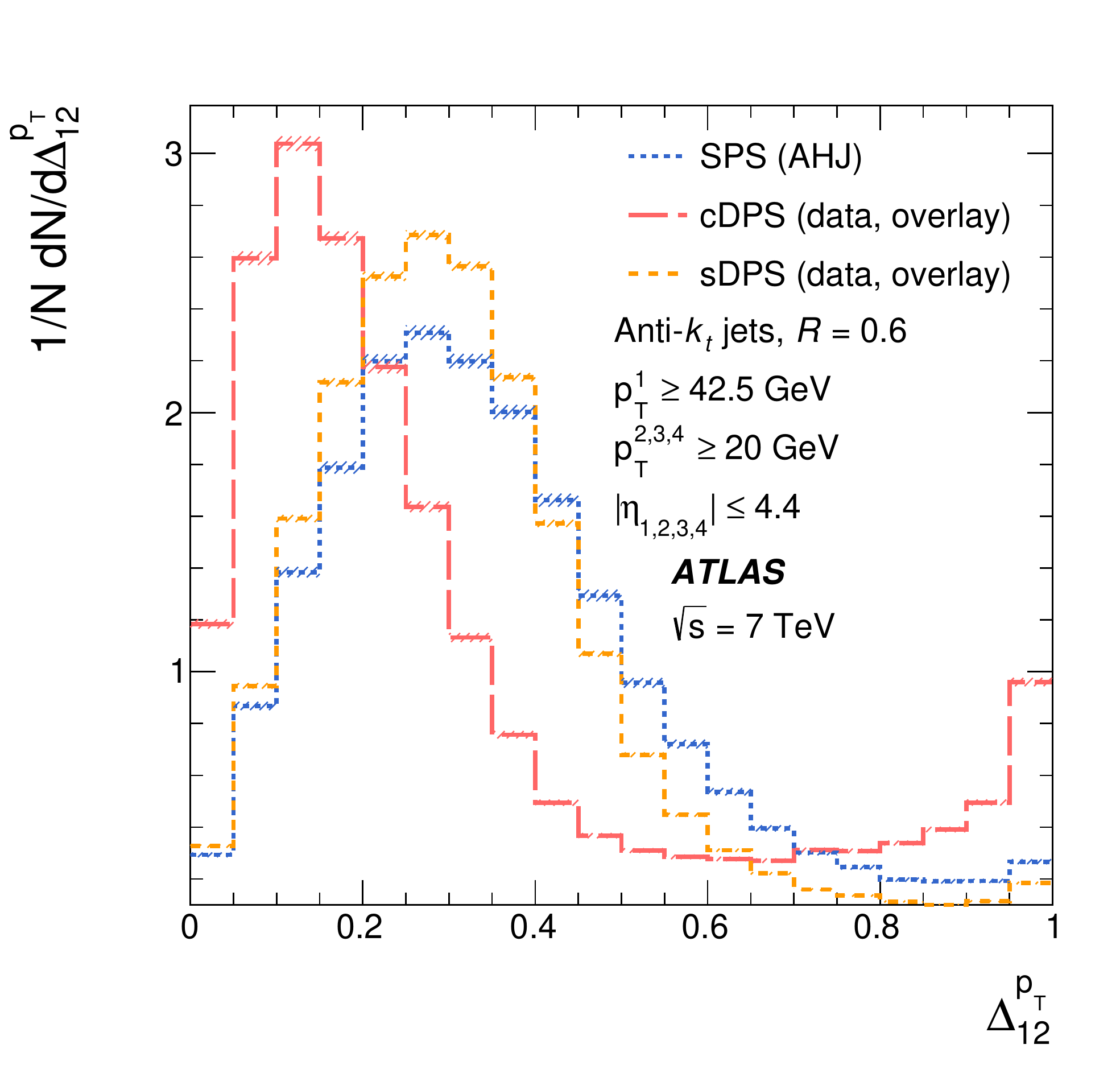}}
\subfloat[]{\label{fig:inputVariables:dpt34}\includegraphics[trim=0mm 0mm 0mm 10mm,clip,width=0.5\textwidth]{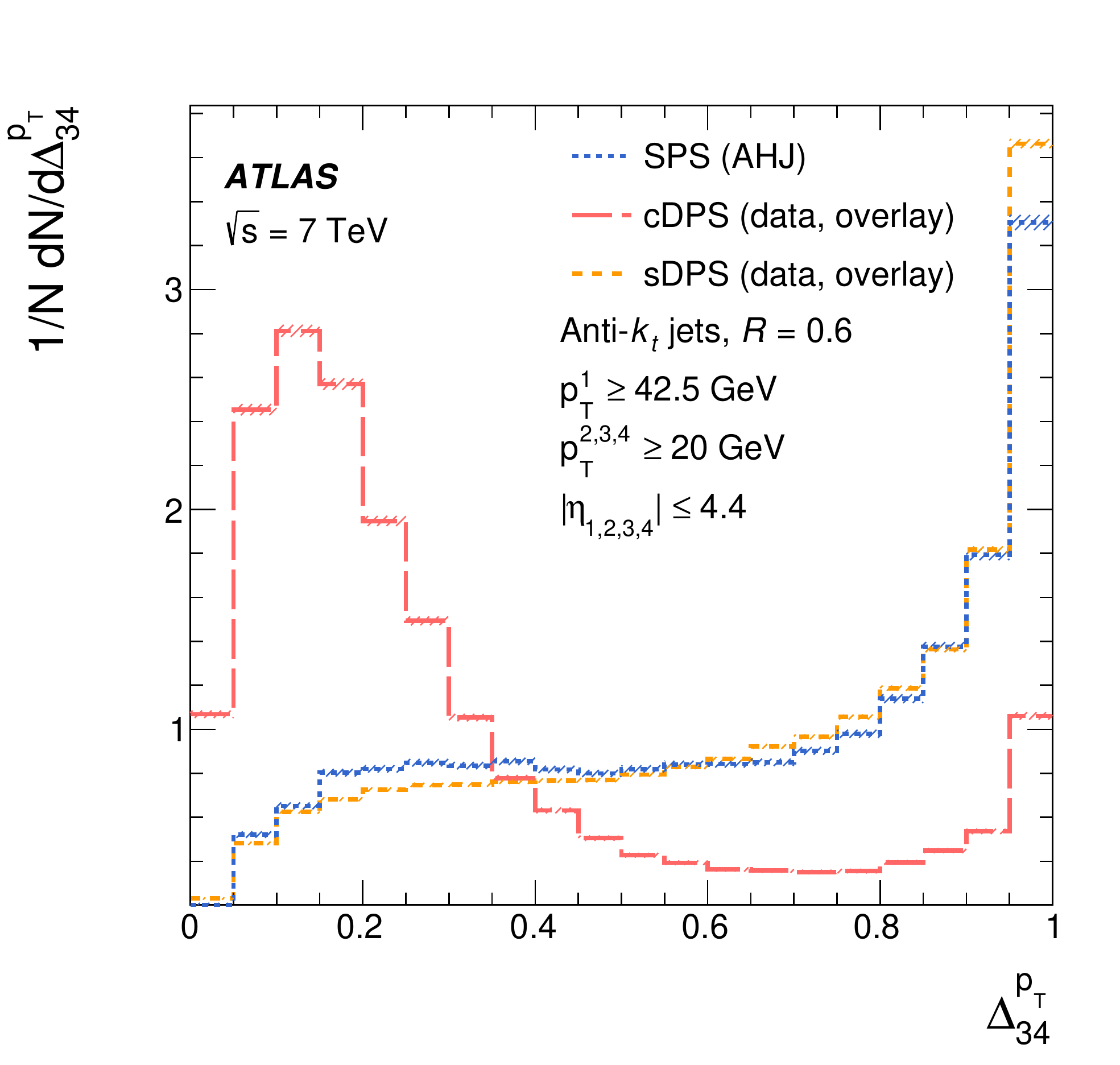}}\\
\subfloat[]{\label{fig:inputVariables:dphi34}\includegraphics[trim=0mm 0mm 0mm 10mm,clip,width=0.5\textwidth]{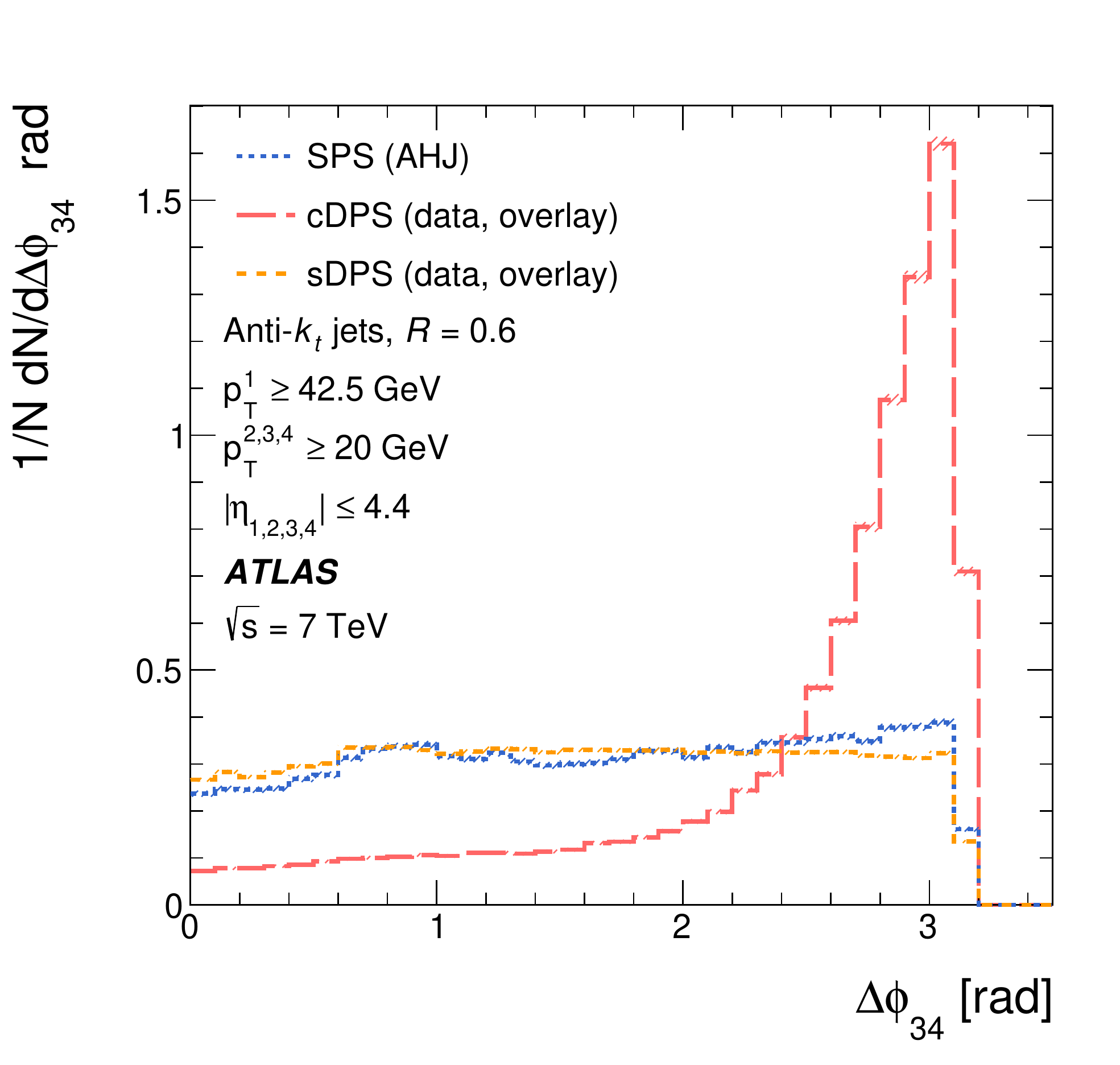}}
\subfloat[]{\label{fig:inputVariables:dy34}\includegraphics[trim=0mm 0mm 0mm 10mm,clip,width=0.5\textwidth]{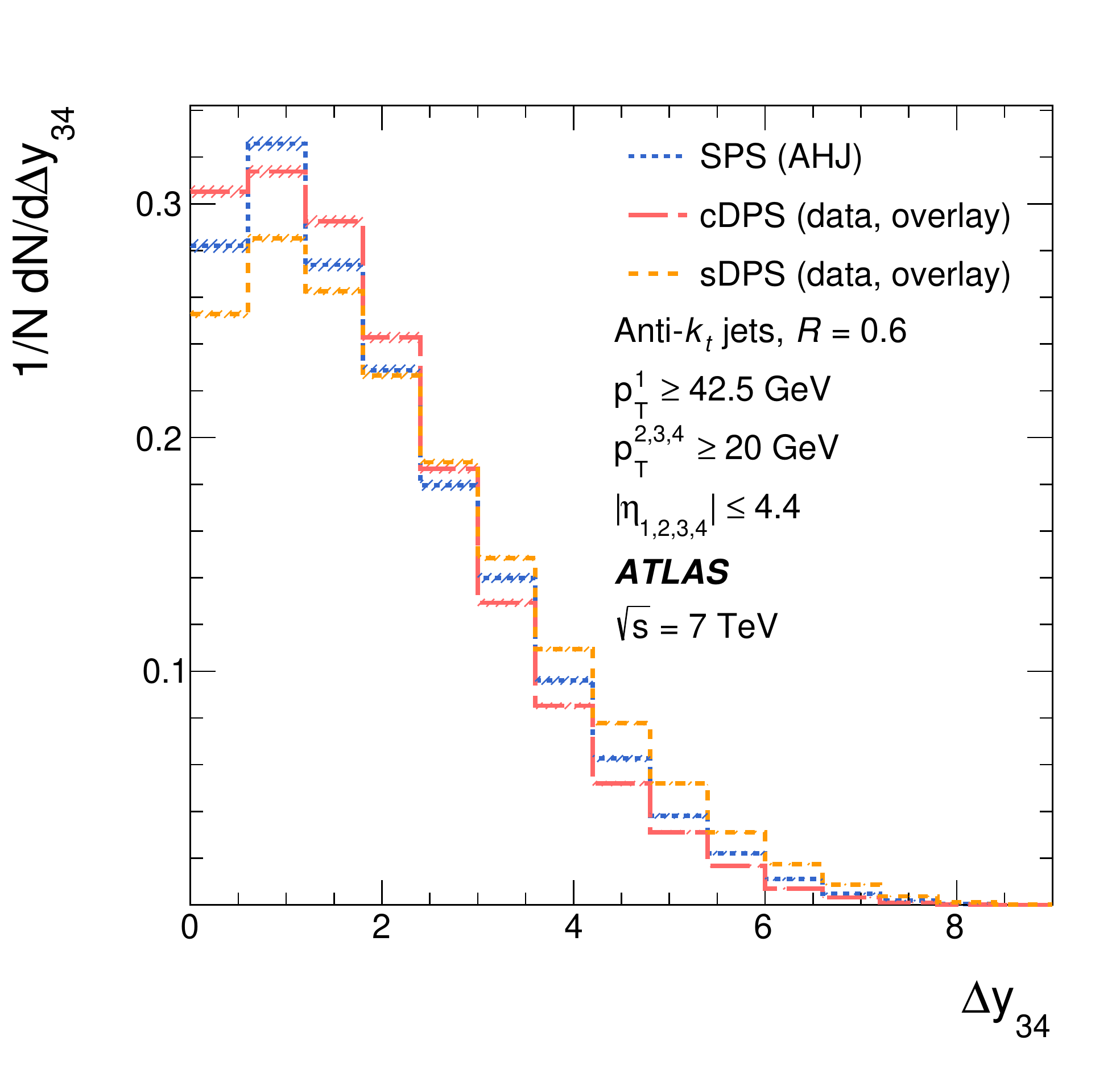}}
\caption{\label{fig:inputVariables}
    Normalized distributions of the variables, \Subref{fig:inputVariables:dpt12} $\dpt{12}$, \Subref{fig:inputVariables:dpt34} $\dpt{34}$, 
    \Subref{fig:inputVariables:dphi34} $\dphi{34}$ and \Subref{fig:inputVariables:dy34} $\dy{34}$,
    defined in \autoref{eq:inputVariablesPCA}, for the \sps, \cDps and \sDps samples as indicated in the legend.
    The hatched areas, where visible, represent the statistical uncertainties for each sample.
}
\end{center}
\end{figure}

Normalized distributions of the \dpt{12} and \dpt{34} variables in the three samples (\sps, \cDps and \sDps) 
are shown in \autorefs{fig:inputVariables:dpt12}~and~\ref{fig:inputVariables:dpt34}.
In the \cDps sample, the \dpt{12} and \dpt{34} distributions peak at low values, indicating that both the leading and the sub-leading jet pairs are balanced in \pt. 
The small peak around unity is due to events in which the appropriate pairing of the jets is $\{ \langle 1,3 \rangle \langle 2,4 \rangle \}$ or $\{ \langle 1,4 \rangle \langle 2,3 \rangle \}$.
In the \sps and \sDps samples, the leading jet-pair exhibits a wider peak at higher values of $\dpt{12}$ compared to that in the \cDps sample.
This indicates that the two leading jets are not well balanced in \pt since a significant fraction of the hard-scatter momentum is carried by additional jets.

The \dphi{34} distributions in the three samples are shown in \autoref{fig:inputVariables:dphi34}.
The \pt balance between the jets seen in the \dpt{34} distribution in the \cDps sample is reflected in the \dphi{34} distribution.
The $\dphi{34}$ distribution is almost uniform for the \sps and \sDps samples. 
The correlation between the distributions of the $\dpt{34}$ and $\dphi{34}$ variables can be readily understood through the following approximation: 
$\pt^{3} \approx \pt^{4} \approx \pt$.
The expression for $\dpt{34}$ then becomes
\begin{equation}
    \dpt{34} = \dfrac{\left|\vec{p}_{\text{T}}^{\,3}+\vec{p}_{\text{T}}^{\,4}\right|}{\pt^{3}+\pt^{4}} 
    \approx \frac{ \sqrt{ 2\pt +2\pt \cos{\!(\dphi{34})} } }{ 2\pt } = \frac{ \sqrt{ 1 + \cos{\!(\dphi{34})} } }{ \sqrt{2} } \,.
\label{eq:approximationPt3eqPt4}
\end{equation}
The peak around unity observed in the $\dpt{34}$ distributions in the \sps and \sDps samples is thus a direct consequence of the Jacobian of the relation between \dpt{34} and \dphi{34}.

The set of variables quantifying the distance between jets in rapidity, \dy{ij}, is particularly important for the \sDps topology.
The colour flow is different in \sps leading to the four-jet final state and results in smaller angles between the sub-leading jets.
Hence, on average, smaller distances between non-leading jets are expected in the \sps sample compared to the \sDps sample.
This is observed in the comparison of the \dy{34} distributions shown in \autoref{fig:inputVariables:dy34}, 
where the distribution in the \sDps sample is slightly wider than in the other two samples.

The study of the various distributions in the three samples is summed up as follows:
\begin{itemize}
    \item Strong correlations between all variables are observed. 
        The \dpt{ij} and \dphi{ij} variables are correlated in a non-linear way, while geometrical constraints correlate the \dy{ij} and \dphi{ij} variables.
        Transverse momentum conservation correlates the \dPhiPlanes{i}{j}{k}{l} variables with the \dpt{ij} and \dphi{ij} variables.
    \item None of the variables displays a clear separation between all three samples.
        The variables in which a large difference is observed between the \sps and \cDps distributions, \eg, \dpt{34}, do not provide any differentiating power between \sps and \sDps.
    \item All variables are important --
         in \cDps events, where the pairing of the jets is different from $\{ \langle 1,2 \rangle \langle 3,4 \rangle \}$, variables relating the other possible pairs, \eg, \dphi{13}, 
         may indicate which is the correct pairing.
\end{itemize}
These conclusions led to the decision to use a multivariate technique in the form of an NN to perform event classification.

\subsection{Extraction of the fraction of \dps events using an artificial neural network\label{subsec:extractionOfFdpsWithNN}}

For the purpose of training the NN, events from each sample were divided into two statistically independent subsamples, the training sample and the test sample. 
The former was used to train the NN and the latter to test the robustness of the result.
To avoid bias during training, the events in the \sps, \cDps and \sDps training samples were reweighted such that each sample contributed a third of the total sum of weights.
In all subsequent figures, only the test samples are shown.

\begin{figure}[tbp]
\captionsetup{aboveskip=5pt, belowskip=-3ex}
\begin{center}
        \begingroup
        \begin{tikzpicture}

    \draw[ultra thick, ->] (0,0) -- (11,0) node(xline)[below left] {{\LARGE $\frac{1}{\sqrt{3}} \xisDps + \frac{2}{\sqrt{3}} \xicDps$}};
    \draw[ultra thick, ->] (0,0) -- (0,{(sqrt(3)/2)*10}) node(yline)[below left] {{\LARGE \xisDps}};

    \coordinate (SPS) at (0,0);
    \coordinate (cDPS) at (10,0);
    \coordinate (sDPS) at ( 5, {(sqrt(3)/2)*10} ); 
    \coordinate (event) at (5, {((sqrt(3)/2)*10)/3} );
    \coordinate (xisps) at (7.5,{((sqrt(3)/2)*10)/2} );
    \coordinate (xicDps) at (2.5,{((sqrt(3)/2)*10)/2} );
    \coordinate (xisDps) at (5,0);

    \coordinate (oneFillSPS) at (2,0);
    \coordinate (oneFillcDPS) at (8,0);
    \coordinate (oneFillsDPS) at ( 4, {((sqrt(3)/2)*10) - (cos(30)*2)} ); 

    \coordinate (twoFillSPS) at (1,{((sqrt(3)/2)*2)});
    \coordinate (twoFillcDPS) at (9,{((sqrt(3)/2)*2)});
    \coordinate (twoFillsDPS) at ( 6, {((sqrt(3)/2)*10) - (cos(30)*2)} ); 
    
    \filldraw[shading = axis, left color=myRootCol2, right color=myRootCol2!0!white, shading angle=135] (SPS)  -- (xicDps) -- (event) -- (xisDps) -- cycle;
    \filldraw[shading = axis, right color=myRootCol1, left color=myRootCol1!0!white, shading angle=45] (cDPS)  -- (xisps) -- (event) -- (xisDps) -- cycle;
    \filldraw[shading = axis, top color=myRootCol3, bottom color=myRootCol3!0!white, shading angle=0] (sDPS) -- (xicDps) -- (event) -- (xisps)  -- cycle;


    
    \coordinate [label=below right:\textcolor{white}{{\large \sps}}] (labelSPS) at (0.4, {((sqrt(3)/2)*10)/13});
    \coordinate [label=below left:{\large \cDps}] (labelcDPS) at (9.7, {((sqrt(3)/2)*10)/13}); 
    \coordinate [label=below:{\large \sDps}] (labelsDPS) at (5, {((sqrt(3)/2)*10)/1.15} );

    \coordinate [label=below:{\LARGE \xisps}] (labelSPS) at (5.95, {((sqrt(3)/2)*10)/1.95});
    \coordinate [label=below:{\LARGE \xicDps}] (labelcDPS) at (4.2, {((sqrt(3)/2)*10)/1.95}); 
    \coordinate [label=below:{\LARGE \xisDps}] (labelsDPS) at (5.9, {((sqrt(3)/2)*10)/4.8} );

    
    \draw[ultra thick] (SPS) -- (cDPS) -- (sDPS) -- cycle;
    
    \draw[ultra thick] (event) -- (xisps);
    \draw[ultra thick] (event) -- (xicDps);
    \draw[ultra thick] (event) -- (xisDps);
    

\end{tikzpicture}%
        \endgroup
    
\caption{\label{fig:dalitzIllustration}
    Illustration of the ternary plot constructed from three NN outputs, \xisps, \xicDps, and \xisDps, with the constraint, $\xisps + \xicDps + \xisDps = 1$.
    The vertical and horizontal axes are defined in the figure.
    The coloured areas illustrate the classes of events expected to populate the corresponding vertices.
}
\end{center}
\end{figure}
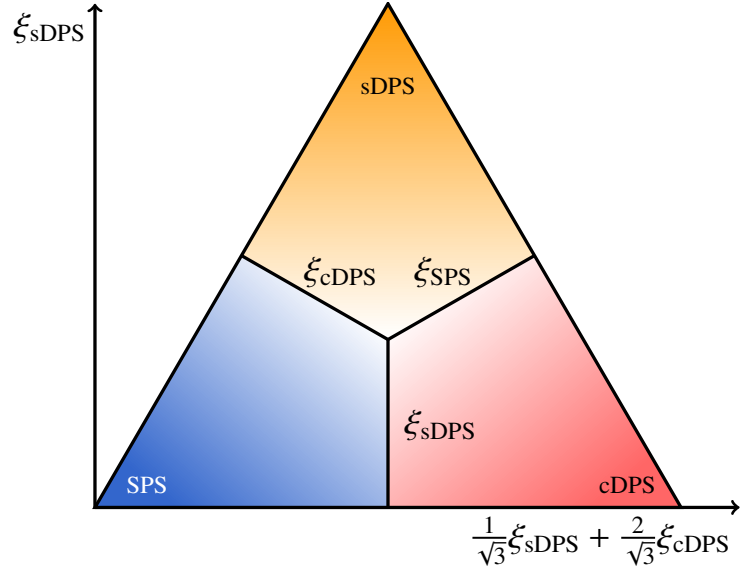

The NN used is a feed-forward multilayer perceptron with two hidden layers, implemented in the ROOT analysis framework~\cite{Brun:1997pa}.
The input layer has 21 neurons, corresponding to the variables defined in \autoref{eq:inputVariablesPCA}, and the first and second hidden layers have 42 and 12 neurons respectively.
These choices represent the product of a study conducted to optimize the performance of the NN and balance the complexity of the network with the computation time of the training.
The output of the NN consists of three variables, which are interpreted as the probability for an event to be more like  \sps (\xisps), \cDps (\xicDps) or \sDps (\xisDps).
During training, each event is marked as belonging to one of the samples; \eg, an event from the \cDps sample is marked as
\begin{equation}
    \xisps = 0, \; \; \xicDps = 1, \; \; \xisDps = 0.
\label{eq:flags}
\end{equation}
For each event, the three outputs are plotted as a single point inside an equilateral triangle (ternary plot) using the constraint $\xisps + \xicDps + \xisDps = 1$.
A point in the triangle expresses the three probabilities as three distances from each of the sides of the triangle.
The vertices would therefore be populated by events with high probability to belong to a single sample.
\Autoref{fig:dalitzIllustration} shows an illustration of the ternary plot, 
where the horizontal axis corresponds to $\frac{1}{\sqrt{3}} \xisDps + \frac{2}{\sqrt{3}} \xicDps$ and the vertical axis to the value of \xisDps.
The coloured areas illustrate where each of the three classes of events is expected to populate the ternary plot.

\begin{figure}[tbp]
    \captionsetup[subfloat]{captionskip=3ex}
\begin{center}
\subfloat[]{\label{fig:nn:dalitz:sps}\includegraphics[trim=0mm 0mm 0mm 10mm,clip,width=0.5\textwidth]{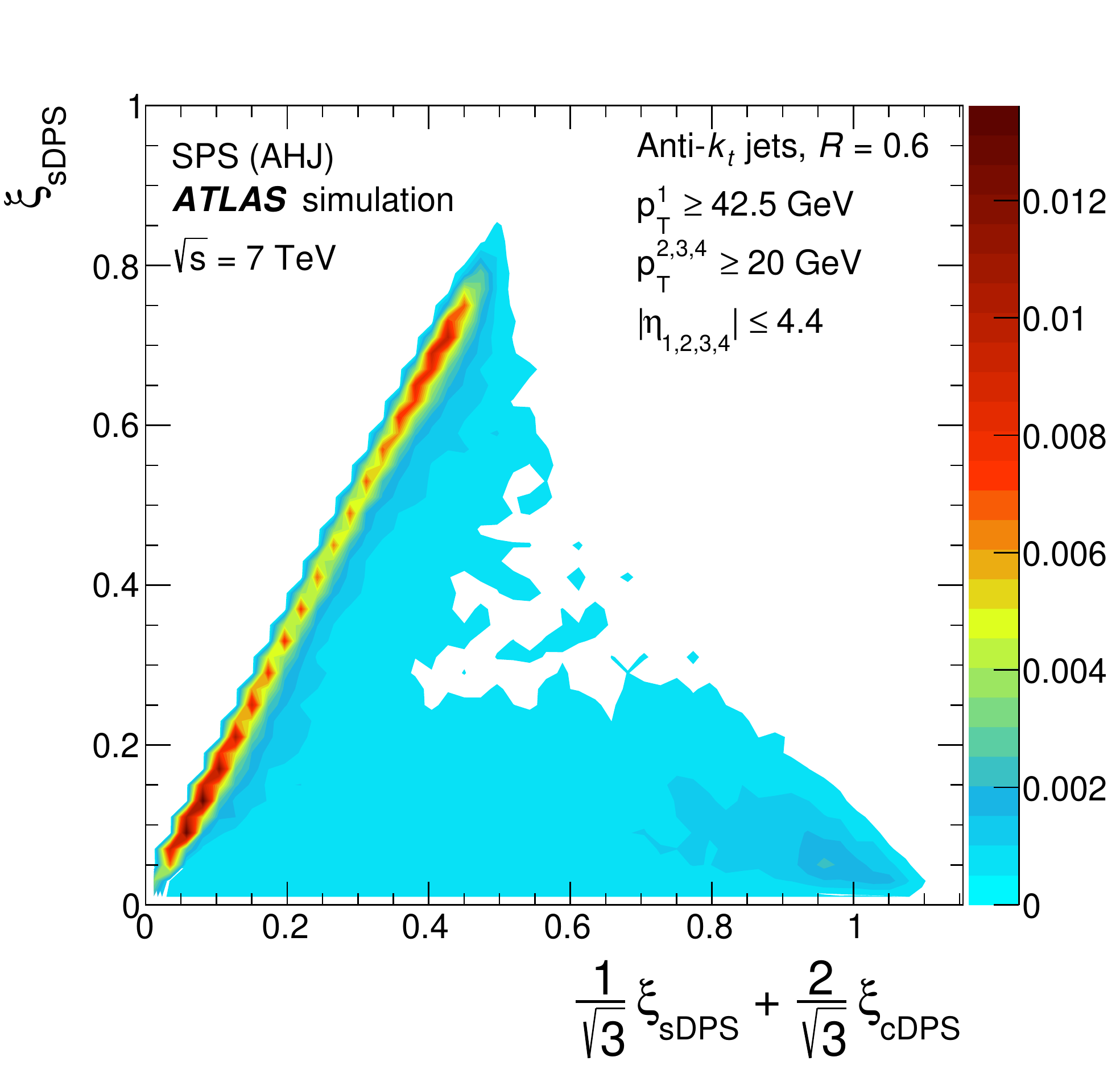}} 
\subfloat[]{\label{fig:nn:dalitz:cDps}\includegraphics[trim=0mm 0mm 0mm 10mm,clip,width=0.5\textwidth]{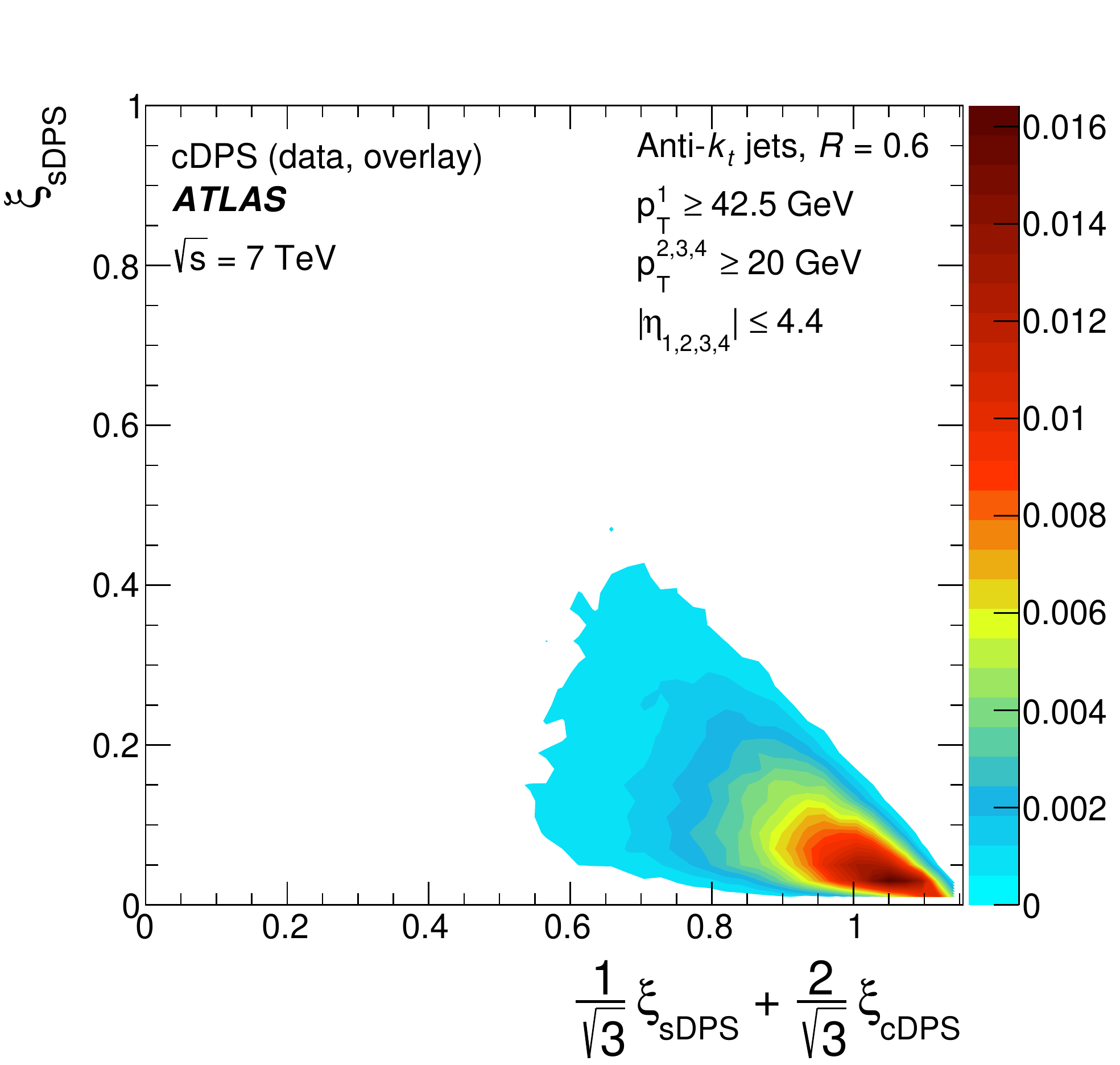}} \\
\subfloat[]{\label{fig:nn:dalitz:sDps}\includegraphics[trim=0mm 0mm 0mm 10mm,clip,width=0.5\textwidth]{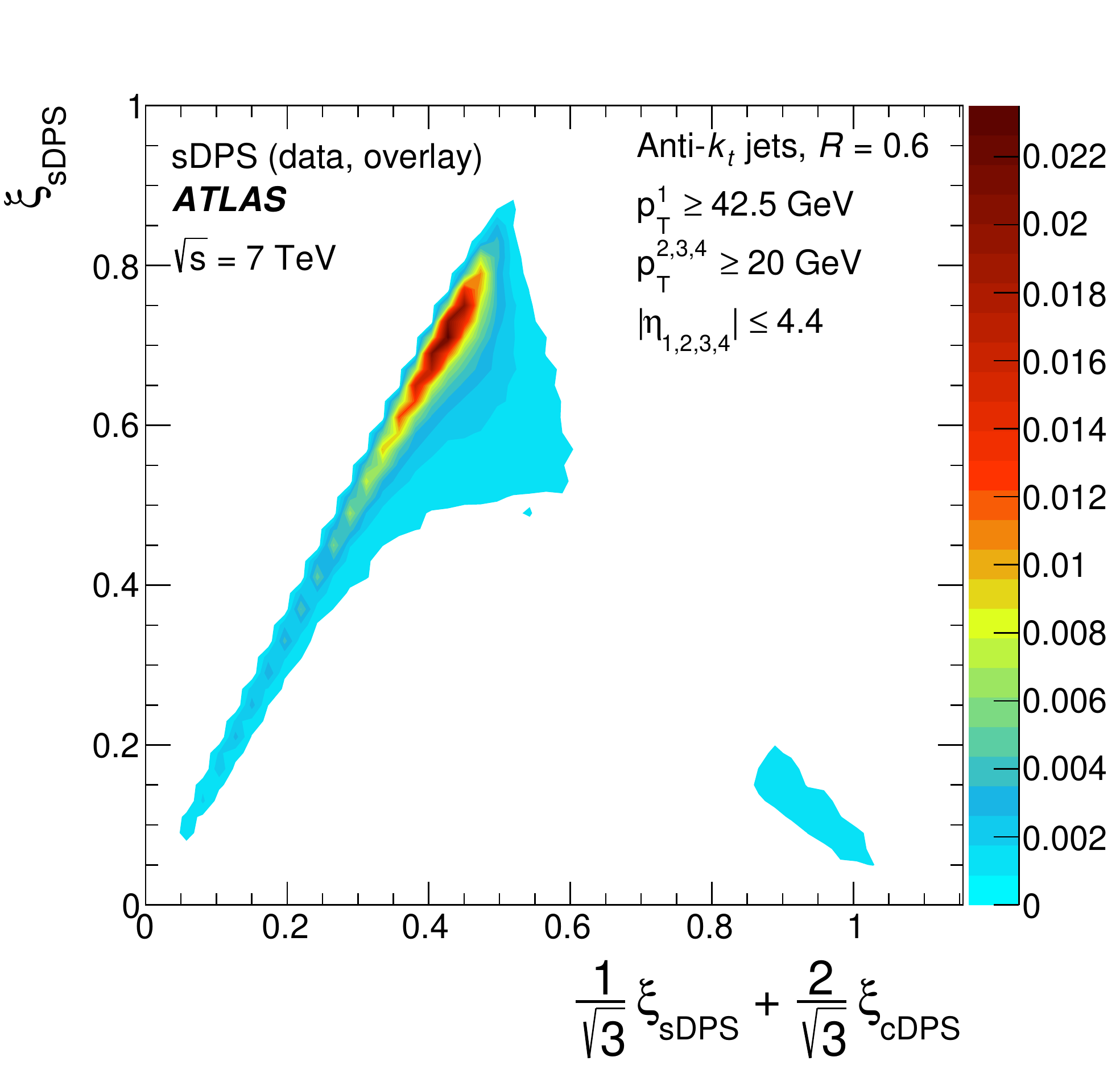}}
\subfloat[]{\label{fig:nn:dalitz:data}\includegraphics[trim=0mm 0mm 0mm 10mm,clip,width=0.5\textwidth]{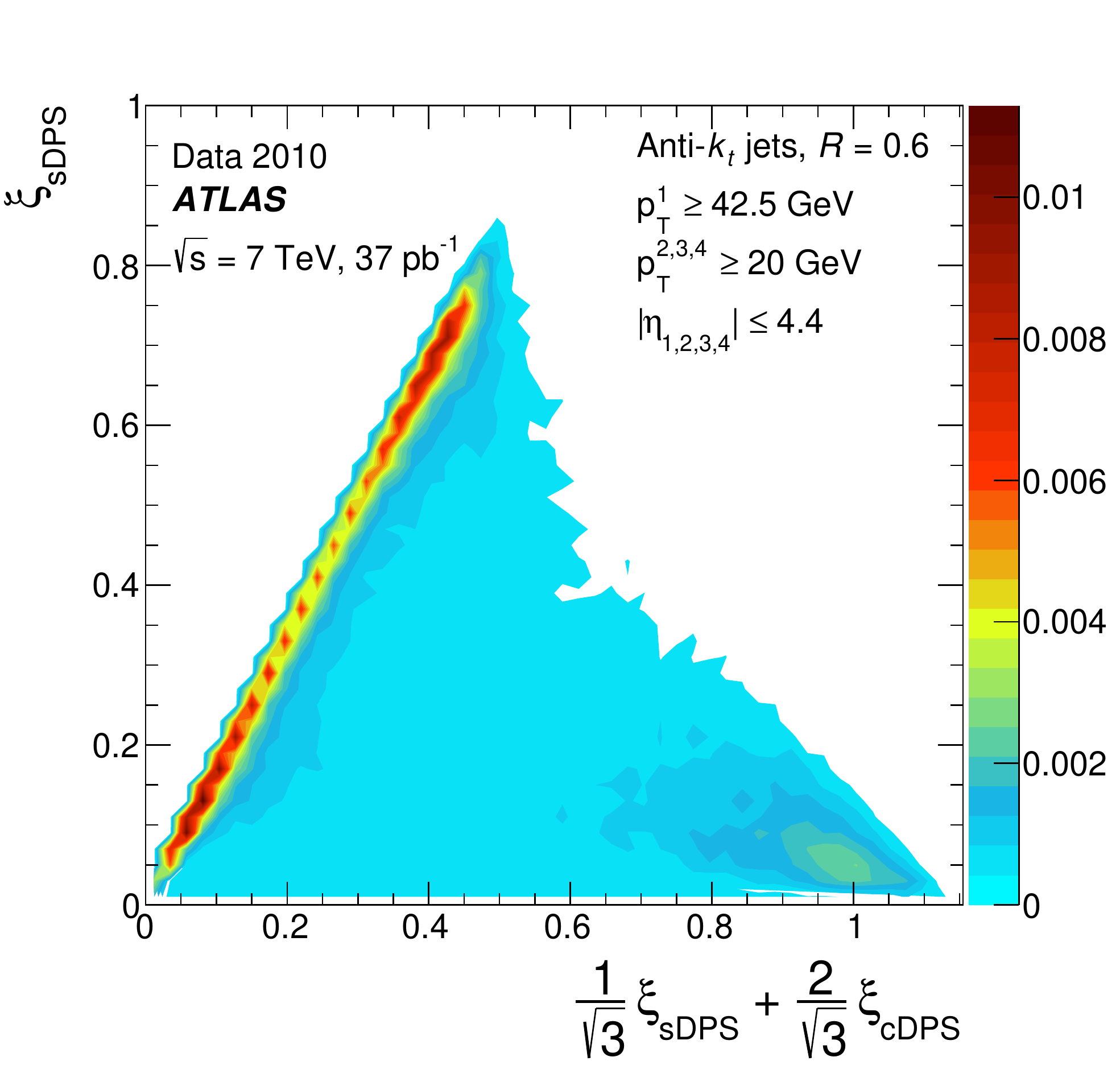}}
\caption{\label{fig:nn:dalitz:all}
    Normalized distributions of the NN outputs, mapped to a ternary plot as described in the text, in the \Subref{fig:nn:dalitz:sps}~\sps, 
    \Subref{fig:nn:dalitz:cDps}~\cDps, \Subref{fig:nn:dalitz:sDps}~\sDps test samples and \Subref{fig:nn:dalitz:data} in the data.
}
\end{center}
\end{figure}

\Autorefs{fig:nn:dalitz:sps},~\ref{fig:nn:dalitz:cDps}~and~\ref{fig:nn:dalitz:sDps} show the NN output distribution for the test samples in the ternary plot, 
presenting the separation power of the NN.
The \sps-type events are mostly found in the bottom left corner in \autoref{fig:nn:dalitz:sps}.
However, a ridge of \sps events extending towards the \sDps corner is observed as well.
A contribution from \sps events is also visible in the bottom right corner.
The clearest peak is seen for events from the \cDps sample in the bottom right corner in \autoref{fig:nn:dalitz:cDps}.
A visible cluster of \sDps events is seen in \autoref{fig:nn:dalitz:sDps} 
concentrated around $\xisDps \sim 0.75$ and there is a tail of events along the side connecting the \sps and \sDps corners.
The NN output distribution in the data, shown in \autoref{fig:nn:dalitz:data}, is visually consistent with a superposition of the three components, \sps, \cDps and \sDps.

Based on these observations, it is clear that event classification on an event-by-event basis is not possible.
However, the differences between the \sps, \cDps and \sDps distributions suggest that an estimation of the different contributions can be performed.
To estimate the \cDps and \sDps fractions in four-jet events, 
the ternary distribution in data ($\mathcal{D}$) is fitted to a weighted sum of the ternary distributions 
in the \sps ($\mathcal{M}_{\sps}$), \cDps ($\mathcal{M}_{\cDps}$) and \sDps ($\mathcal{M}_{\sDps}$) samples, 
each normalized to the measured four-jet cross-section in data, with the fractions as free parameters.
The optimal fractions were obtained using a fit of the form,
\begin{equation}
    \mathcal{D} = (1-\fcDps - \fsDps) \mathcal{M}_{\sps} \,+\, \fcDps \mathcal{M}_{\cDps} \,+\, \fsDps \mathcal{M}_{\sDps} \;,
\label{eq:fdpsFit}
\end{equation}
where a $\chi^{2}$ minimization was performed, as implemented in the {\sc Minuit}~\cite{James:1975dr} package in ROOT, 
taking into account the statistical uncertainties of all the samples in each bin.
The results of the fit are presented in \autoref{sec:detSigEff}, after the methodology validation and discussion of systematic uncertainties.


\subsection{Methodology validation\label{sec:validation}}

A sizeable discrepancy was found in the \dpt{34} and \dphi{34} distributions between the data and \ahj (See \autoref{sec:unfolding} for details),
suggesting that there are more sub-leading jets (jets~3~and~4) that are back-to-back in \ahj than in the data.
In order to test that the discrepancies are not from mis-modelling of \sps in \ahj, 
the \dpt{34} and \dphi{34} distributions in the \sps sample extracted from \ahj were compared to the distributions in the \sps sample generated in \sherpa.
Good agreement between the shapes of the distributions was observed for both variables.
This and further studies indicate that the excess of events with jets~3~and~4 in the back-to-back topology is due to an excess of \dps events in the \ahj sample compared to the data.

\begin{figure}[htbp]
\captionsetup[subfloat]{captionskip=-1ex}
\begin{center}
\subfloat[]{\label{fig:validation:cDps}\includegraphics[trim=0mm 0mm 0mm 17mm,clip,width=0.5\textwidth]{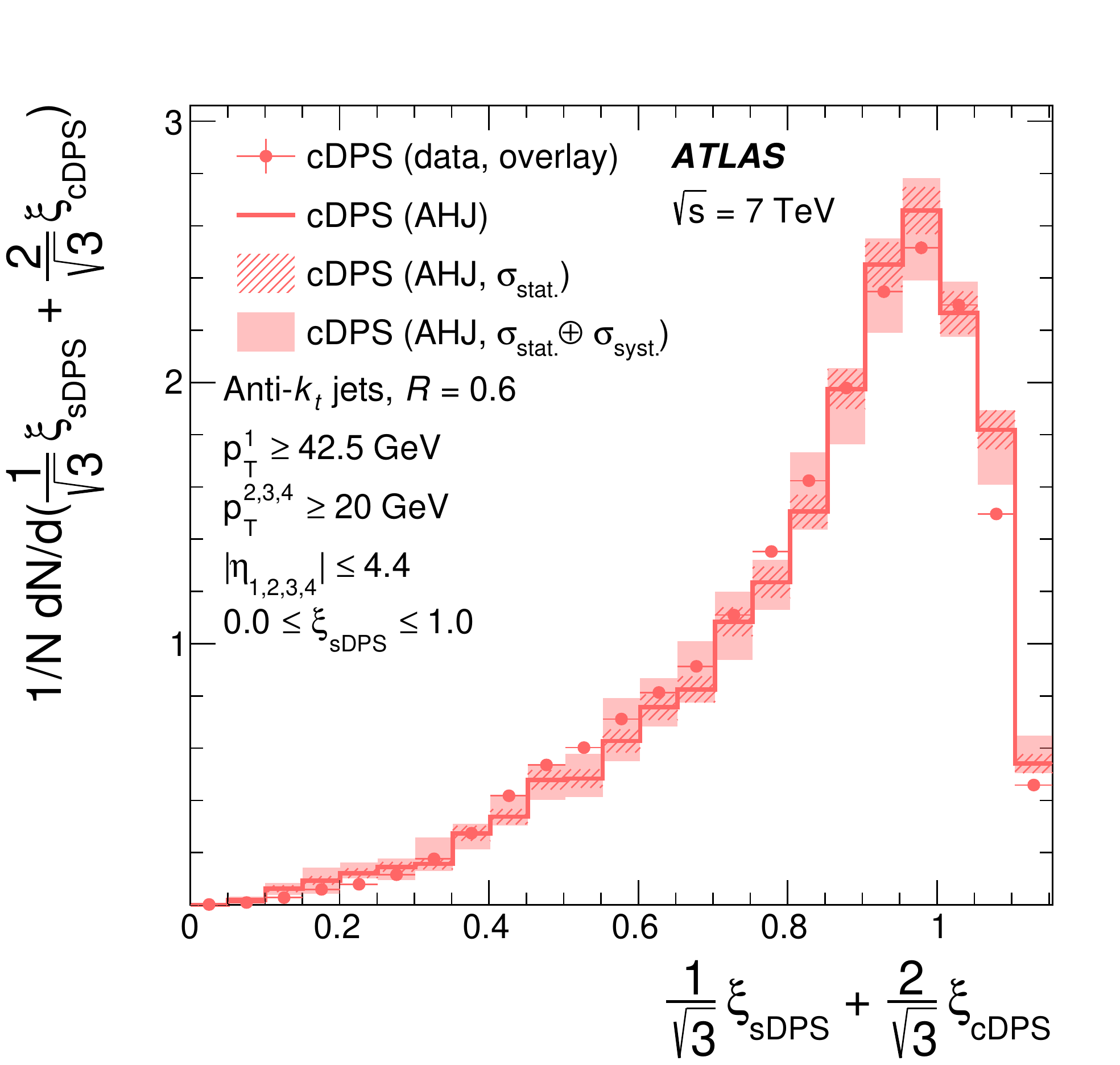}}
\subfloat[]{\label{fig:validation:sDps}\includegraphics[trim=0mm 0mm 0mm 17mm,clip,width=0.5\textwidth]{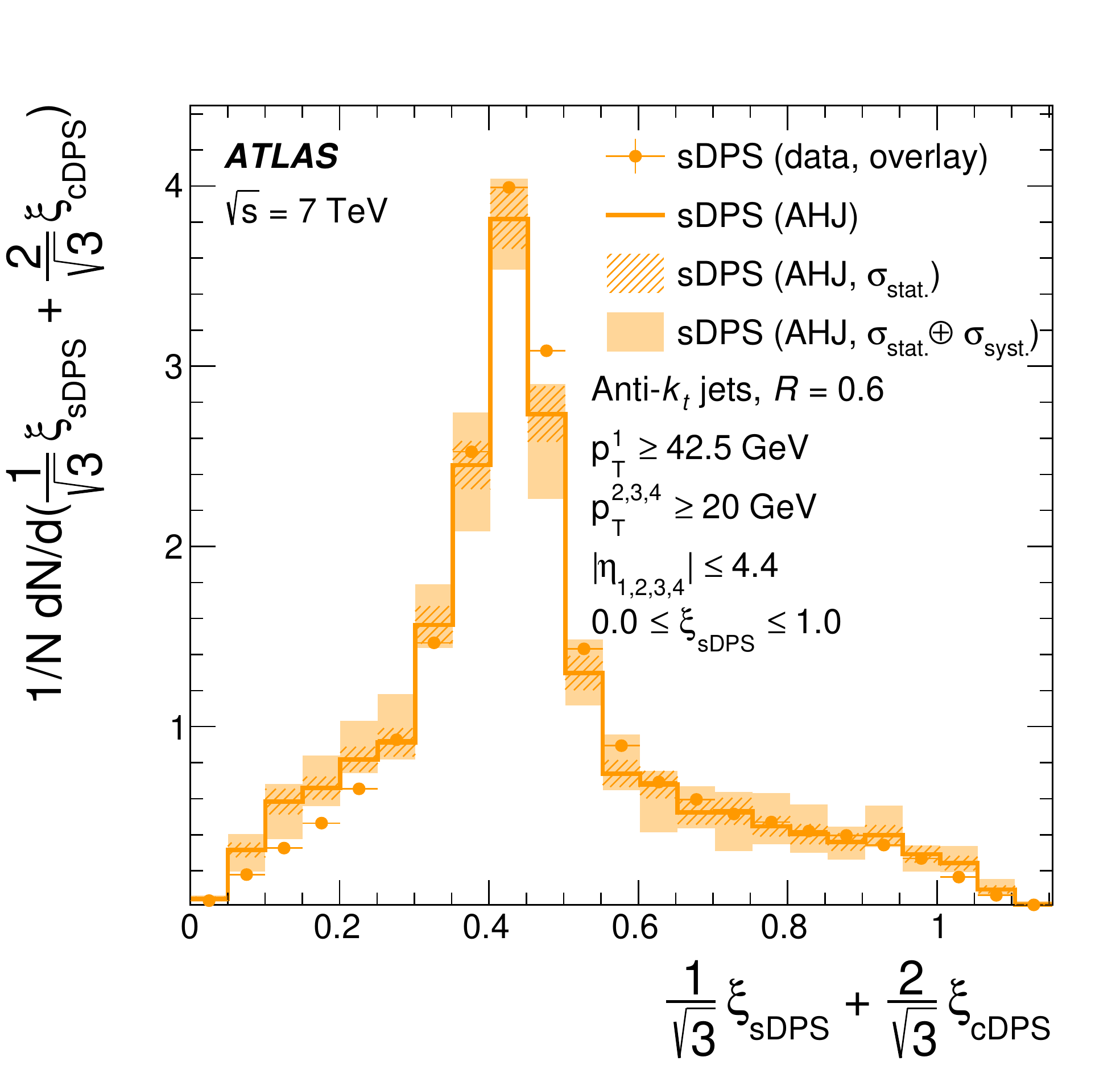}}
\caption{\label{fig:validation}
    Comparison between the normalized distributions of the NN outputs $\frac{1}{\sqrt{3}} \xisDps + \frac{2}{\sqrt{3}} \xicDps$,
    integrated over all \xisDps values $0.0 \le \xisDps \le 1.0$, in \dps events extracted from \ahj and in the \dps samples constructed by overlaying events from data, 
    for \Subref{fig:validation:cDps} \cDps events and \Subref{fig:validation:sDps} \sDps events.
    In the \ahj distributions, statistical uncertainties are shown as the hatched area and the shaded area represents the sum in quadrature of the statistical and systematic uncertainties.
}
\end{center}
\end{figure}

In order to verify that the topologies of \cDps and \sDps events can be reproduced by overlaying two events, the overlay samples are compared to the \cDps and \sDps samples extracted from \ahj.
An extensive comparison between the distributions of the variables used as input to the NN in the overlay samples and in \ahj was performed and good agreement was observed.
This can be summarized by comparing the NN output distributions.
The NN is applied to the \cDps and \sDps samples extracted from \ahj 
and the output distributions are compared to the output distributions in the corresponding samples constructed by overlaying events selected from data.
Normalized distributions of the projection of the full ternary plot on the horizontal axis are shown in \autorefs{fig:validation:cDps}~and~\ref{fig:validation:sDps} 
for the \cDps and \sDps samples respectively. 
Good agreement is observed between the distributions.
Based on these results, it is concluded that the topology of the four jets in the overlaid events is comparable to that of the four leading jets in \dps events extracted from \ahj.
The added advantage of using overlaid events from data to construct the \dps samples is that the jets are at the same JES as the jets in four-jet events in data,
leading to a smaller systematic uncertainty in the final result.

As an additional validation step, the NN is applied to the inclusive \ahj sample and the resulting distribution is fitted with the NN output distributions of the \sps, \cDps and \sDps samples. 
The fraction obtained from the fit, \fdps[MC], is compared to the fraction at parton level, \fdps[P], extracted from the event record,
\begin{equation}
    \fdps[MC]\,=\, 0.129 \pm 0.007 \stat \, , \quad \quad  \fdps[P]\,=\, 0.142 \pm 0.001 \stat \, .
\label{eq:compareFractions} 
\end{equation} 
Fair agreement is observed between the value obtained from the fit and that at parton level.
The larger statistical uncertainty in \fdps[MC] compared to \fdps[P] reflects the loss of statistical power due to the use of a template fit to estimate the former.

\section{Systematic uncertainties\label{sec:uncertainties}}

For jets with $20 \leq \pt < 30$ \gev\xspace, the fractional JES uncertainty is about~4.5\% in the central region of the detector, 
rising to about~10\% in the forward region~\cite{Aad:2011he}.
The overall impact of the JES on the distributions, \fdps and $\alpha_{\text{2j}}^{\text{4j}}$ 
was estimated by shifting the jet energy upwards and downwards in the MC samples by the JES uncertainty and repeating the analysis.
Similarly, the overall impact of the jet energy and angular resolution was determined 
by varying the jet energy and angular resolution in the MC samples by the corresponding resolution uncertainty~\cite{Aad:2012ag}.

The systematic uncertainties in the measured cross-sections due to the integrated luminosity measurement uncertainty ($\pm 3.5\%$),
the jet reconstruction efficiency uncertainty ($\pm 2\%$) and the uncertainty as a result of selecting single-vertex events ($\pm 0.5\%$) were propagated to the uncertainty in \sigeff.

The statistical uncertainty in the \ahj sample was translated to a systematic uncertainty in \fdps by varying the reweighting function used to reweight \ahj and repeating the analysis.

The statistical uncertainty in $\alpha_{\text{2j}}^{\text{4j}}$ ($\sim$1\%) was propagated as a systematic uncertainty in \sigeff.
The systematic uncertainty in $\alpha_{\text{2j}}^{\text{4j}}$ arising from model-dependence ($\pm 2\%$) was determined from deriving $\alpha_{\text{2j}}^{\text{4j}}$ using \sherpa.

The stability of the value of \sigeff relative to the various parameter values used in the measurement was studied.
Parameters such as $\pt^{\mathrm{parton}}$ and $\Delta R_{\,\mathrm{jet-jet}}$ were varied and the requirement $\Delta R_{\mathrm{parton-jet}} \leq 0.6$ was applied,
leading to a relative change in \sigeff of the order of a few percent.
Since the observed relative changes are small compared to the statistical uncertainty in \sigeff, no systematic uncertainty was assigned due to these parameters.

The relative systematic uncertainties in $\fdps$, $\alpha_{\text{2j}}^{\text{4j}}$ and \sigeff are summarized in \autoref{table:sysUncertainties}.
The dominant systematic uncertainty on \fdps originates from the JES variation.
A variation in the JES results in a modification of the NN output distribution for the \sps template used in the fit, 
which directly impacts the value of $\fdps$.

\begin{table}[htbp]
    \captionsetup{aboveskip=-1ex, belowskip=-2ex}
    \begin{center}
        \begin{tabular}{%
                l r r r 
        }
          \toprule
          Source of systematic uncertainty           & \multicolumn{1}{c}{$\Delta\fdps$} & \multicolumn{1}{c}{$\Delta\alpha_{\text{2j}}^{\text{4j}}$} & \multicolumn{1}{c}{$\Delta\sigeff $} \\
          \midrule
          Luminosity                                 &                          &                       & $\pm 3.5 \, \%$        \\ 
          Model dependence for detector corrections  &                          & $\pm 2 \, \%$         & $\pm 2   \, \%$        \\ 
          Reweighting of \ahj                       &  $\pm 6 \, \%$           &                       & $\pm 6   \, \%$        \\ 
          Jet reconstruction efficiency              &                          &                       & $\pm 0.1 \, \%$        \\ 
          Single-vertex events selection             &                          &                       & $\pm 0.1 \, \%$        \\ 
          Jet energy and angular resolution          &  $\pm 15 \, \%$          & $\pm 3 \, \%$         & $\pm 15  \, \%$        \\ 
          JES uncertainty                            &  $^{+32}_{-37} \, \%$    & $\pm 12  \, \%$       & $^{+31}_{-19} \, \%$   \\ 
          \midrule
          Total systematic uncertainty               &  $^{+36}_{-40} \, \%$    & $\pm 13  \, \%$       & $^{+35}_{-25} \, \%$   \\
          \bottomrule
        \end{tabular}
    \end{center}
    \caption{\label{table:sysUncertainties} 
        Summary of the relative systematic uncertainties in \fdps, $\alpha_{\text{2j}}^{\text{4j}}$ and \sigeff.
    }
\end{table}


\section{Determination of \sigeff\label{sec:detSigEff}}

To determine \fdps and \sigeff and their statistical uncertainties taking into account all of the correlations, many replica fits were performed by random sampling from the NN output distributions.
The systematic uncertainties were obtained by propagating the expected variations into this analysis, and the resulting shifts were added in quadrature.
The result for \fdps is
\begin{equation}
\renewcommand{\arraystretch}{1.5}
\begin{array}{ccccccc}
    \fdps = \measFdps \, , &
\end{array}
\label{eq:dataFractions}
\end{equation}
where the contribution of \fsDps to \fdps was found to be about 40\%.
The fraction of \dps estimated in data is \fracOfAHJ of the fraction in \ahj as extracted from the event record (see \autoref{eq:compareFractions}).
Taking into account the systematic uncertainties in the calculation of the goodness-of-fit $\chi^{2}$, 
a value for $\chi^{2}/N_{\mathrm{DF}}$ of \measChiSys is obtained, where $N_{\mathrm{DF}}$ is the number of degrees of freedom in the fit.

\begin{figure}[htbp]
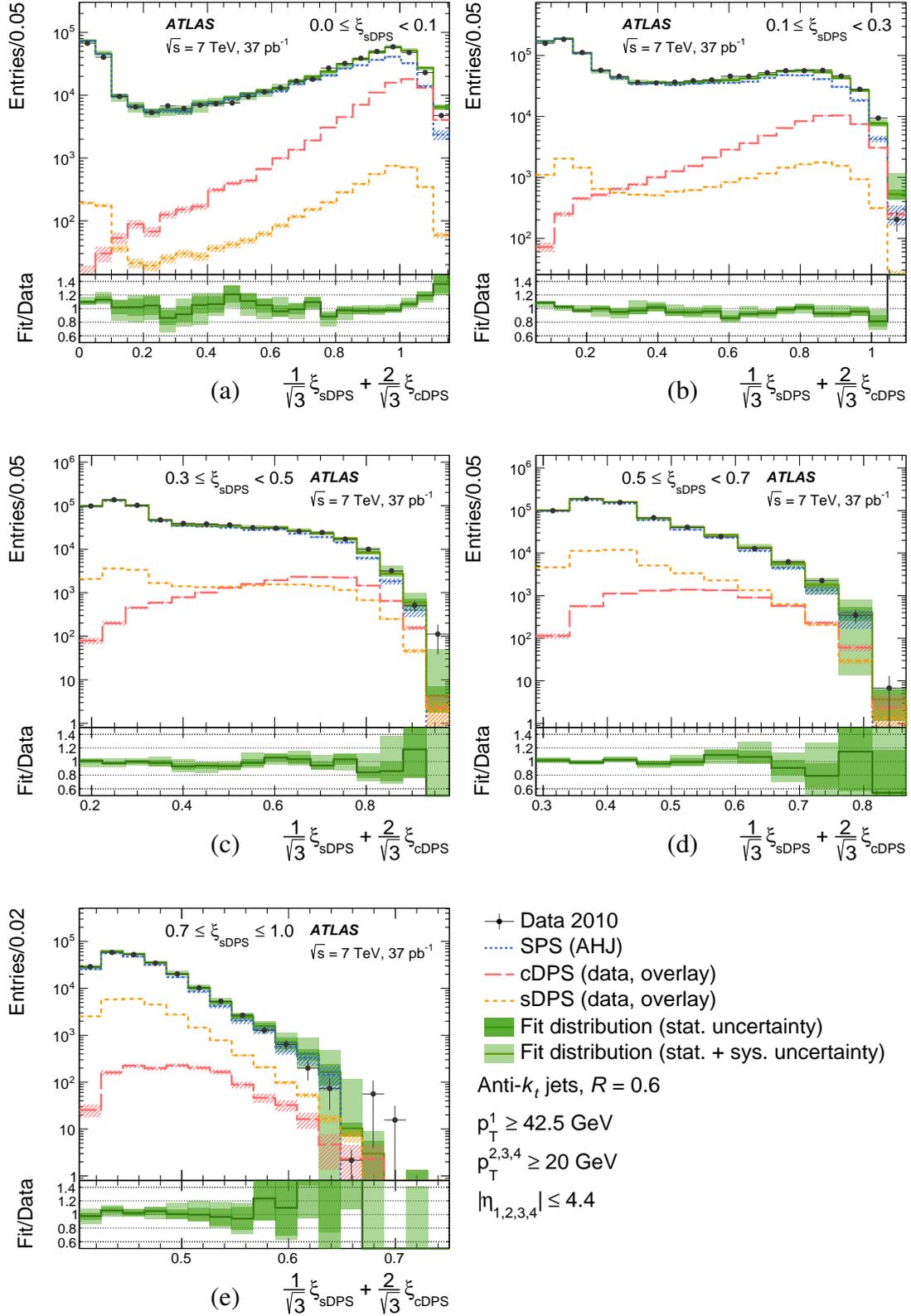

\captionsetup[subfigure]{labelformat=empty}
\captionsetup{aboveskip=-8pt}
\begin{center}
    \subfloat[]{\label{fig:dataFitWithSys:slice1}\sfig[trim=10mm 0mm 0mm 24mm,clip,width=0.45\textwidth,vsep=6.15cm, hsep=3.00cm]{\large (a)}{figures/fitWithSys1}}
    \subfloat[]{\label{fig:dataFitWithSys:slice2}\sfig[trim=10mm 0mm 0mm 24mm,clip,width=0.45\textwidth,vsep=6.15cm, hsep=3.00cm]{\large (b)}{figures/fitWithSys2}} \\ [-5mm]
    \subfloat[]{\label{fig:dataFitWithSys:slice3}\sfig[trim=10mm 0mm 0mm 11mm,clip,width=0.45\textwidth,vsep=6.65cm, hsep=3.00cm]{\large (c)}{figures/fitWithSys3}} 
    \subfloat[]{\label{fig:dataFitWithSys:slice4}\sfig[trim=10mm 0mm 0mm 11mm,clip,width=0.45\textwidth,vsep=6.65cm, hsep=3.00cm]{\large (d)}{figures/fitWithSys4}} \\ [-5mm]
    \subfloat[]{\label{fig:dataFitWithSys:slice5}\sfig[trim=10mm 0mm 0mm 11mm,clip,width=0.45\textwidth,vsep=6.65cm, hsep=3.00cm]{\large (e)}{figures/fitWithSys5}}
    \subfloat[]{\label{fig:dataFitWithSys:legend}\sfig[trim=10mm 0mm 0mm 11mm,clip,width=0.45\textwidth,vsep=0.00cm, hsep=0.0cm]{          }{figures/fitWithSys6}}
\caption{\label{fig:dataFitWithSys}
    Distributions of the NN outputs, $\frac{1}{\sqrt{3}} \xisDps + \frac{2}{\sqrt{3}} \xicDps$,
    in the \xisDps ranges indicated in the panels, for four-jet events in data, selected in the phase space defined in the legend, 
    compared to the result of fitting a combination of the \sps, \cDps and \sDps contributions,
    the sum of which is shown as the solid line.
    In the fit distribution, statistical uncertainties are shown as the dark shaded area and the light shaded area represents 
    the sum in quadrature of the statistical and systematic uncertainties.
    The ratio of the fit distribution to the data is shown in the bottom panels.
}
\end{center}
\end{figure}

\begin{figure}[htbp]
\captionsetup[subfigure]{labelformat=empty}
\captionsetup{aboveskip=-11pt}
\begin{center}
\subfloat[]{\label{fig:relativeCont:dpt34}
    \begin{tabular}{p{0.66\textwidth} p{0.25\textwidth}}
      \vspace{0pt} \sfig[trim=0mm 0mm 10mm 15mm,clip,width=0.66\textwidth,vsep=9.50cm, hsep=6.00cm]{\large (a)}{figures/relativeCont2} &
      \vspace{2.0ex} \includegraphics[trim=9mm 8mm 79mm 20mm,clip,width=0.25\textwidth]{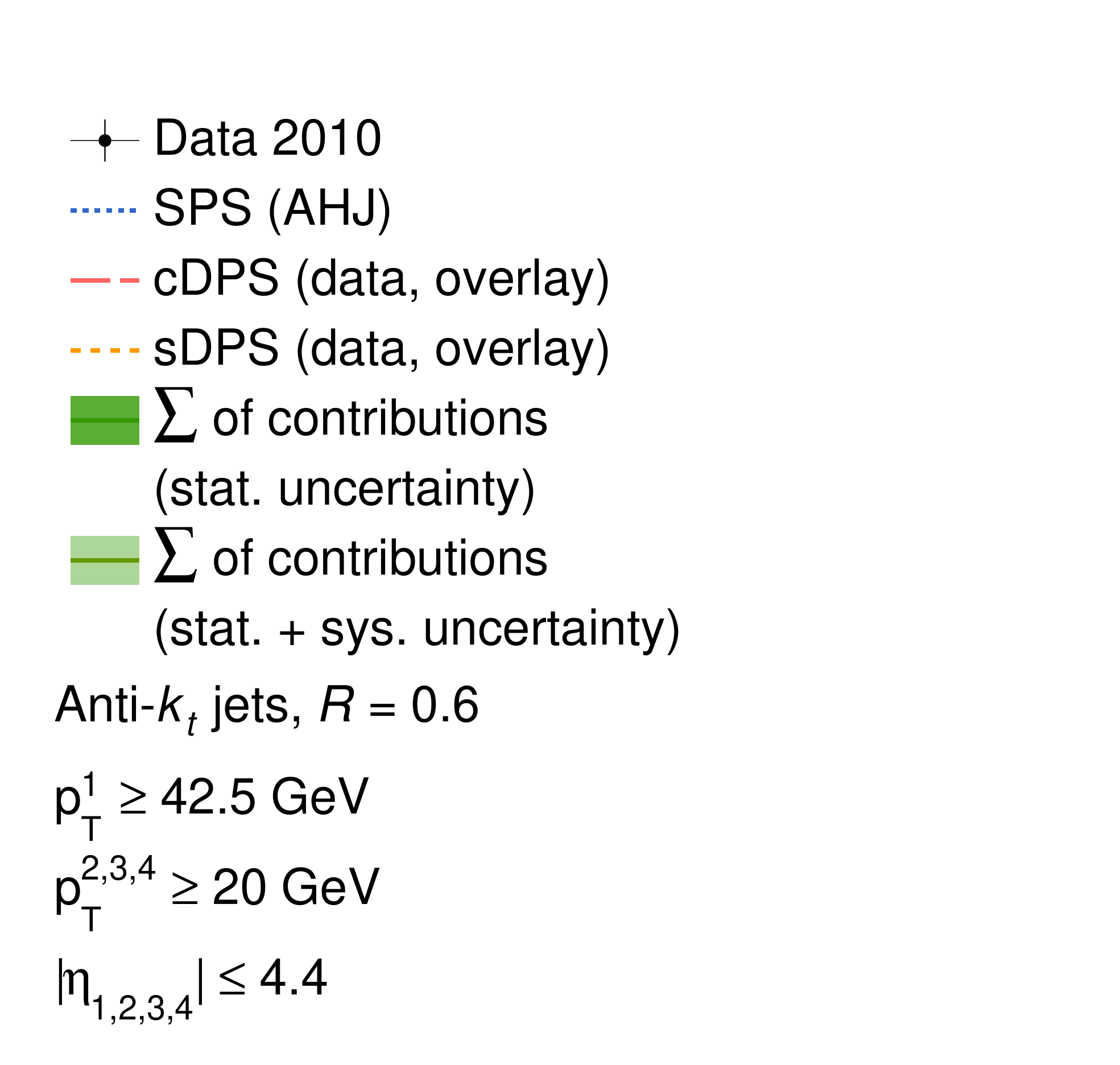}
    \end{tabular}
} \\ [-2.2ex]
\subfloat[]{\label{fig:relativeCont:dphi34}
    \begin{tabular}{p{0.66\textwidth} p{0.25\textwidth}}
        \vspace{0pt} \sfig[trim=0mm 0mm 10mm 15mm,clip,width=0.66\textwidth,vsep=9.50cm, hsep=6.00cm]{\large (b)}{figures/relativeCont8} &
        \vspace{2.0ex} \includegraphics[trim=9mm 8mm 79mm 20mm,clip,width=0.25\textwidth]{figures/relativeCont22}
    \end{tabular}
}
\caption{\label{fig:relativeCont}
    Comparison between the distributions of the variables \Subref{fig:relativeCont:dpt34} \dpt{34} and \Subref{fig:relativeCont:dphi34} \dphi{34}, 
    defined in \autoref{eq:inputVariablesPCA}, in four-jet events in data and the sum of the \sps, \cDps and \sDps contributions, as indicated in the legend.
    The sum of the contributions is normalized to the cross-section measured in data and 
    the various contributions are normalized to their respective fractions obtained from the fit.
    In the sum of contributions, statistical uncertainties are shown as the dark shaded area and the light shaded area represents 
    the sum in quadrature of the statistical and systematic uncertainties.
    The ratio of the sum of contributions to the data is shown in the bottom panels.
}
\end{center}
\end{figure}

In order to visualize the results of the fit, the ternary distribution is divided into five slices,
\begin{itemize}
    \item $0.0 \le \xisDps < 0.1$,
    \item $0.1 \le \xisDps < 0.3$,
    \item $0.3 \le \xisDps < 0.5$,
    \item $0.5 \le \xisDps < 0.7$,
    \item $0.7 \le \xisDps \le 1.0$.
\end{itemize}
A comparison of the fit distributions with the distributions in data in the five slices of the ternary plot is shown in \autoref{fig:dataFitWithSys}.
Considering the systematic uncertainties,
the most significant difference between the data and the fit is seen for the two left-most bins in the range $0.0 \le \xisDps < 0.1$ (\autoref{fig:dataFitWithSys:slice1}) of the ternary plot.
These bins are dominated by the \sps contribution. 
Thus, a discrepancy between the data and the fit result in these bins is expected to have a negligible effect on the measurement of the \dps rate.
A discrepancy between the data and the fit result is also observed in the three rightmost bins in \autoref{fig:dataFitWithSys:slice1}. 
These bins have about~a~30\% contribution from \cDps.
To test the effect of this discrepancy on the description of observables in data, 
the distributions of the various variables in data were compared to a combination of the distributions in the \sps, \cDps and \sDps samples,
normalizing the latter three distributions to their respective fractions in the data as obtained in the fit.
This comparison for the~\dpt{34} and \dphi{34} variables is shown in \autoref{fig:relativeCont}, where a good description of the data is observed.
The same level of agreement is seen for all the variables.


Before calculating \sigeff, the symmetry factor in \autoref{eq:sigEffFourJet} has to be adjusted because there is an overlap in the cross-sections 
$\sigma_{\text{2j}}^{\text{A}}$ and~$\sigma_{\text{2j}}^{\text{B}}$ when the leading jet in sample $\text{A}$ has \ptGe{42.5} (see \autoref{eq:eventSelection}). 
The adjusted symmetry factor is
\begin{equation}
    \frac{1}{1 + \delta_{\!\text{AB}}} \longrightarrow 1 - \frac{1}{2} \frac{ \sigma_{\text{2j}}^{\text{B}} }{ \sigma_{\text{2j}}^{\text{A}} } = 0.9353 \pm 0.0003 \stat \,,
    \label{eq:symmetryFactor}
\end{equation}
as determined from the measured dijet cross-sections.
This factor was also determined using \pythia{6} and good agreement was observed between the two values.
The relative difference in the value of \sigeff obtained by using the symmetry factors extracted from the data and from \pythia{6} was of the order of~0.2\%, 
a negligible difference compared to the statistical uncertainty of \sigeff.

An additional correction of~$+4$\% is applied to the measured \dps cross-section 
due to the probability of jets from the secondary interaction overlapping with jets from the primary interaction.
In this configuration, the \AKT algorithm merges the two overlapping jets into one, and hence the event cannot pass the four-jet requirement.
The value of this correction was determined from the fraction of phase space occupied by a jet. 
It was also determined directly in \ahj and good agreement between the two values was observed.

\begin{figure}[htbp]
    \captionsetup{aboveskip=0pt}
\begin{center}
\includegraphics[trim=0mm 0mm 0mm 4mm,clip,width=0.80\textwidth]{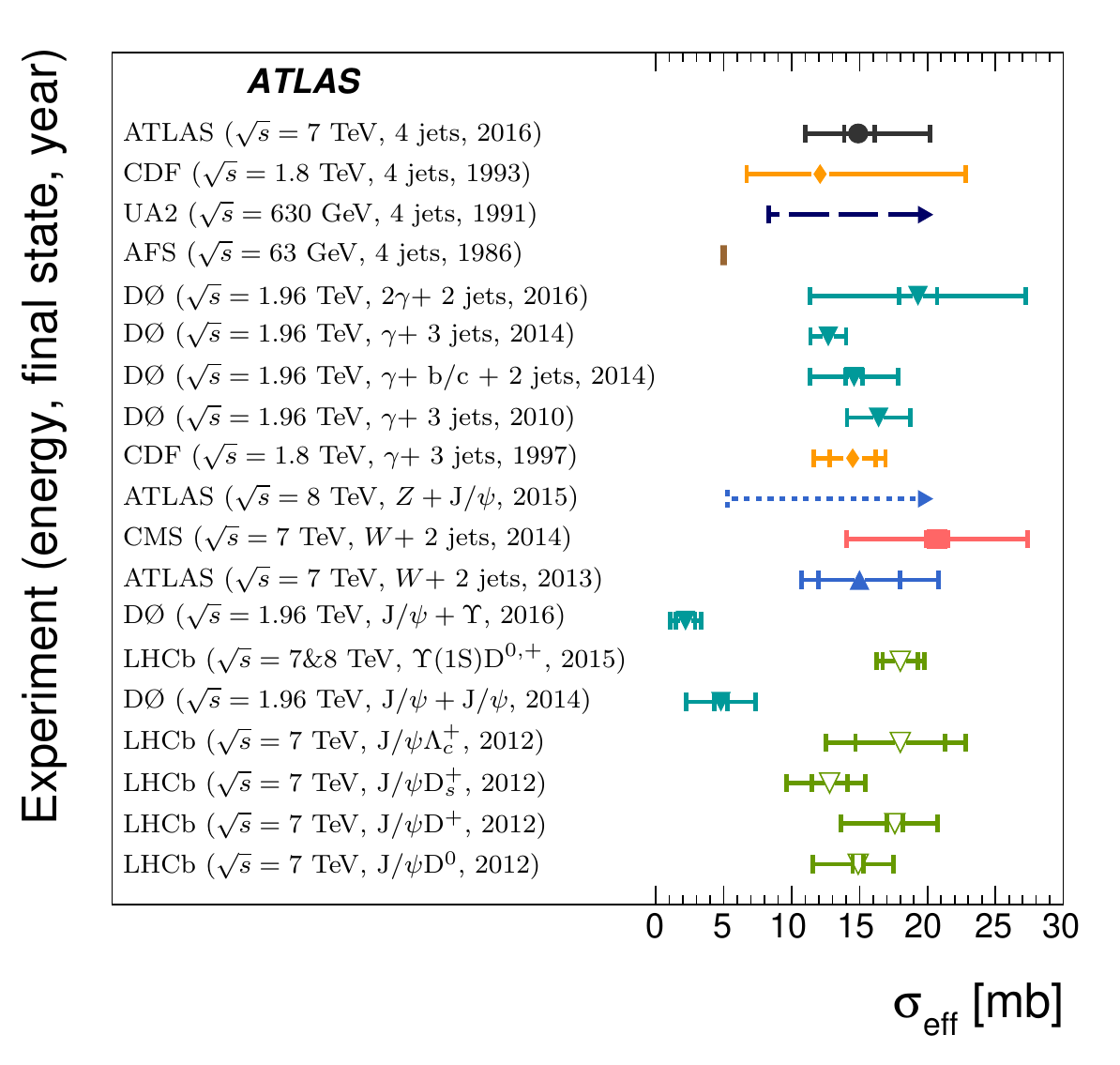}
\caption{\label{fig:vertSigmaEffPlot}
    The effective cross-section, \sigeff, determined in various final states and in different
    experiments~\protect\cite{
    Akesson:173908,Alitti1991145,PhysRevD.47.4857,Abe:1997bp,PhysRevD.56.3811, Abazov:2009gc,Aaij:2012dz,Aad:2013bjm,Chatrchyan:2013xxa,Abazov:2014fha,Abazov:2014qba,Aad:2014kba,Aaij:2015wpa,Abazov:2015fbl,Abazov:2015nnn}.
    The inner error bars (where visible) correspond to the statistical uncertainties and the outer error bars represent the sum in quadrature of the statistical and systematic uncertainties.
    Dashed arrows indicate lower limits and the vertical line represents the AFS measurement published without uncertainties.
}
\end{center}
\end{figure}

\begin{figure}[htbp]
    \captionsetup{aboveskip=0pt}
\begin{center}
\begin{tabular}{p{0.65\textwidth} p{0.35\textwidth}}
  \vspace{0pt}\includegraphics[trim=0mm 0mm 7mm 15mm,clip,width=0.65\textwidth]{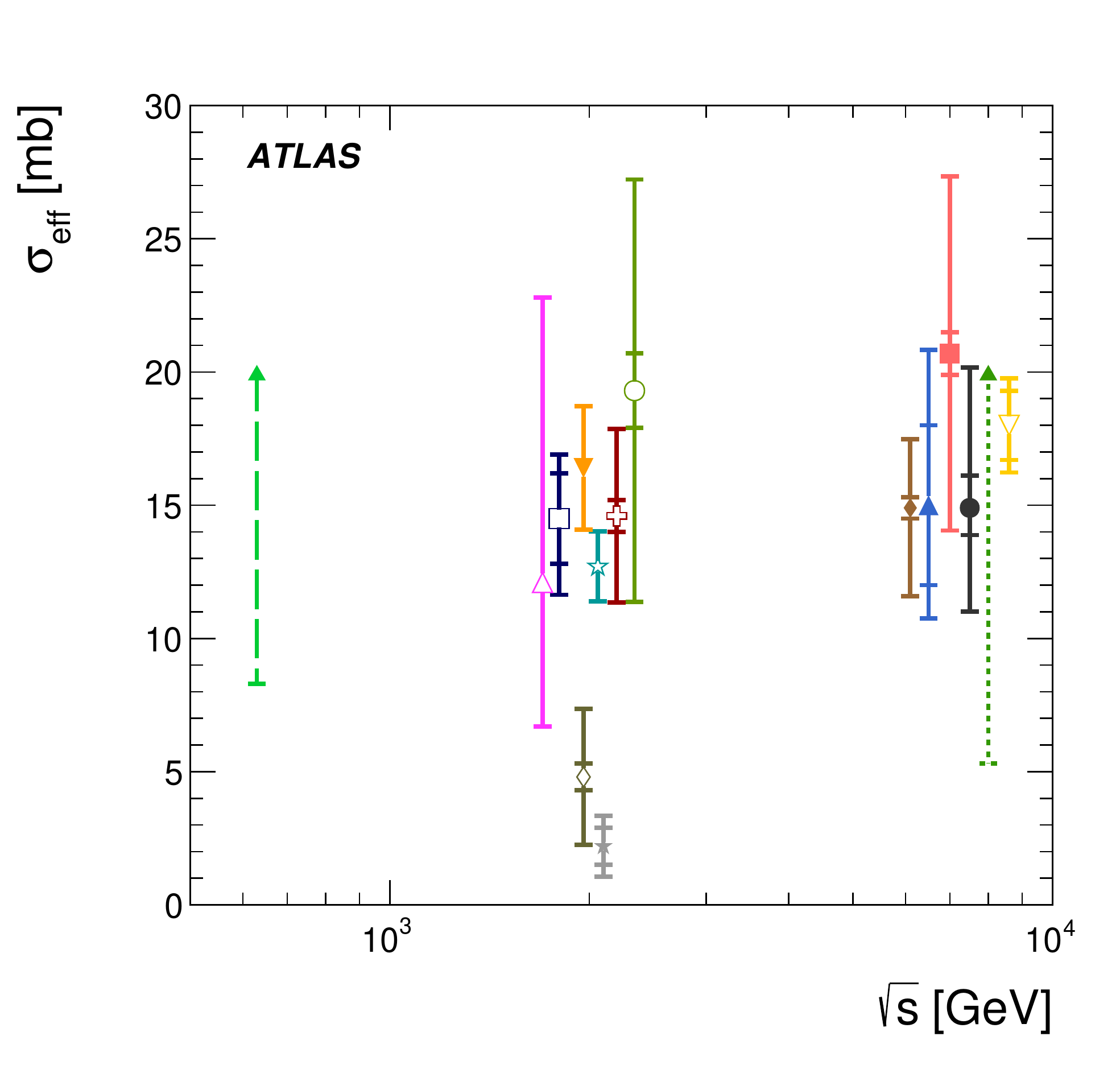} &
  \vspace{2.5ex}\includegraphics[trim=12mm 17mm 40mm 7mm,clip,width=0.35\textwidth]{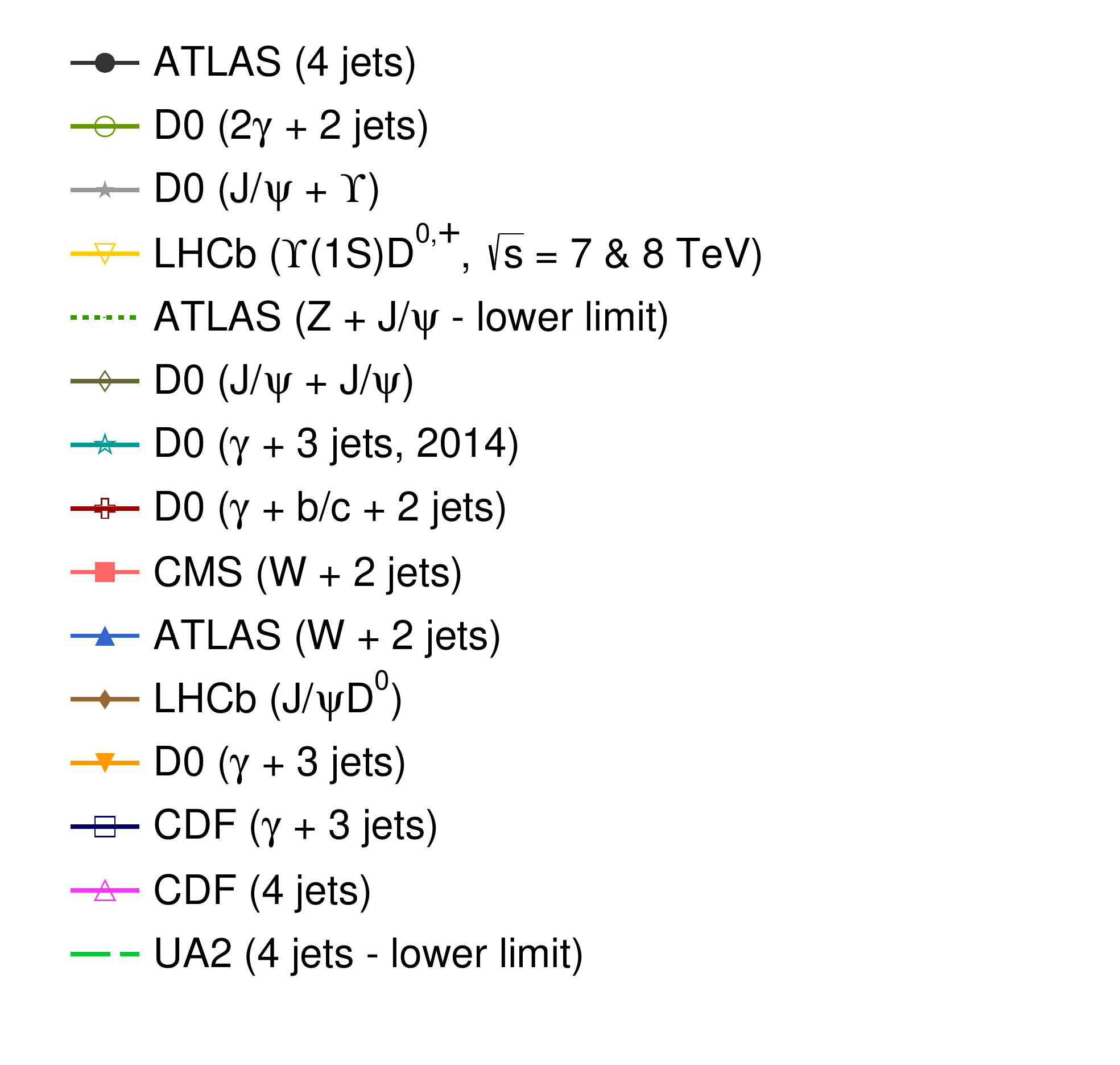}
\end{tabular}
\caption{\label{fig:sigmaEffPlot}
    The effective cross-section, \sigeff, as a function of the centre-of-mass energy, $\sqrt{s}$, for a representative set of 
    measurements~\protect\cite{Alitti1991145,PhysRevD.47.4857,Abe:1997bp,PhysRevD.56.3811,Abazov:2009gc,Aaij:2012dz,Aad:2013bjm,Chatrchyan:2013xxa,Abazov:2014fha,Abazov:2014qba,Aad:2014kba,Aaij:2015wpa,Abazov:2015fbl,Abazov:2015nnn}.
    The inner error bars (where visible) correspond to the statistical uncertainties and the outer error bars represent the sum in quadrature of the statistical and systematic uncertainties.
    Dashed arrows indicate lower limits.
    For clarity, measurements at identical centre-of-mass energies are slightly offset in $\sqrt{s}$.
}
\end{center}
\end{figure}

Finally, the measurements of the dijet and four-jet cross-sections can be used to calculate the effective cross-section, yielding
\begin{equation}
    \sigeff = \measSigeff \, .
\label{eq:finalSigmaEff}
\end{equation}
This value is consistent within the quoted uncertainties with previous measurements, performed by the ATLAS collaboration and by other experiments~\cite{
Akesson:173908,Alitti1991145,PhysRevD.47.4857,Abe:1997bp,PhysRevD.56.3811, Abazov:2009gc,Aaij:2012dz,Aad:2013bjm,Chatrchyan:2013xxa,Abazov:2014fha,Abazov:2014qba,Aad:2014kba,Aaij:2015wpa,Abazov:2015fbl,Abazov:2015nnn},
all of which are summarized in \autoref{fig:vertSigmaEffPlot}.
\Autoref{fig:sigmaEffPlot} shows \sigeff as a function of $\sqrt{s}$, where the AFS result and some of the LHCb results are omitted for clarity.
Within the large uncertainties, the measurements are consistent with no $\sqrt{s}$ dependence of \sigeff.
The \sigeff value obtained is \fracInelXsecPer of the inelastic cross-section, $\sigma_{\mathrm{inel}}$, measured by ATLAS at $\sqrt{s} = 7 \tev$~\cite{Aad:2014dca}.

%


\section{Normalized differential cross-sections\label{sec:unfolding}}

To allow the results of this study to be used in future comparisons with MPI models, 
the distributions of the variables used as input to the NN were corrected for detector effects.
The corrections were derived using an iterative unfolding, producing an unfolding matrix for each observable, relating the particle-level and reconstructed-level quantities.
These matrices were derived using samples of four-jet events selected from the \ahj and \pythia{}6 samples by imposing the cuts detailed in \autoref{eq:eventSelection} on particle jets.
The \ahj sample generated with the AUET1 tune was used to derive the unfolding matrix. 
The distributions were unfolded with the Bayesian unfolding algorithm, implemented in the RooUnfold package~\cite{Adye:2011gm}, using two iterations.

\begin{figure}[htbp]
    \captionsetup[subfloat]{captionskip=-3ex}
\begin{center}
\subfloat[]{\label{fig:dataUnfold:dpt34}\includegraphics[trim=0mm 0mm 0mm 0mm,clip,width=0.5\textwidth]{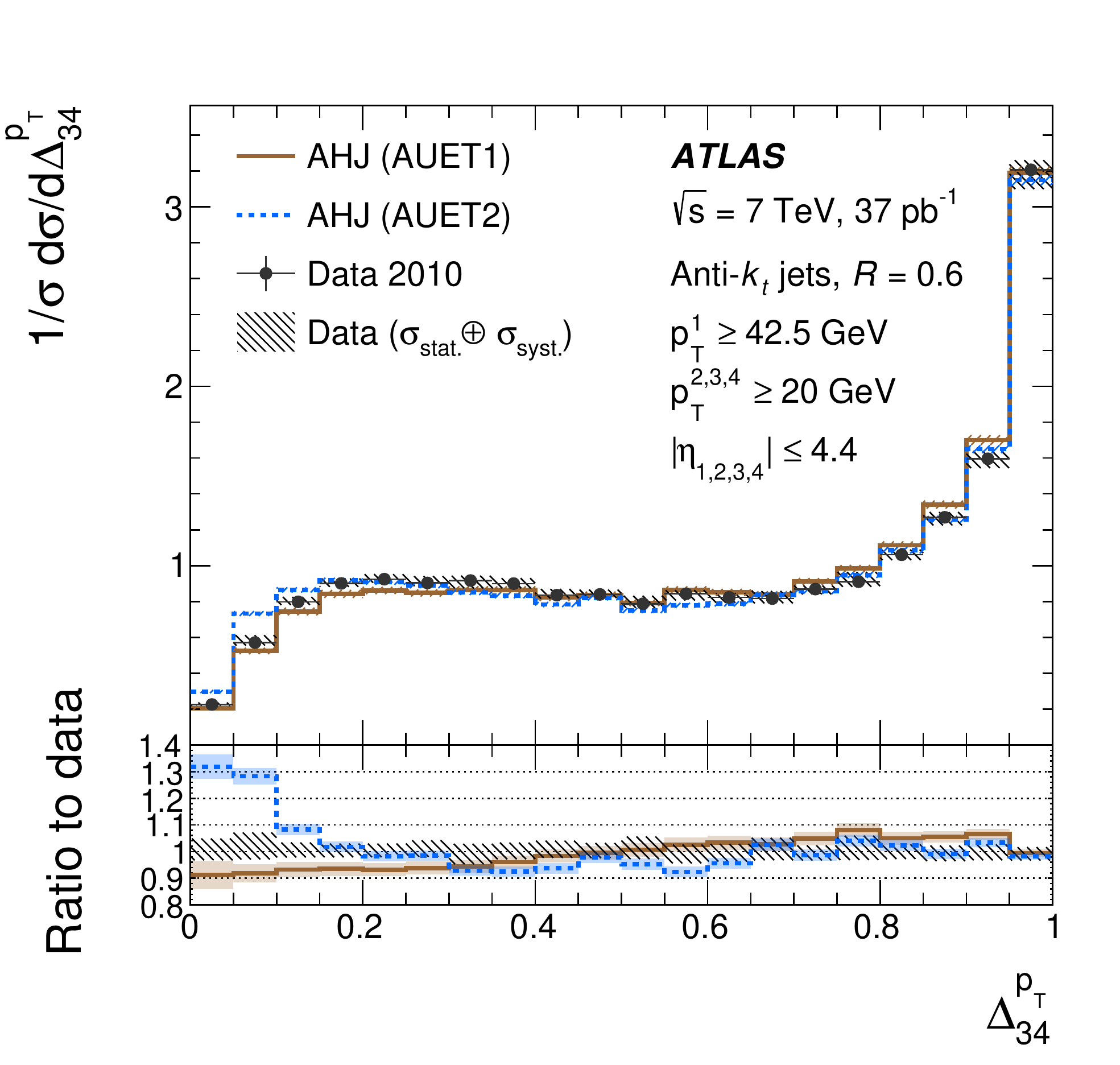}}
\subfloat[]{\label{fig:dataUnfold:dphi34}\includegraphics[trim=0mm 0mm 0mm 0mm,clip,width=0.5\textwidth]{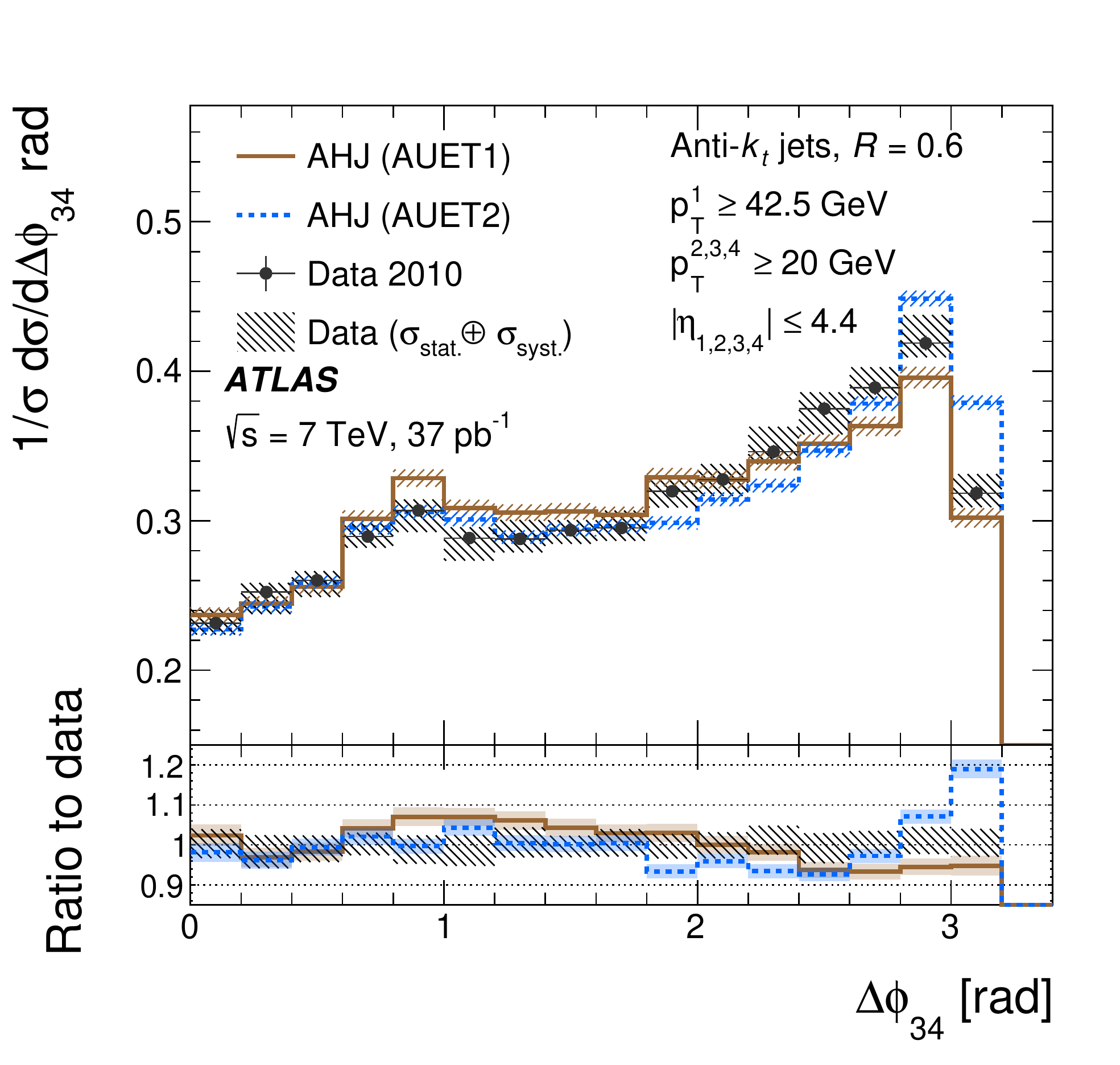}}
\caption{\label{fig:dataUnfold}
    Distributions of the variables \Subref{fig:dataUnfold:dpt34} \dpt{34} and \Subref{fig:dataUnfold:dphi34} \dphi{34}, 
    defined in \autoref{eq:inputVariablesPCA}, in data after unfolding to particle level, compared to the MC prediction from \ahj at the particle level, 
    generated using the AUET1 and AUET2 tunes, as indicated in the legend.
    The hatched area represents the sum in quadrature of the statistical and systematic uncertainties in the normalized differential cross-sections 
    and all histograms are normalized to unity.
    The ratio of the particle-level distribution to the normalized differential cross-section is shown in the bottom panels, where the shaded areas represent statistical uncertainties.
}
\end{center}
\end{figure}

The unfolding matrices derived from \ahj were taken as the nominal matrices
and the differences observed when using the matrices derived from \pythia{}6 were used as an additional systematic uncertainty, typically of the order of a few percent in each bin.
The total systematic uncertainty of the differential distributions in data was obtained by summing in quadrature the uncertainty due to MC modelling in a given bin 
with the systematic uncertainties in this bin due to the JES and jet energy and angular resolution uncertainties, while preserving correlations between bins.
\Autoref{fig:dataUnfold} shows the normalized differential cross-section distribution in data for the \dpt{34} and \dphi{34} variables 
compared to the particle-level distributions in the \ahj samples generated with the AUET1 and AUET2 tunes.
The particle-level distributions in the AUET2 \ahj sample overestimate the normalized differential cross-section distributions in data in the regions $\dpt{34} \le 0.15$ and $\dphi{34} \ge 2.8$, 
demonstrating the excess of the \dps contribution in this sample compared to the data.
On the other hand, the \dps contribution in the data is underestimated by the prediction obtained with the AUET1 tune.
These comparisons demonstrate the power of these distributions to constrain MPI models and tunes.
In \autoref{sec:appendix}, 
the normalized differential cross-sections in data for the remaining variables are compared to the particle-level distributions in the \ahj samples generated using the AUET1 and AUET2 tunes.

%


\section{Summary and conclusions\label{sec:Conclusions}}

A measurement of the rate of hard \DPS in four-jet events was performed using a sample of data collected with the ATLAS experiment at the LHC in 2010, 
with an average of approximately~0.4 proton--proton interactions per bunch crossing, corresponding to an integrated luminosity of $37.3 \pm 1.3~\mathrm{pb}^{-1}$.
Three different samples were selected, all consisting of single-vertex events from proton--proton collisions at a centre-of-mass energy of \sqs.
Four-jet events were defined as those containing at least four reconstructed jets with \ptGe{20} and \etaLe{4.4}, 
and at least one jet having \ptGe{42.5}.
Two additional dijet samples were selected with the requirement of having at least two jets with \ptGe{20} and \etaLe{4.4}.
One of the dijet samples was further constrained such that it contained at least one jet with \ptGe{42.5}.

The contribution of hard \DPS to the production of four-jet events was extracted using an artificial neural network.  
The four-jet topology originating from hard \DPS was represented by a random combination of events selected in data.
The fraction of events corresponding to the contribution made by hard \DPS in four-jet events was determined to be,
\begin{equation}
    \fdps = \measFdps \,. 
\label{eq:finalFdpsConc} 
\end{equation}
After combining this result with measurements of the dijet and four-jet cross-sections in the appropriate phase space regions,
the effective cross-section was determined to be 
\begin{equation*}
    \sigeff = \measSigeff \, .
\label{eq:finalSigmaEffConc}
\end{equation*}
This value is \fracInelXsecPer of the measured value of $\sigma_{\mathrm{inel}}$ at $\sqrt{s} = 7 \tev$ 
and is consistent with previous measurements performed at various centre-of-mass energies and in various final states.
It is compatible with a model in which \sigeff is a universal parameter that does not depend on the process or phase space.
To facilitate future studies of the dynamics of multi-parton interactions, distributions of observables sensitive to the presence of hard \DPS are also presented.

\FloatBarrier

\section*{Acknowledgements} %


We thank CERN for the very successful operation of the LHC, as well as the
support staff from our institutions without whom ATLAS could not be
operated efficiently.

We acknowledge the support of ANPCyT, Argentina; YerPhI, Armenia; ARC, Australia; BMWFW and FWF, Austria; ANAS, Azerbaijan; SSTC, Belarus; CNPq and FAPESP, Brazil; NSERC, NRC and CFI, Canada; CERN; CONICYT, Chile; CAS, MOST and NSFC, China; COLCIENCIAS, Colombia; MSMT CR, MPO CR and VSC CR, Czech Republic; DNRF and DNSRC, Denmark; IN2P3-CNRS, CEA-DSM/IRFU, France; GNSF, Georgia; BMBF, HGF, and MPG, Germany; GSRT, Greece; RGC, Hong Kong SAR, China; ISF, I-CORE and Benoziyo Center, Israel; INFN, Italy; MEXT and JSPS, Japan; CNRST, Morocco; FOM and NWO, Netherlands; RCN, Norway; MNiSW and NCN, Poland; FCT, Portugal; MNE/IFA, Romania; MES of Russia and NRC KI, Russian Federation; JINR; MESTD, Serbia; MSSR, Slovakia; ARRS and MIZ\v{S}, Slovenia; DST/NRF, South Africa; MINECO, Spain; SRC and Wallenberg Foundation, Sweden; SERI, SNSF and Cantons of Bern and Geneva, Switzerland; MOST, Taiwan; TAEK, Turkey; STFC, United Kingdom; DOE and NSF, United States of America. In addition, individual groups and members have received support from BCKDF, the Canada Council, CANARIE, CRC, Compute Canada, FQRNT, and the Ontario Innovation Trust, Canada; EPLANET, ERC, FP7, Horizon 2020 and Marie Sk{\l}odowska-Curie Actions, European Union; Investissements d'Avenir Labex and Idex, ANR, R{\'e}gion Auvergne and Fondation Partager le Savoir, France; DFG and AvH Foundation, Germany; Herakleitos, Thales and Aristeia programmes co-financed by EU-ESF and the Greek NSRF; BSF, GIF and Minerva, Israel; BRF, Norway; Generalitat de Catalunya, Generalitat Valenciana, Spain; the Royal Society and Leverhulme Trust, United Kingdom.

The crucial computing support from all WLCG partners is acknowledged gratefully, in particular from CERN, the ATLAS Tier-1 facilities at TRIUMF (Canada), NDGF (Denmark, Norway, Sweden), CC-IN2P3 (France), KIT/GridKA (Germany), INFN-CNAF (Italy), NL-T1 (Netherlands), PIC (Spain), ASGC (Taiwan), RAL (UK) and BNL (USA), the Tier-2 facilities worldwide and large non-WLCG resource providers. Major contributors of computing resources are listed in Ref.~\cite{ATL-GEN-PUB-2016-002}.

\clearpage
\appendix  

\part*{Appendix}
\vspace*{-3ex}
\addcontentsline{toc}{part}{Appendix}
\section{Normalized differential cross-sections\label{sec:appendix}}
\vspace*{-1.5ex}
%

\begin{figure}[htbp]
\captionsetup[subfloat]{captionskip=-3ex}
\begin{center}
\subfloat[]{\label{fig:app:unfoldedDist:dpt12}\includegraphics[trim=0mm 0mm 0mm 15mm,clip,width=0.5\textwidth]{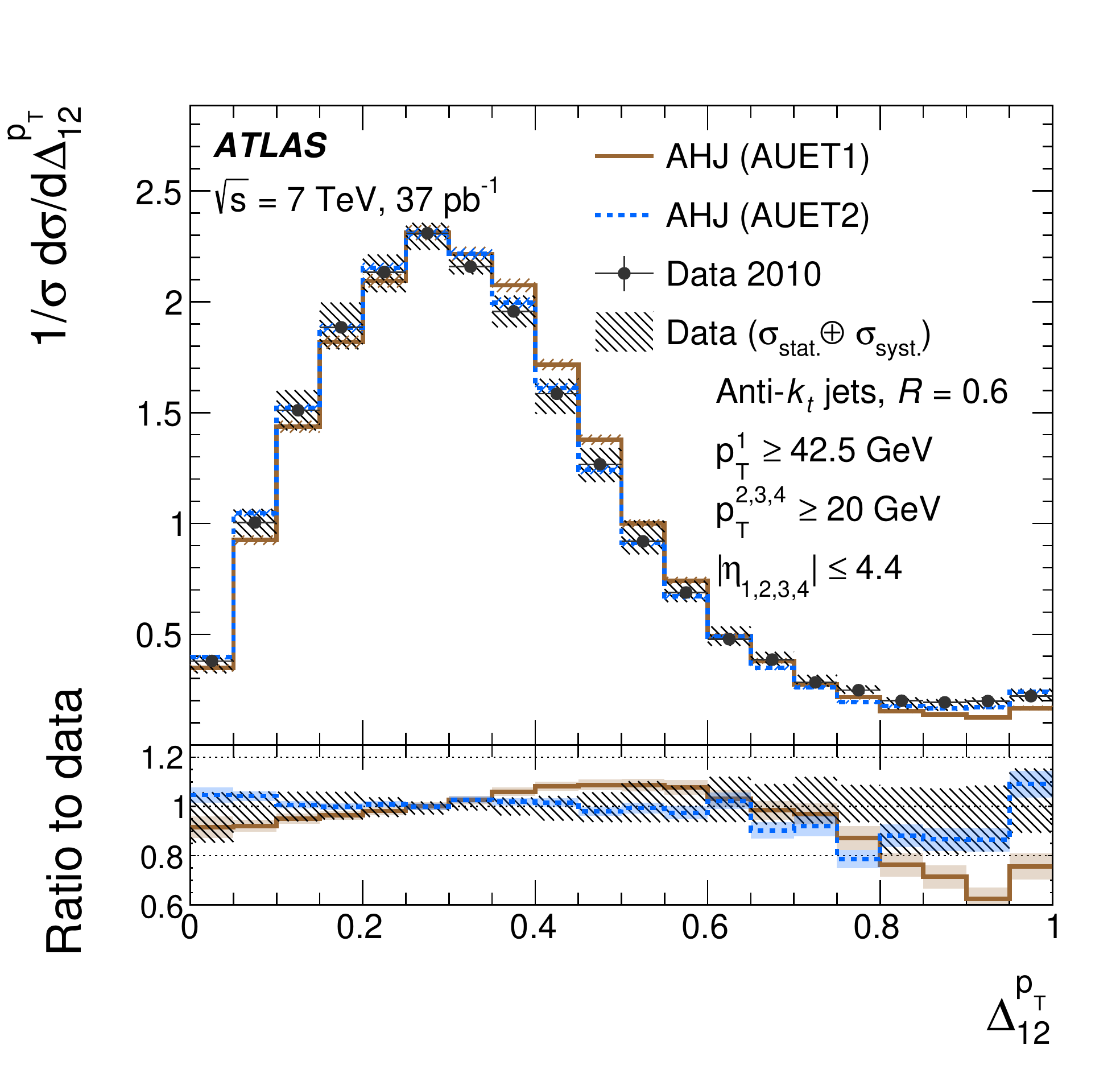}}
\subfloat[]{\label{fig:app:unfoldedDist:dpt13}\includegraphics[trim=0mm 0mm 0mm 15mm,clip,width=0.5\textwidth]{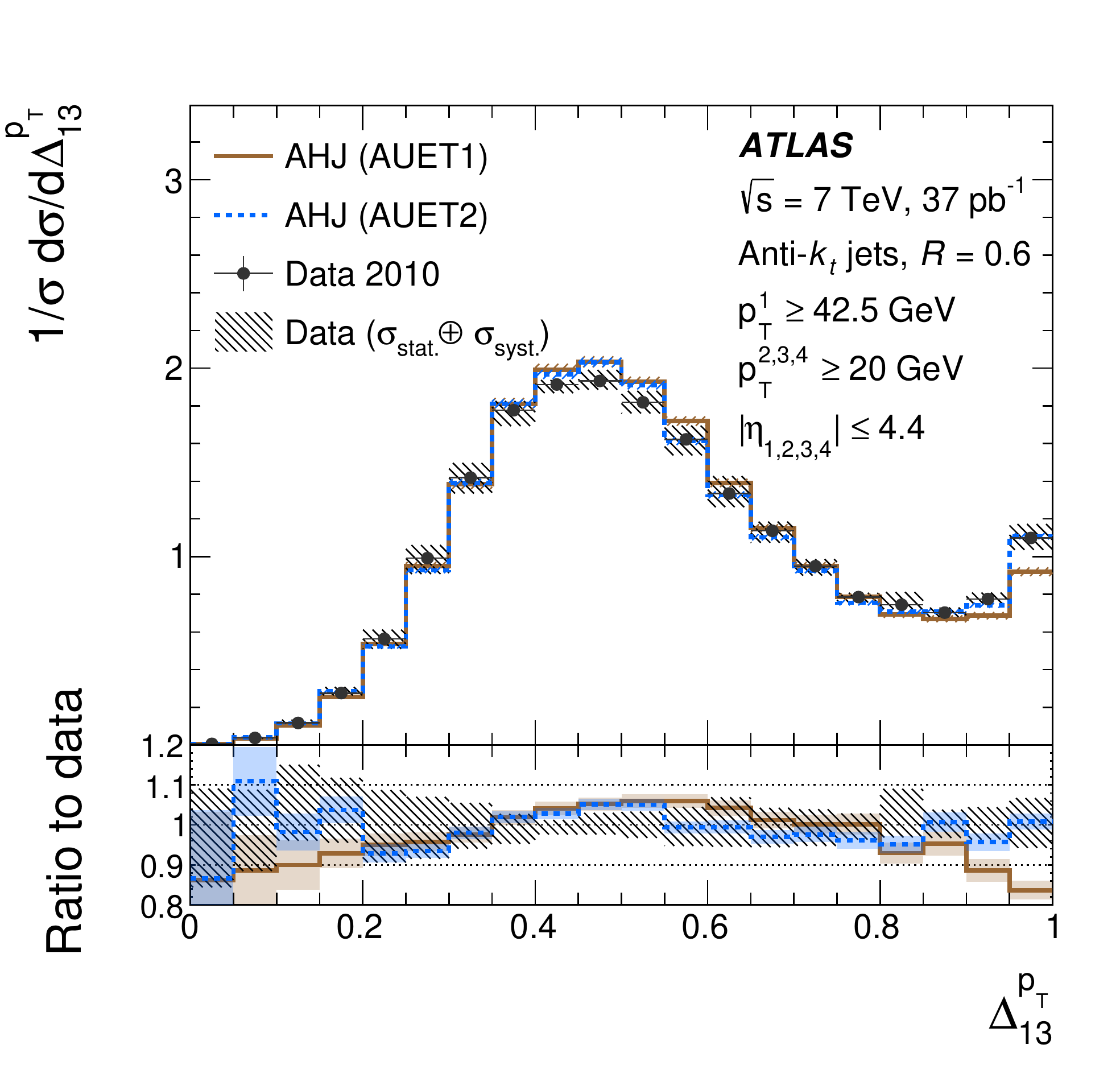}} \\
\subfloat[]{\label{fig:app:unfoldedDist:dpt23}\includegraphics[trim=0mm 0mm 0mm 15mm,clip,width=0.5\textwidth]{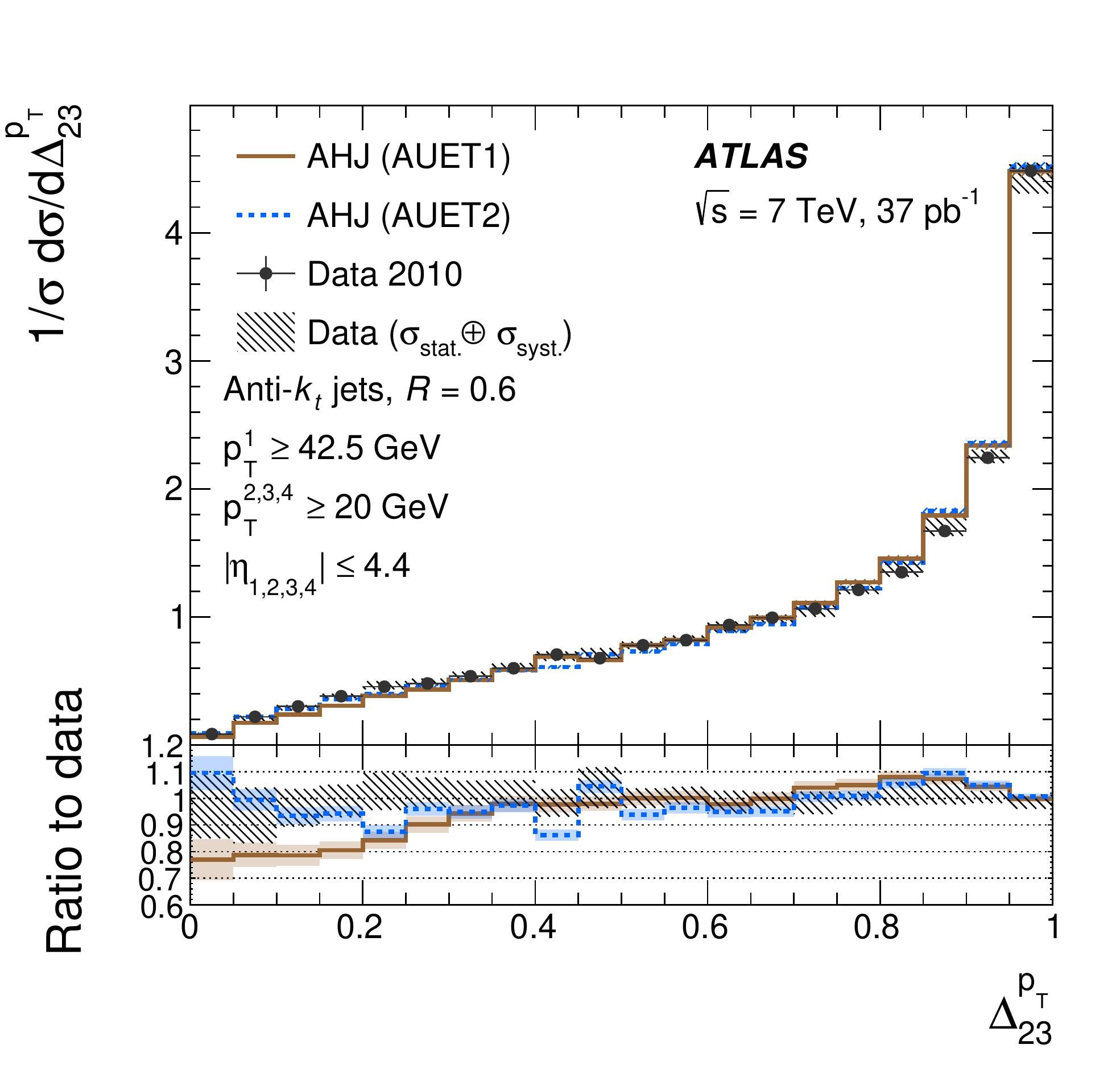}}
\subfloat[]{\label{fig:app:unfoldedDist:dpt14}\includegraphics[trim=0mm 0mm 0mm 15mm,clip,width=0.5\textwidth]{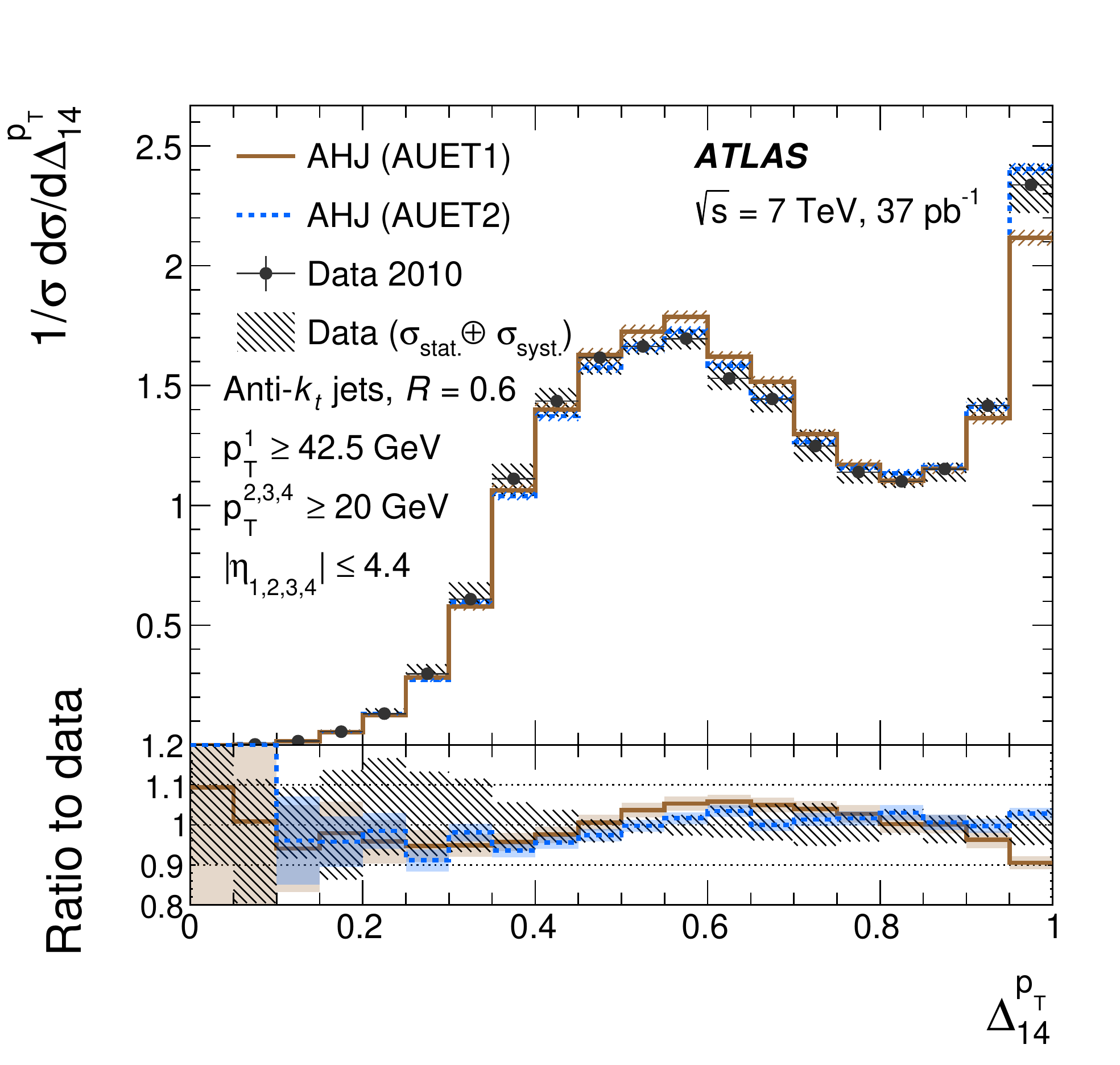}}
\caption{\label{fig:app:unfoldedDist:dpt:12:34}
    Distributions of the variables \Subref{fig:app:unfoldedDist:dpt12} $\dpt{12}$, \Subref{fig:app:unfoldedDist:dpt13} $\dpt{13}$, \Subref{fig:app:unfoldedDist:dpt23} $\dpt{23}$
    and \Subref{fig:app:unfoldedDist:dpt14} $\dpt{14}$, defined in \autoref{eq:inputVariablesPCA}, in data after unfolding to particle level, compared to the MC prediction from \ahj 
    at the particle level, generated using the AUET1 and AUET2 tunes, as indicated in the legend.
    The hatched areas represent the sum in quadrature of the statistical and systematic uncertainties in the normalized differential cross-sections and all histograms are normalized to unity.
    The ratio of the particle-level distribution to the normalized differential cross-section is shown in the bottom panels, where the shaded areas represent statistical uncertainties.
}
\end{center}
\end{figure}

\begin{figure}[p]
\captionsetup[subfloat]{captionskip=-3ex}
\begin{center}
\subfloat[]{\label{fig:app:unfoldedDist:dpt24}\includegraphics[trim=0mm 0mm 0mm 15mm,clip,width=0.5\textwidth]{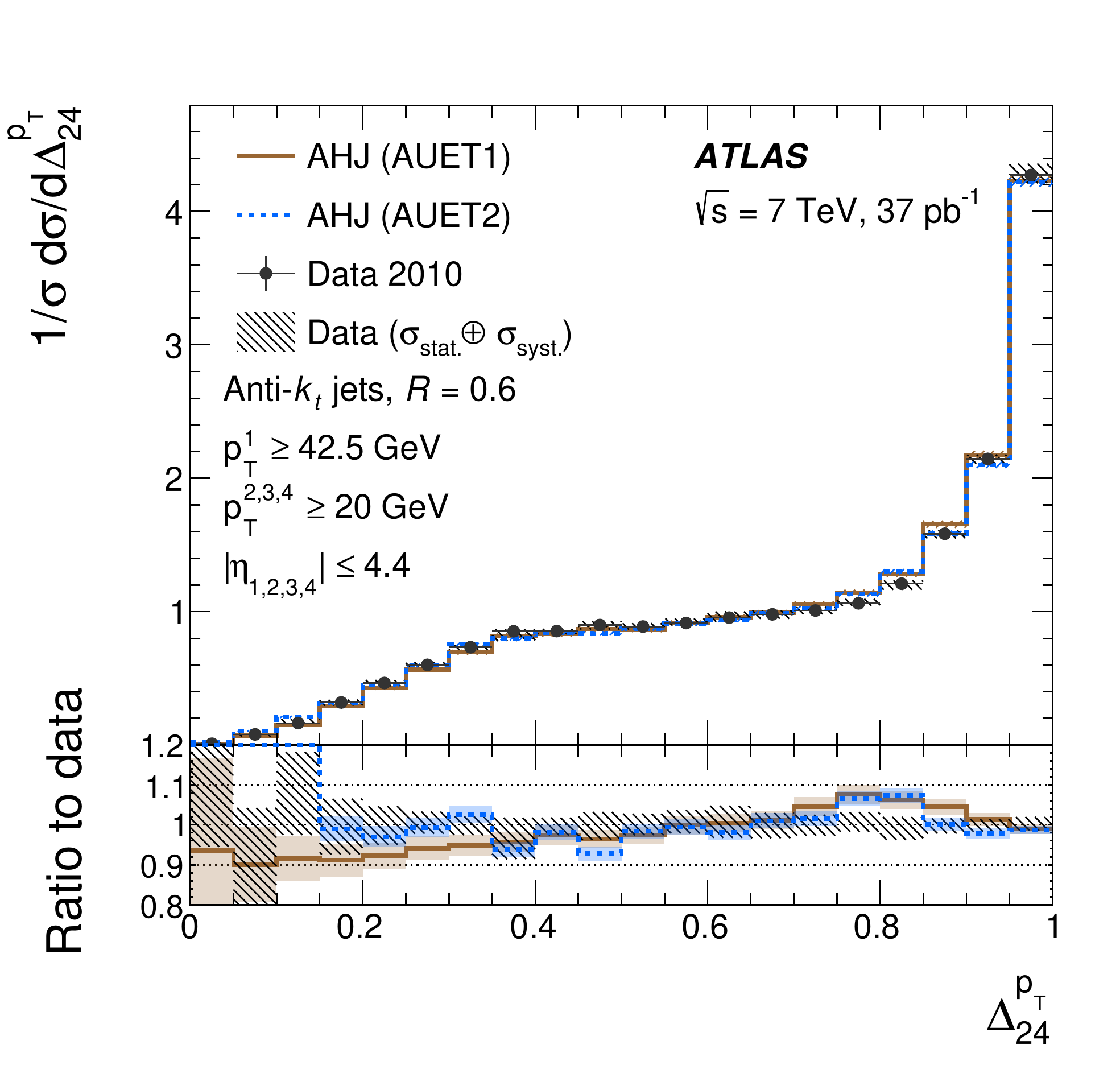}} 
\subfloat[]{\label{fig:app:unfoldedDist:dphi12}\includegraphics[trim=0mm 0mm 0mm 15mm,clip,width=0.5\textwidth]{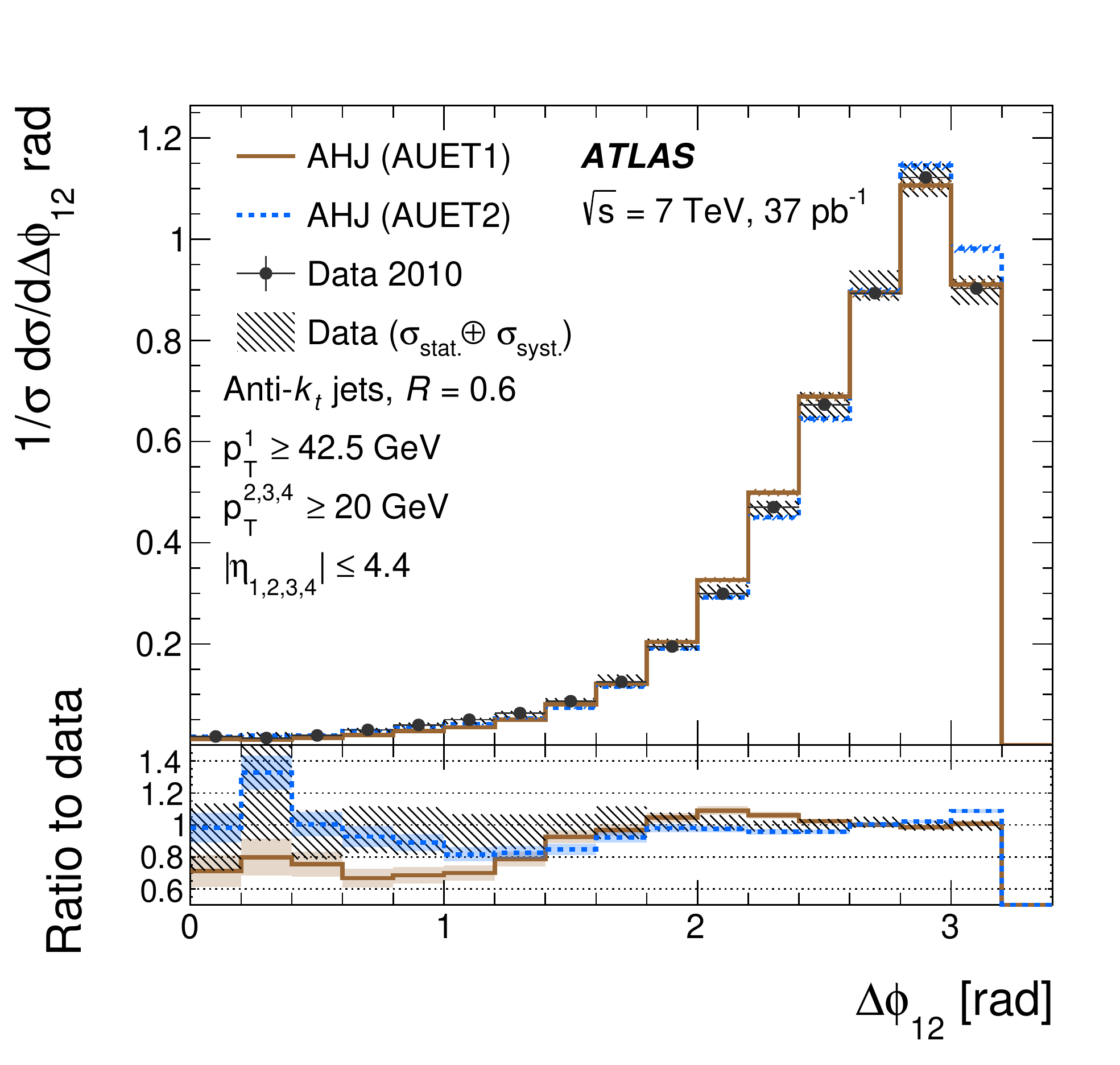}} \\
\subfloat[]{\label{fig:app:unfoldedDist:dphi13}\includegraphics[trim=0mm 0mm 0mm 15mm,clip,width=0.5\textwidth]{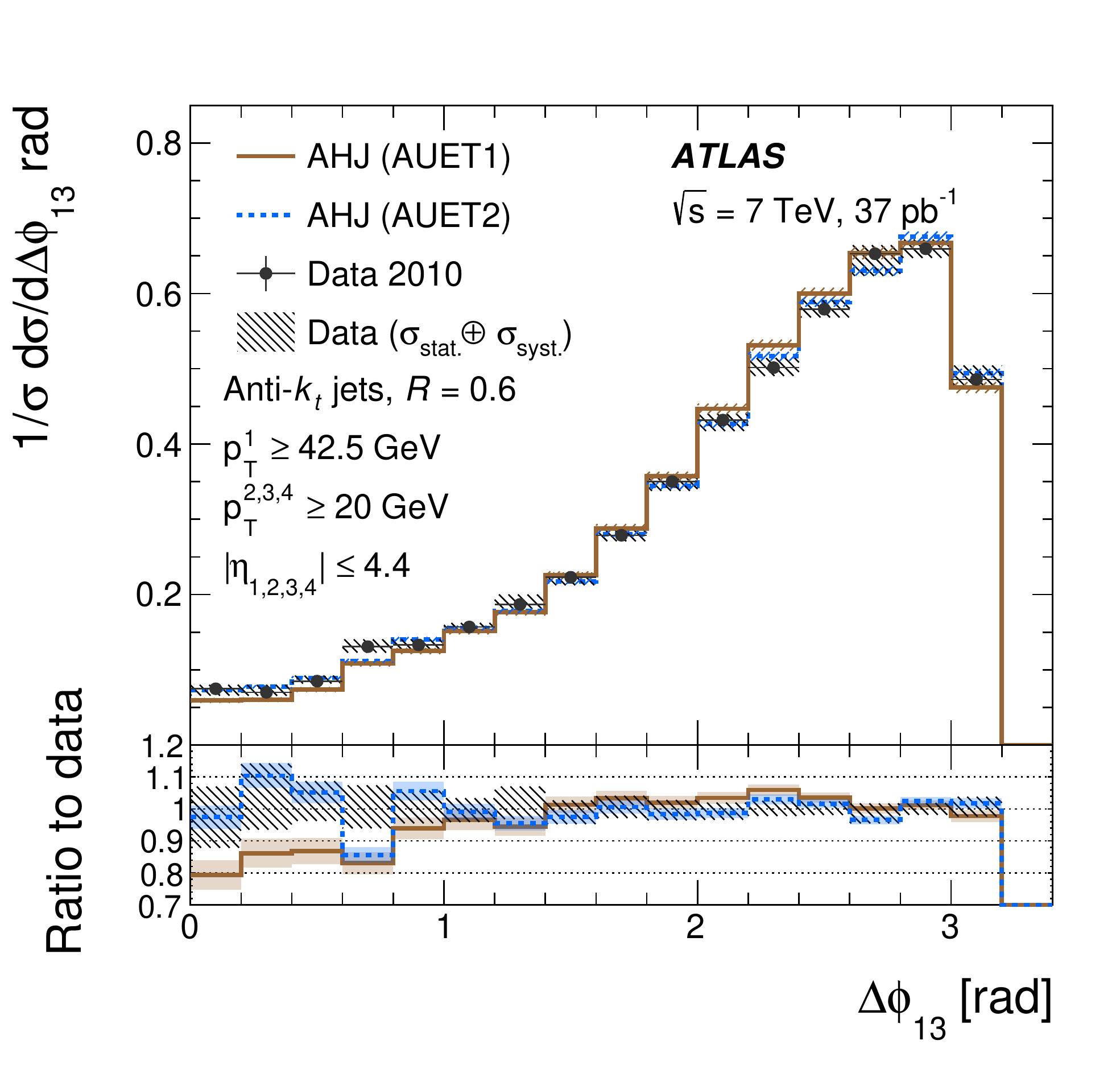}}
\subfloat[]{\label{fig:app:unfoldedDist:dphi23}\includegraphics[trim=0mm 0mm 0mm 15mm,clip,width=0.5\textwidth]{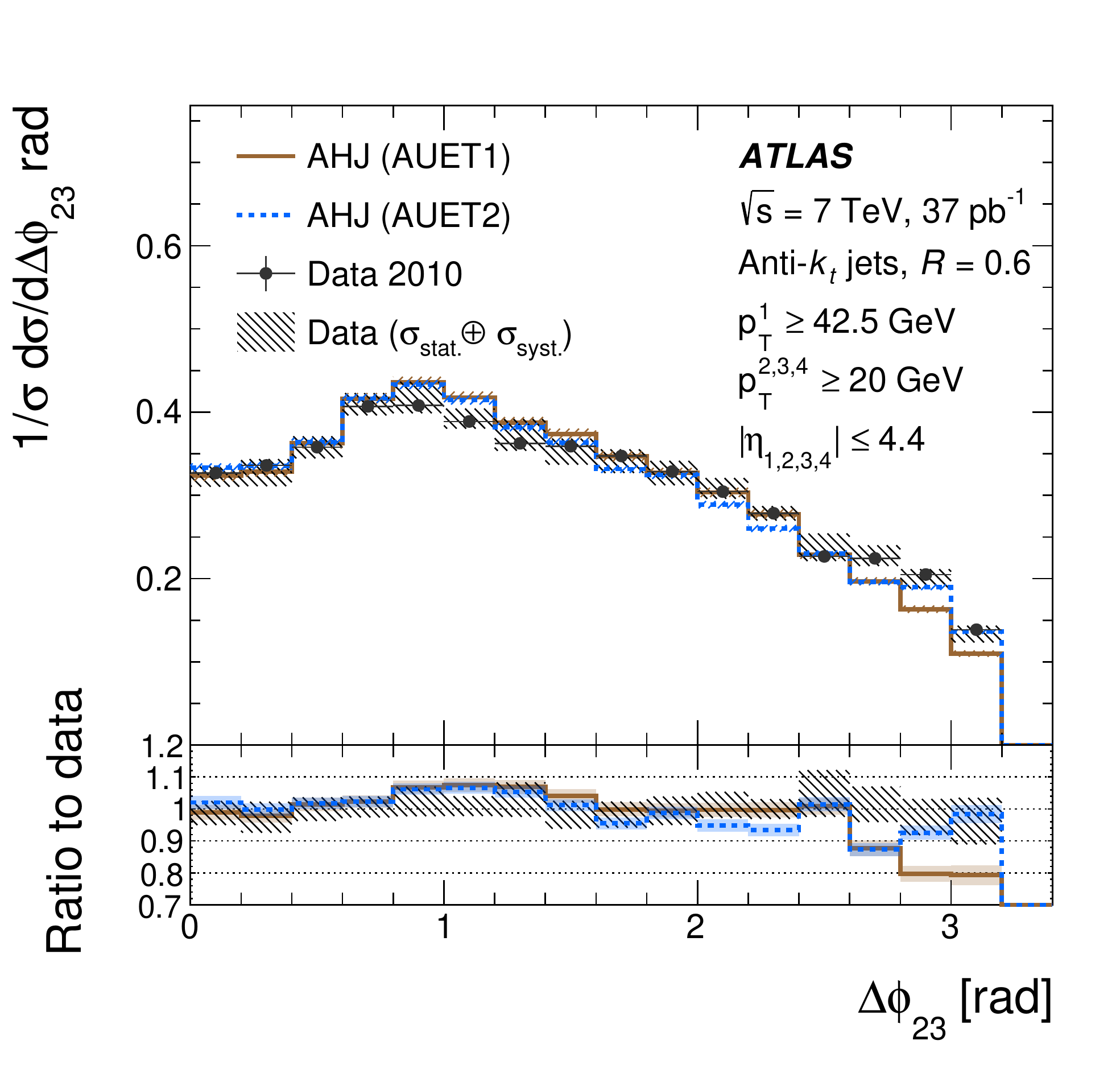}}
\caption{\label{fig:app:unfoldedDist:dpt:14:24}
    Distributions of the variables \Subref{fig:app:unfoldedDist:dpt24} $\dpt{24}$, 
    \Subref{fig:app:unfoldedDist:dphi12} $\dphi{12}$, \Subref{fig:app:unfoldedDist:dphi13} $\dphi{13}$ and \Subref{fig:app:unfoldedDist:dphi23} $\dphi{23}$,
    defined in \autoref{eq:inputVariablesPCA}, in data after unfolding to particle level, compared to the MC prediction from \ahj 
    at the particle level, generated using the AUET1 and AUET2 tunes, as indicated in the legend.
    The hatched areas represent the sum in quadrature of the statistical and systematic uncertainties in the normalized differential cross-sections and all histograms are normalized to unity.
    The ratio of the particle-level distribution to the normalized differential cross-section is shown in the bottom panels, where the shaded areas represent statistical uncertainties.
}
\end{center}
\end{figure}

\begin{figure}[htbp]
\captionsetup[subfloat]{captionskip=-3ex}
\begin{center}
\subfloat[]{\label{fig:app:unfoldedDist:dphi14}\includegraphics[trim=0mm 0mm 0mm 15mm,clip,width=0.5\textwidth]{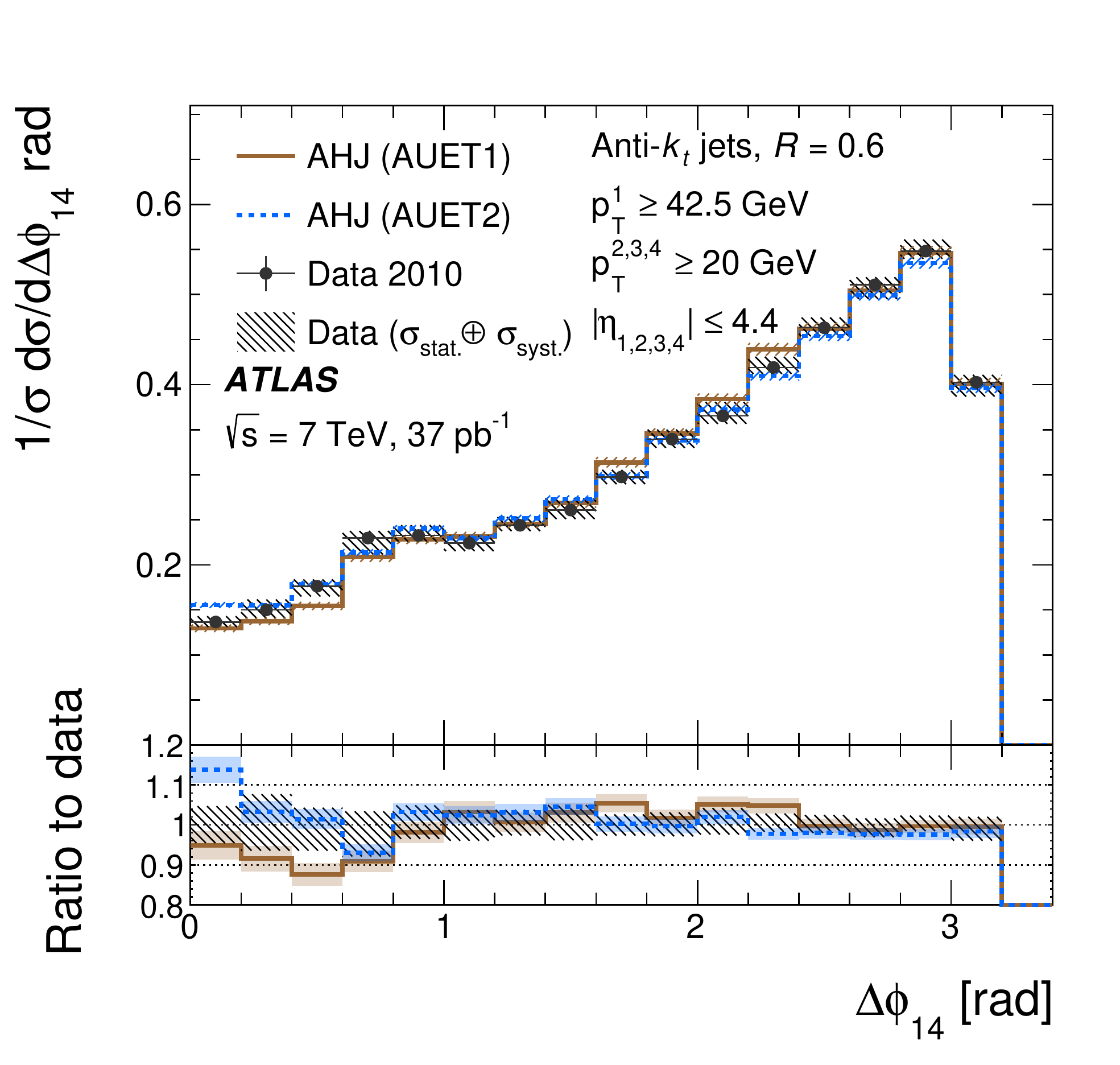}} 
\subfloat[]{\label{fig:app:unfoldedDist:dphi24}\includegraphics[trim=0mm 0mm 0mm 15mm,clip,width=0.5\textwidth]{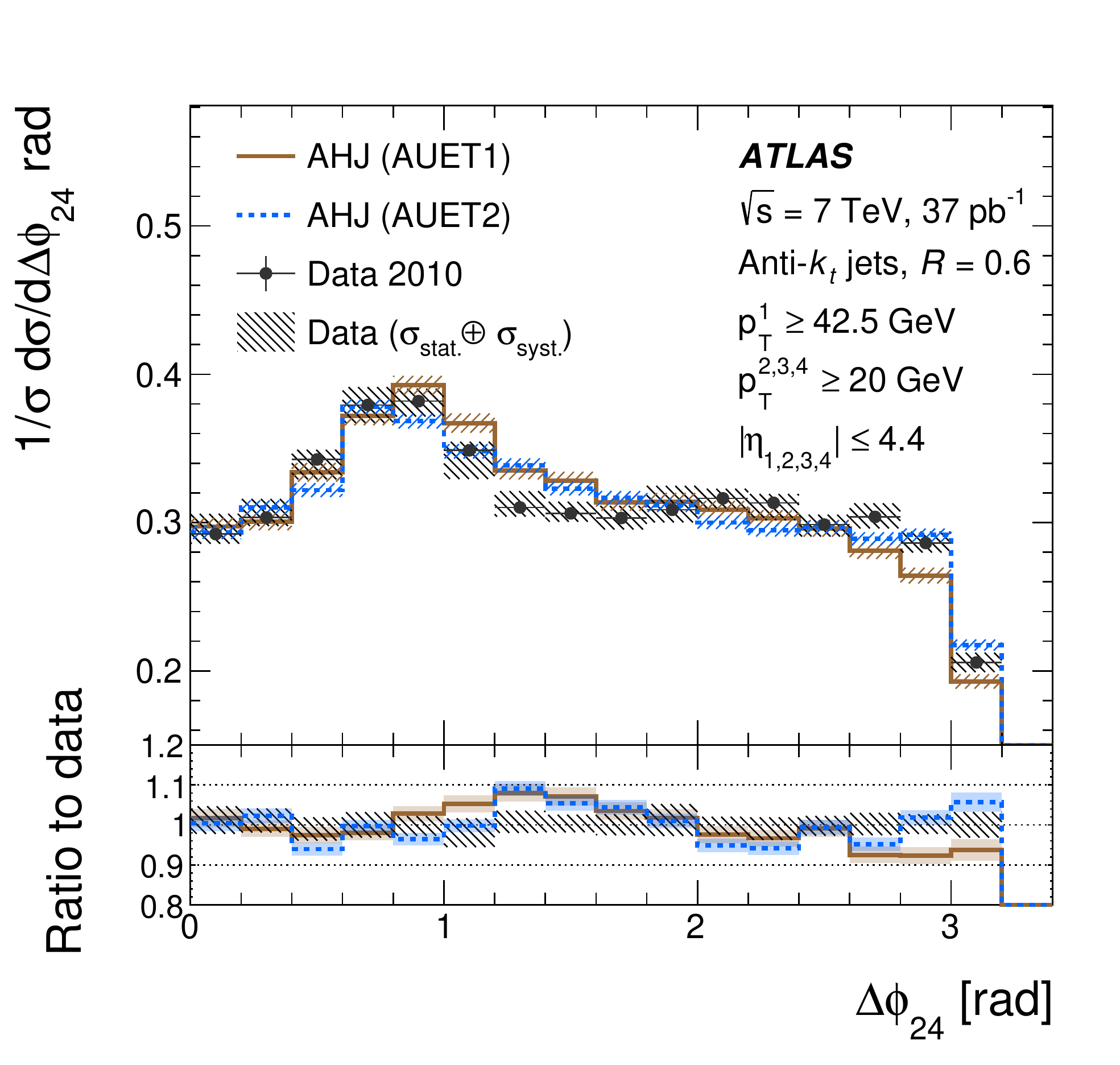}} \\ 
\subfloat[]{\label{fig:app:unfoldedDist:dy12}\includegraphics[trim=0mm 0mm 0mm 15mm,clip,width=0.5\textwidth]{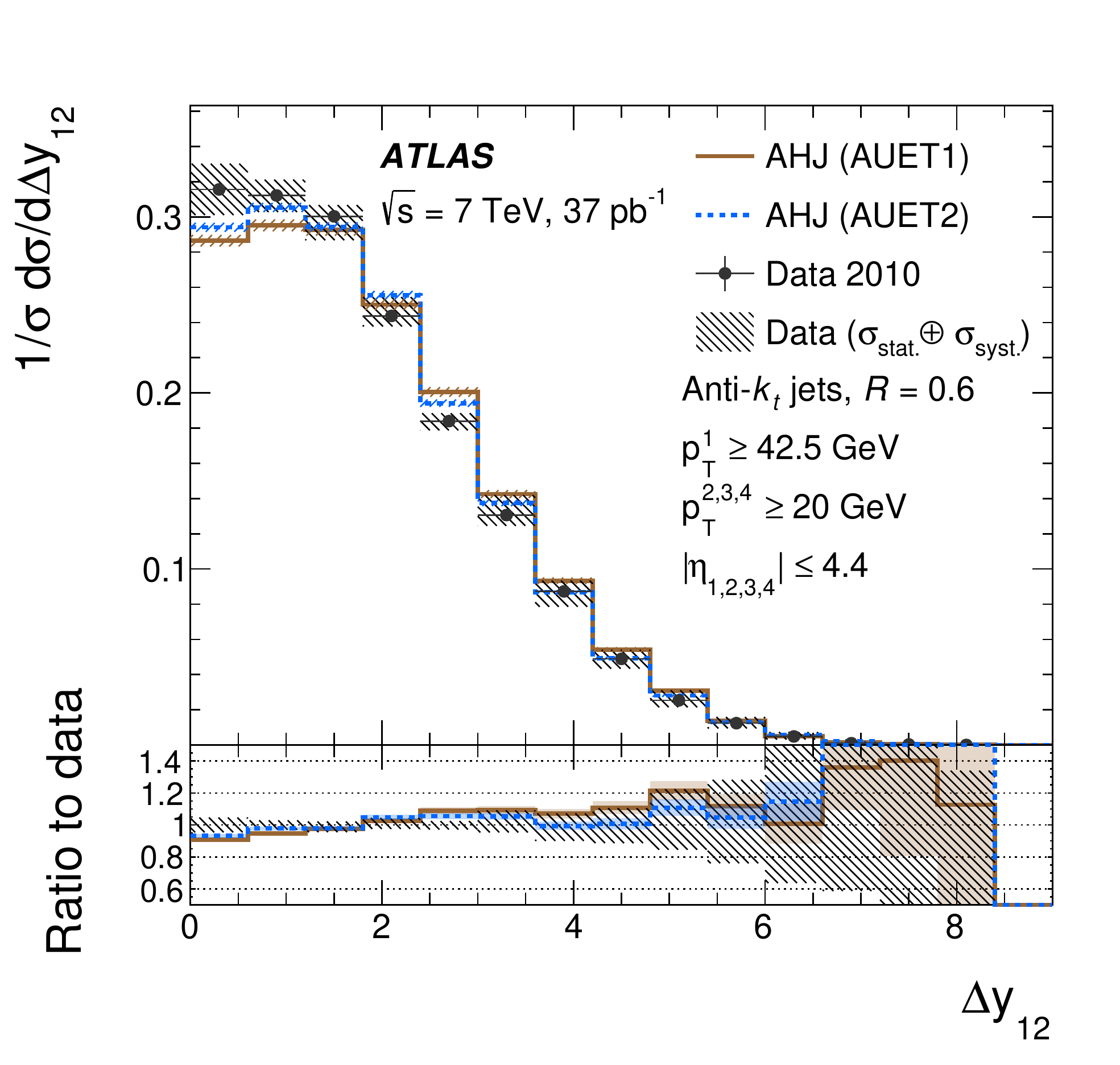}}
\subfloat[]{\label{fig:app:unfoldedDist:dy34}\includegraphics[trim=0mm 0mm 0mm 15mm,clip,width=0.5\textwidth]{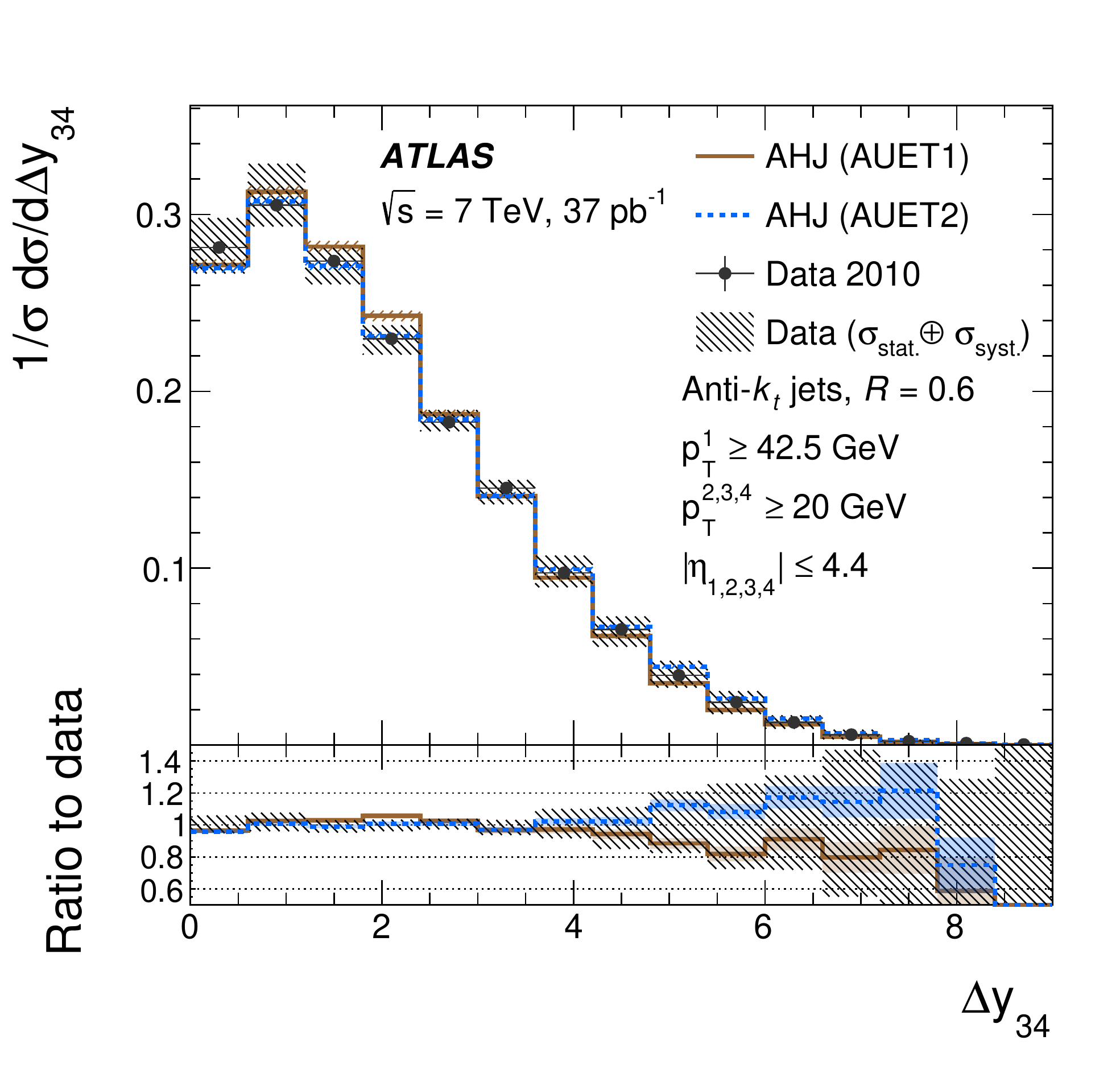}}
\caption{\label{fig:app:unfoldedDist:dphi:13:23}
    Distributions of the variables \Subref{fig:app:unfoldedDist:dphi14} $\dphi{14}$, \Subref{fig:app:unfoldedDist:dphi24} $\dphi{24}$,
    \Subref{fig:app:unfoldedDist:dy12} $\dy{12}$ and \Subref{fig:app:unfoldedDist:dy34} $\dy{34}$,
    defined in \autoref{eq:inputVariablesPCA}, in data after unfolding to particle level, compared to the MC prediction from \ahj 
    at the particle level, generated using the AUET1 and AUET2 tunes, as indicated in the legend.
    The hatched areas represent the sum in quadrature of the statistical and systematic uncertainties in the normalized differential cross-sections and all histograms are normalized to unity.
    The ratio of the particle-level distribution to the normalized differential cross-section is shown in the bottom panels, where the shaded areas represent statistical uncertainties.
}
\end{center}
\end{figure}

\begin{figure}[htbp]
\captionsetup[subfloat]{captionskip=-3ex}
\begin{center}
\subfloat[]{\label{fig:app:unfoldedDist:dy13}\includegraphics[trim=0mm 0mm 0mm 15mm,clip,width=0.5\textwidth]{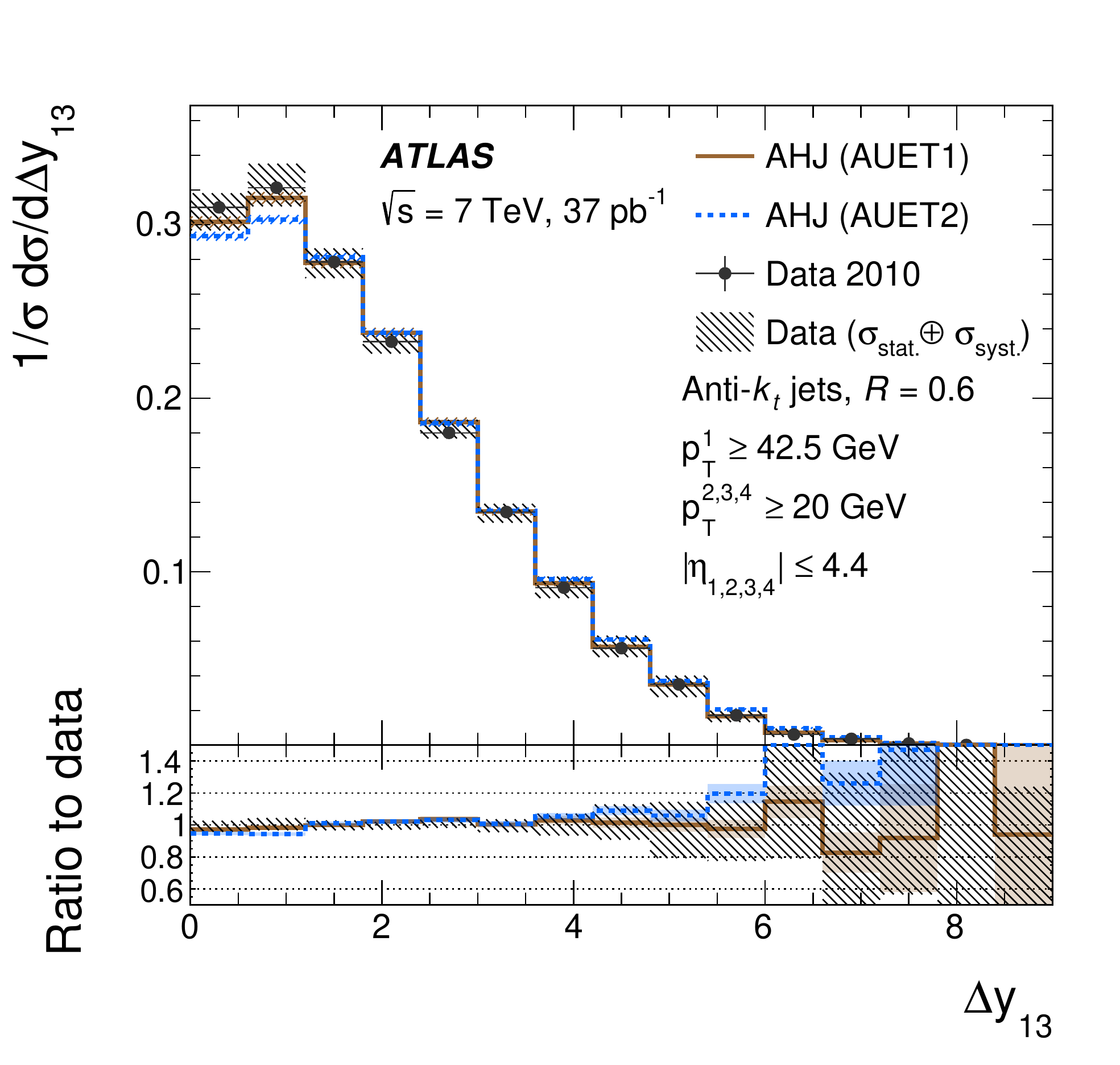}}
\subfloat[]{\label{fig:app:unfoldedDist:dy23}\includegraphics[trim=0mm 0mm 0mm 15mm,clip,width=0.5\textwidth]{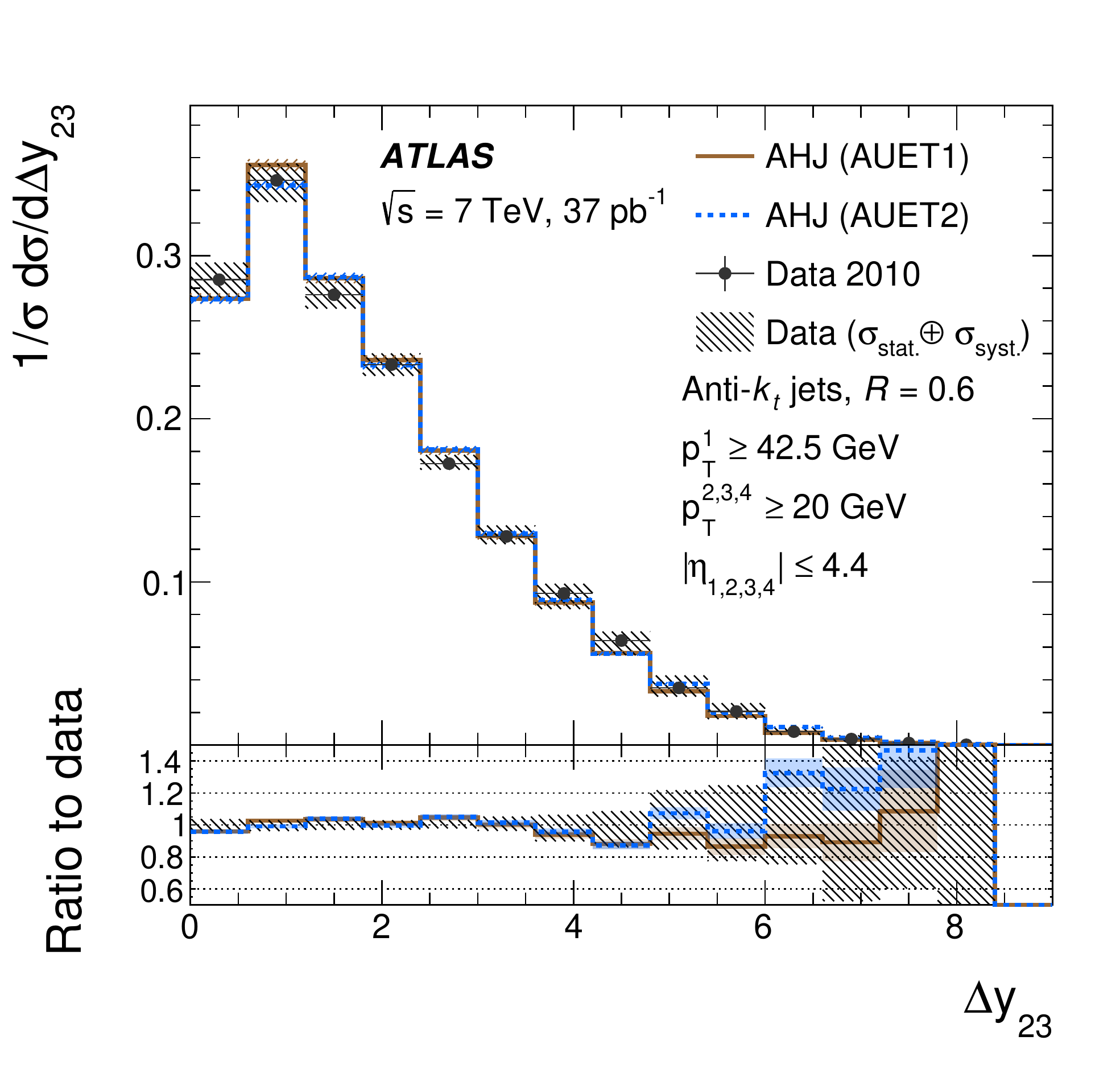}} \\
\subfloat[]{\label{fig:app:unfoldedDist:dy14}\includegraphics[trim=0mm 0mm 0mm 15mm,clip,width=0.5\textwidth]{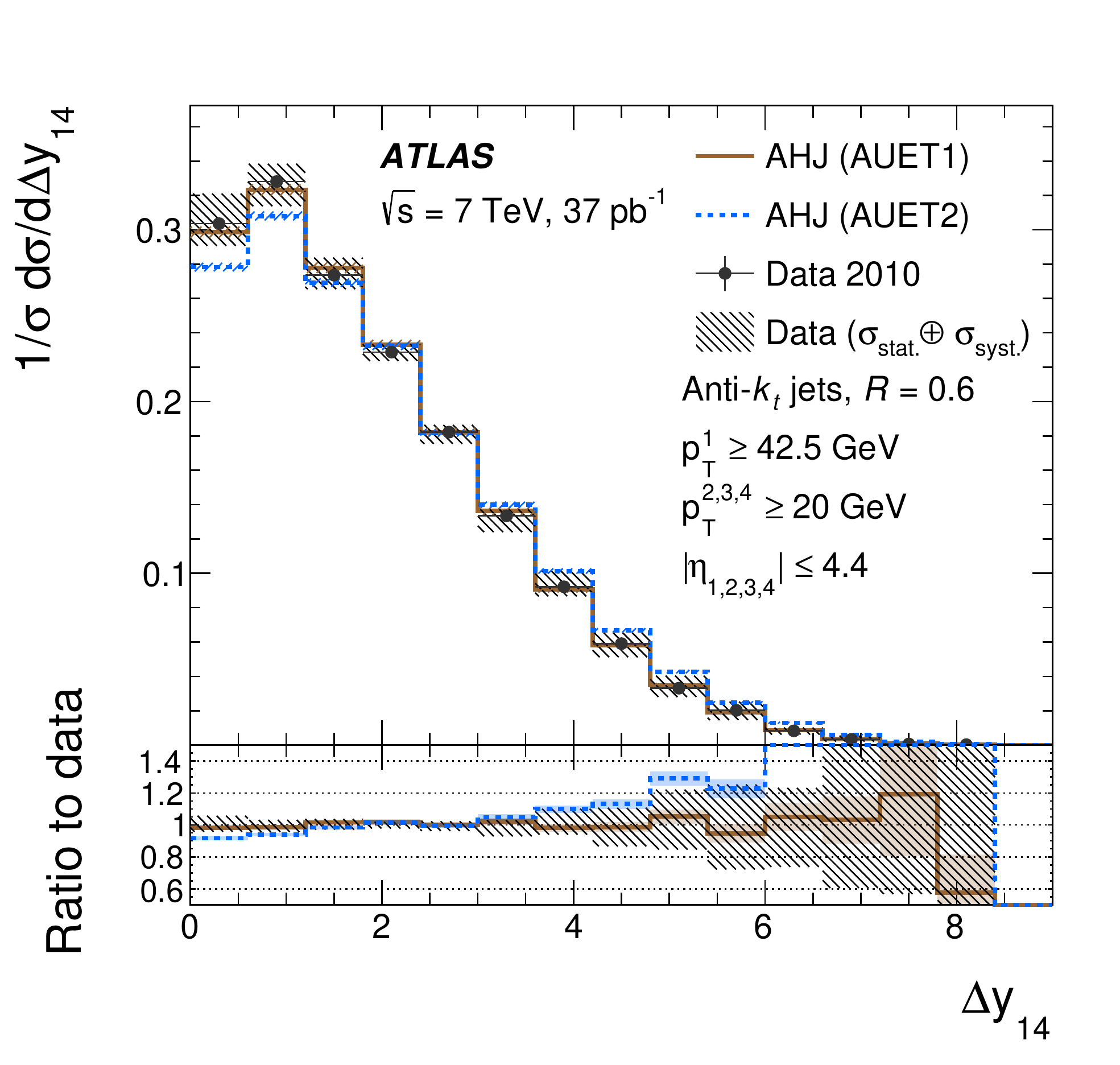}}
\subfloat[]{\label{fig:app:unfoldedDist:dy24}\includegraphics[trim=0mm 0mm 0mm 15mm,clip,width=0.5\textwidth]{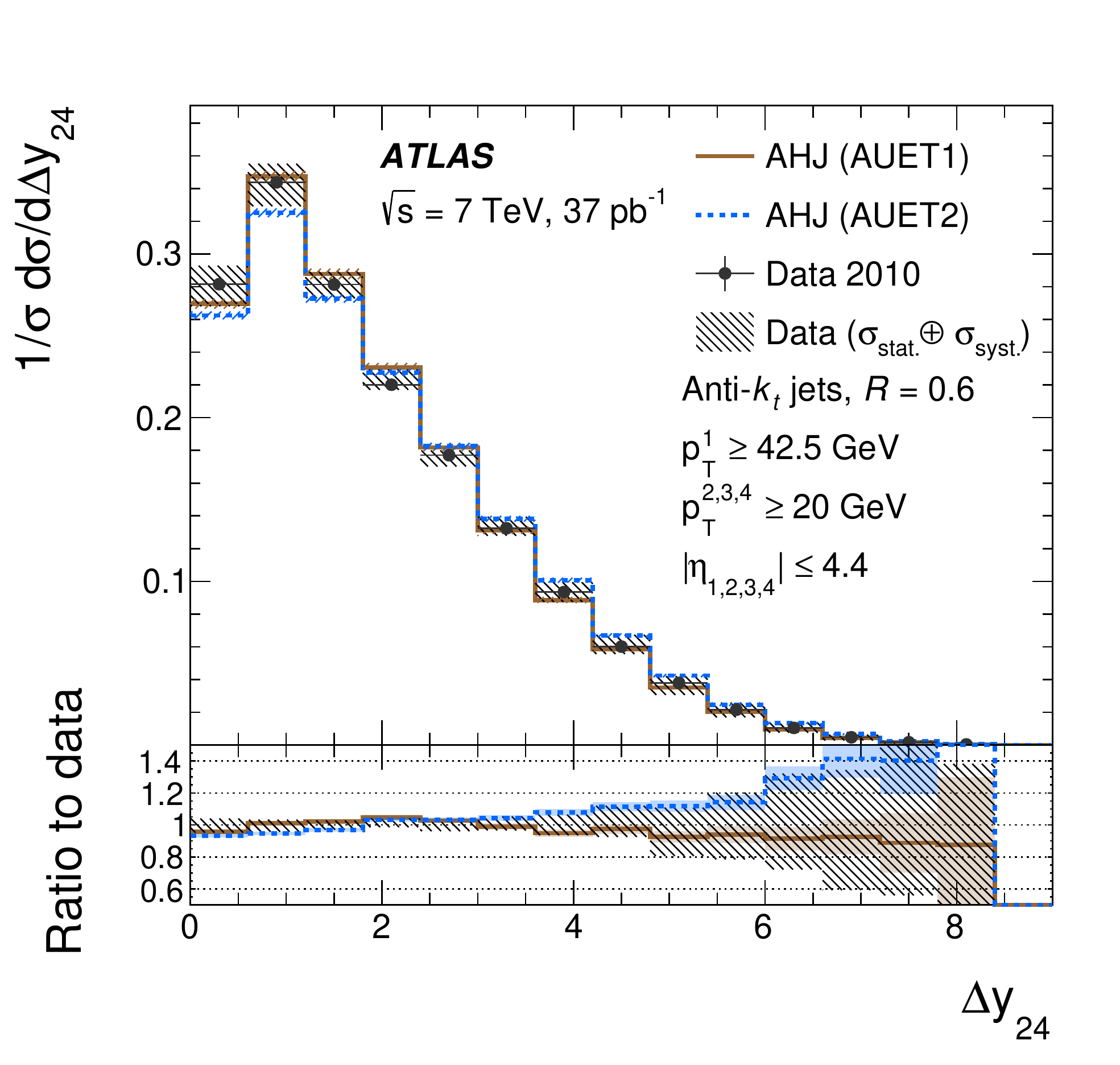}}
\caption{\label{fig:app:unfoldedDist:dy:12:34}
    Distributions of the variables \Subref{fig:app:unfoldedDist:dy13} $\dy{13}$, \Subref{fig:app:unfoldedDist:dy23} $\dy{23}$,
    \Subref{fig:app:unfoldedDist:dy14} $\dy{14}$ and \Subref{fig:app:unfoldedDist:dy24} $\dy{24}$
    defined in \autoref{eq:inputVariablesPCA}, in data after unfolding to particle level, compared to the MC prediction from \ahj 
    at the particle level, generated using the AUET1 and AUET2 tunes, as indicated in the legend.
    The hatched areas represent the sum in quadrature of the statistical and systematic uncertainties in the normalized differential cross-sections and all histograms are normalized to unity.
    The ratio of the particle-level distribution to the normalized differential cross-section is shown in the bottom panels, where the shaded areas represent statistical uncertainties.
}
\end{center}
\end{figure}

\begin{figure}[htbp]
\captionsetup[subfloat]{captionskip=-3ex}
\begin{center}
\subfloat[]{\label{fig:app:unfoldedDist:dPhiPlanes:12}\includegraphics[trim=0mm 0mm 0mm 15mm,clip,width=0.5\textwidth]{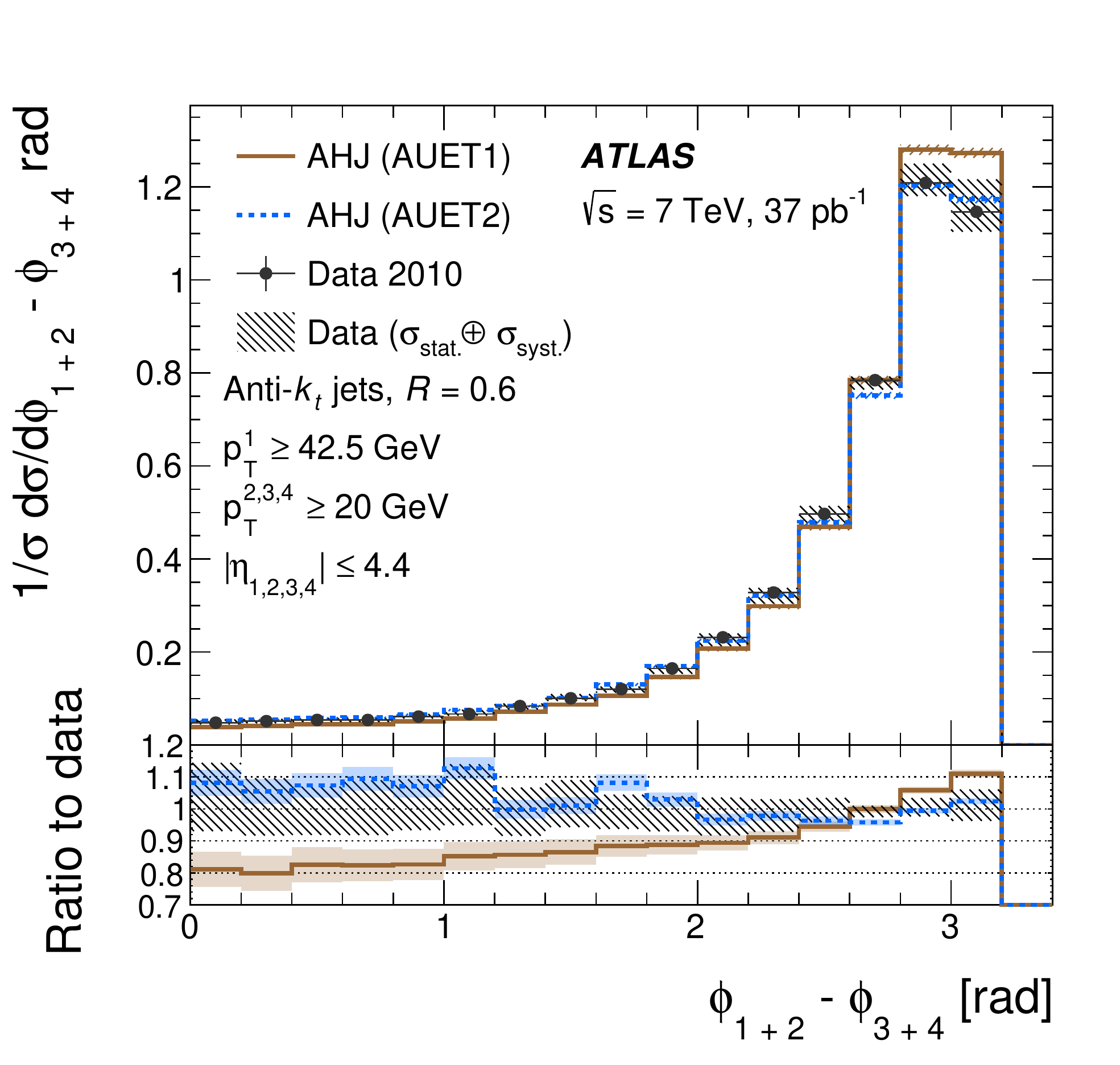}}
\subfloat[]{\label{fig:app:unfoldedDist:dPhiPlanes:13}\includegraphics[trim=0mm 0mm 0mm 15mm,clip,width=0.5\textwidth]{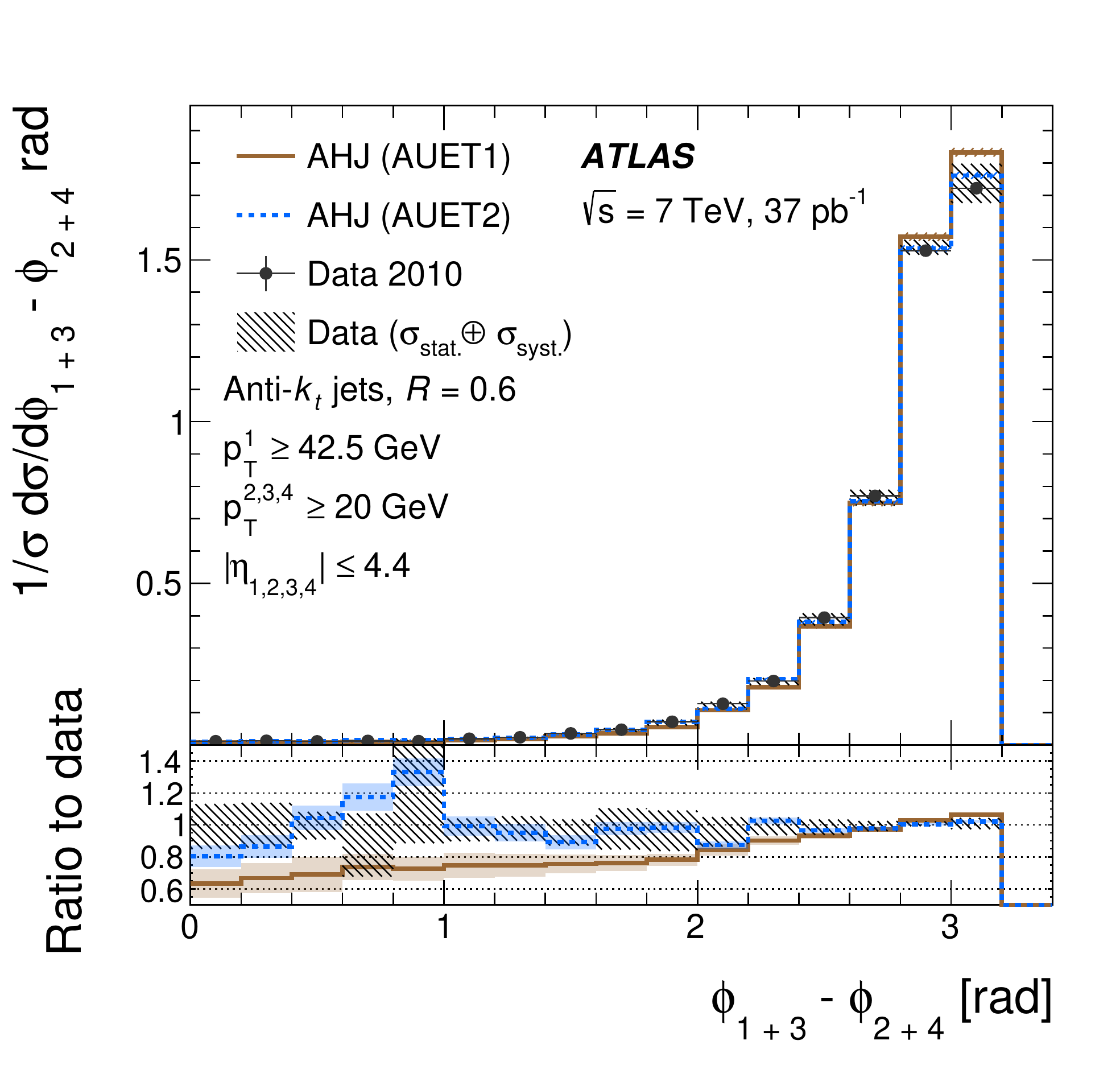}} \\ 
\subfloat[]{\label{fig:app:unfoldedDist:dPhiPlanes:14}\includegraphics[trim=0mm 0mm 0mm 15mm,clip,width=0.5\textwidth]{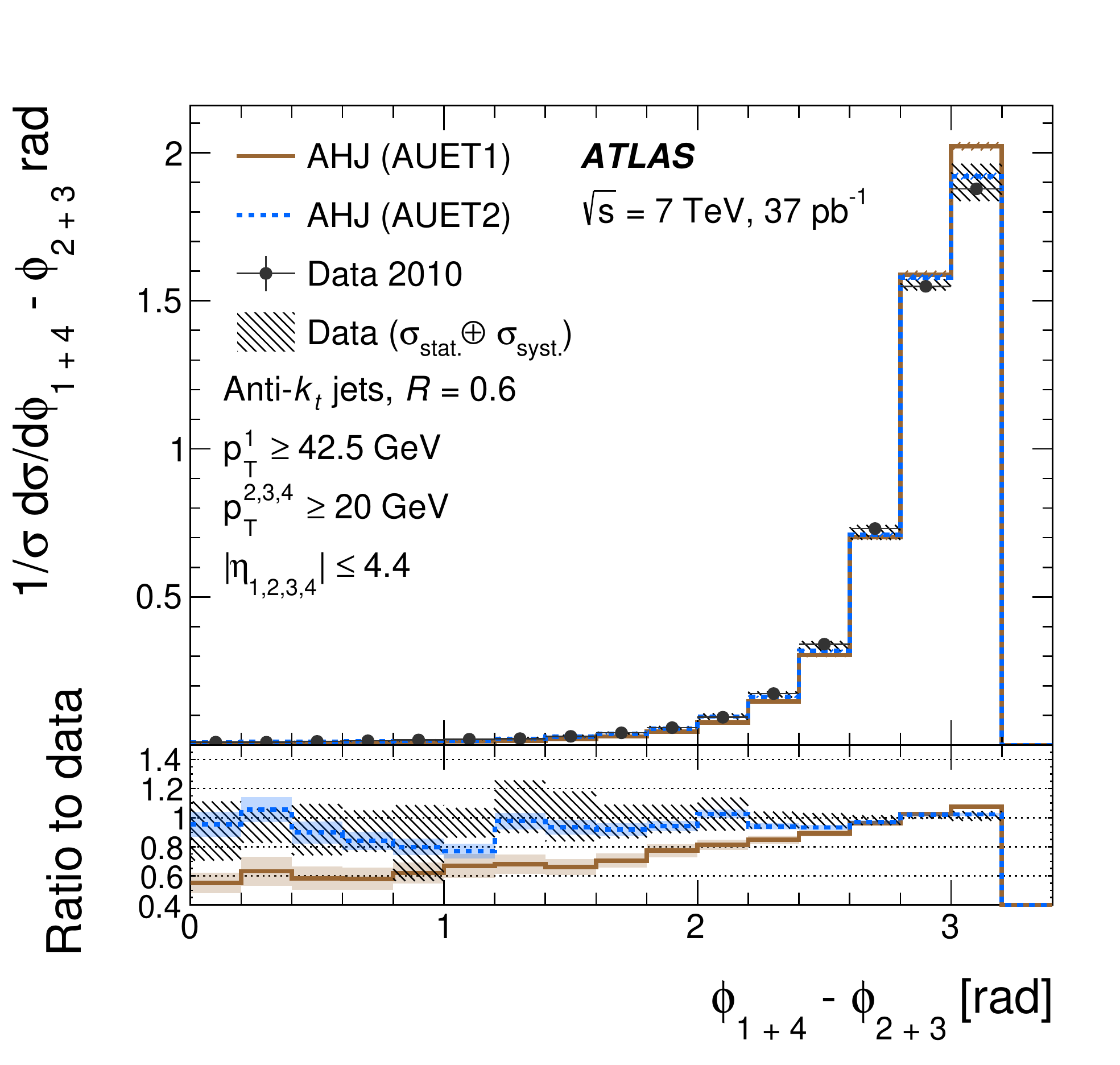}}
\caption{\label{fig:app:unfoldedDist:dy:14:24}
    Distributions of the variables \Subref{fig:app:unfoldedDist:dPhiPlanes:12} \dPhiPlanes{1}{2}{3}{4}, \Subref{fig:app:unfoldedDist:dPhiPlanes:13} \dPhiPlanes{1}{3}{2}{4} 
    and \Subref{fig:app:unfoldedDist:dPhiPlanes:14} \dPhiPlanes{1}{4}{2}{3},
    defined in \autoref{eq:inputVariablesPCA}, in data after unfolding to particle level, compared to the MC prediction from \ahj 
    at the particle level, generated using the AUET1 and AUET2 tunes, as indicated in the legend.
    The hatched areas represent the sum in quadrature of the statistical and systematic uncertainties in the normalized differential cross-sections and all histograms are normalized to unity.
    The ratio of the differential distribution to the particle-level distributions is shown in the bottom panels, where the shaded areas represent statistical uncertainties.
}
\end{center}
\end{figure}

\Autorefs{fig:app:unfoldedDist:dpt:12:34}--\ref{fig:app:unfoldedDist:dy:14:24} show the normalized differential cross-sections in data for all the observables used as input to the NN, 
compared to the particle-level distributions in the \ahj samples generated using the AUET1 and AUET2 tunes.
The hatched areas in the distributions represent the total uncertainty of the normalized differential cross-section, obtained by adding in quadrature the statistical and systematic uncertainties.

%

\bibliographystyle{atlasPackage/bibtex/bst/atlasBibStyleWithTitle}
\bibliography{dps4J,atlasPackage/bibtex/bib/ATLAS,Ackno-Bibtex_2Aug-2016}

\newpage 

\begin{flushleft}
{\Large The ATLAS Collaboration}

\bigskip

M.~Aaboud$^\textrm{\scriptsize 136d}$,
G.~Aad$^\textrm{\scriptsize 87}$,
B.~Abbott$^\textrm{\scriptsize 114}$,
J.~Abdallah$^\textrm{\scriptsize 8}$,
O.~Abdinov$^\textrm{\scriptsize 12}$,
B.~Abeloos$^\textrm{\scriptsize 118}$,
R.~Aben$^\textrm{\scriptsize 108}$,
O.S.~AbouZeid$^\textrm{\scriptsize 138}$,
N.L.~Abraham$^\textrm{\scriptsize 152}$,
H.~Abramowicz$^\textrm{\scriptsize 156}$
% ...
\bigskip
\\
% ...
$^{8}$ Department of Physics, The University of Texas at Arlington, Arlington TX, United States of America\\
% ...
$^{12}$ Institute of Physics, Azerbaijan Academy of Sciences, Baku, Azerbaijan\\
% ...
$^{87}$ CPPM, Aix-Marseille Universit{\'e} and CNRS/IN2P3, Marseille, France\\
% ...
$^{108}$ Nikhef National Institute for Subatomic Physics and University of Amsterdam, Amsterdam, Netherlands\\
% ...
$^{114}$ Homer L. Dodge Department of Physics and Astronomy, University of Oklahoma, Norman OK, United States of America\\
% ...
$^{118}$ LAL, Univ. Paris-Sud, CNRS/IN2P3, Universit{\'e} Paris-Saclay, Orsay, France\\
% ...
$^{136}$ $^{(a)}$ Facult{\'e} des Sciences Ain Chock, R{\'e}seau Universitaire de Physique des Hautes Energies - Universit{\'e} Hassan II, Casablanca; $^{(b)}$ Centre National de l'Energie des Sciences Techniques Nucleaires, Rabat; $^{(c)}$ Facult{\'e} des Sciences Semlalia, Universit{\'e} Cadi Ayyad, LPHEA-Marrakech; $^{(d)}$ Facult{\'e} des Sciences, Universit{\'e} Mohamed Premier and LPTPM, Oujda; $^{(e)}$ Facult{\'e} des sciences, Universit{\'e} Mohammed V, Rabat, Morocco\\
% ...
$^{138}$ Santa Cruz Institute for Particle Physics, University of California Santa Cruz, Santa Cruz CA, United States of America\\
% ...
$^{152}$ Department of Physics and Astronomy, University of Sussex, Brighton, United Kingdom\\
% ...
$^{156}$ Raymond and Beverly Sackler School of Physics and Astronomy, Tel Aviv University, Tel Aviv, Israel\\
% ...
$^{a}$ Also at Department of Physics, King's College London, London, United Kingdom\\
$^{b}$ Also at Institute of Physics, Azerbaijan Academy of Sciences, Baku, Azerbaijan\\
$^{c}$ Also at Novosibirsk State University, Novosibirsk, Russia\\
$^{d}$ Also at TRIUMF, Vancouver BC, Canada\\
$^{e}$ Also at Department of Physics {\&} Astronomy, University of Louisville, Louisville, KY, United States of America\\
% ...
\end{flushleft}

\end{document}